\documentclass[a4paper,useAMS,usenatbib,usegraphicx]{mn2eadapt}

\usepackage[latin1]{inputenc}
\usepackage{amsmath}
\usepackage{amsfonts}
\usepackage{amssymb}
\usepackage[english]{babel}

\usepackage{times}
\usepackage{color}
\usepackage{rotating}
\usepackage[colorlinks=false,dvips]{hyperref}
\hypersetup{pdfborder=0 0 0}
\usepackage{longtable}
\usepackage{afterpage}
\usepackage{textcomp}
\usepackage{multirow}

\newcommand{\etal}{et~al.\ }
\newcommand{\eg}{e.g.\ }
\newcommand{\cf}{cf.\ }
\newcommand{\ie}{i.e.\ }
\newcommand{\Msun}{\ensuremath{\textrm{M}_{\odot}}}

\newcommand{\Lsun}{\ensuremath{\textrm{L}_{\odot}}}
\newcommand{\kms}{\mbox{km\vgv{}s$^{-1}$}}

\newcommand{\OI}{\mbox{O\vgv{}{\sc i}}}

\newcommand{\CII}{\mbox{C\vgv{}{\sc ii}}}
\newcommand{\NaI}{\mbox{Na\vgv{}{\sc i}}}

\newcommand{\SII}{\mbox{S\vgv{}{\sc ii}}}

\newcommand{\SiII}{\mbox{Si\vgv{}{\sc ii}}}
\newcommand{\SiIII}{\mbox{Si\vgv{}{\sc iii}}}

\newcommand{\CrII}{\mbox{Cr\vgv{}{\sc ii}}}
\newcommand{\CrIII}{\mbox{Cr\vgv{}{\sc iii}}}

\newcommand{\FeII}{\mbox{Fe\vgv{}{\sc ii}}}
\newcommand{\FeIII}{\mbox{Fe\vgv{}{\sc iii}}}
\newcommand{\CoII}{\mbox{Co\vgv{}{\sc ii}}}
\newcommand{\CoIII}{\mbox{Co\vgv{}{\sc iii}}}

\newcommand{\Fefs}{$^{56}$Fe}
\newcommand{\Cofs}{$^{56}$Co}
\newcommand{\Nifs}{$^{56}$Ni}

\newcommand{\phao}{$\phantom{\textrm{1}}$}

\newcommand{\quadbynine}{\ensuremath{\!\;}}

\newcommand{\verticalonehalf}{\!\!\!\protect\begin{array}{c} \scriptstyle \textrm{1} \protect\vspace{-0.22cm} \\ - \protect\vspace{-0.23cm} \\ \scriptstyle \textrm{2} \protect\vspace{0.08cm} \protect\end{array}\!\!\!}
\newcommand{\vgv}{\protect\hspace{0.25em}}
\newcommand{\myto}{\hspace{0.18em}--\hspace{0.18em}}

\definecolor{myredder}{rgb}{1.0,0.0,0.0}

\definecolor{myorange}{rgb}{0.75,0.55,0.0}

\begin{document}

\title[Spectroscopic analysis of SN~2009dc]{Spectral modelling of the ``Super-Chandra'' Type Ia SN~2009dc -- \\testing a 2\vgv\Msun\ white dwarf explosion model and alternatives}
\author[Hachinger et al.]{Stephan Hachinger$^{1,2}$, Paolo A. Mazzali$^{1,2}$, Stefan Taubenberger$^{2}$, Michael Fink$^{3,2}$
\protect\vspace{0.2cm}\\{\upshape\LARGE R\"udiger Pakmor$^{4,2}$, Wolfgang Hillebrandt$^{2}$, Ivo R. Seitenzahl$^{3,2}$ }\\
$^1$Istituto Nazionale di Astrofisica-OAPd, vicolo dell'Osservatorio 5, 35122 Padova, Italy\\
$^2$Max-Planck-Institut f\"ur Astrophysik, Karl-Schwarzschild-Str.\ 1, 85748 Garching, Germany\\
$^3$Universit\"at W\"urzburg, Emil-Fischer-Str. 31, 97074 W\"urzburg, Germany\\
$^4$Heidelberg Institute for Theoretical Studies, Schloss-Wolfsbrunnenweg 35, 69118 Heidelberg, Germany}
\date{arXiv v2, 2013 Jan 31. The published version is available at the MNRAS web site.}

\pubyear{2012}
\volume{}
\pagerange{}

\maketitle

\begin{abstract}
Extremely luminous, ``super-Chandrasekhar'' (SC) Type Ia Supernovae (SNe Ia) are as yet an unexplained phenomenon. We analyse a well-observed SN of this class, SN~2009dc, by modelling its photospheric spectra with a spectral synthesis code, using the technique of ``Abundance Tomography''. We present spectral models based on different density profiles, corresponding to different explosion scenarios, and discuss their consistency. First, we use a density structure of a simulated explosion of a 2\vgv\Msun\ rotating C-O white dwarf, which is often proposed as a possibility to explain SC SNe Ia. Then, we test a density profile empirically inferred from the evolution of line velocities (blueshifts). This model may be interpreted as a core-collapse SN with an ejecta mass of  $\sim$\hspace{0.1em}3\vgv\Msun. Finally, we calculate spectra assuming an ``interaction scenario''. In such a scenario, SN~2009dc would be a standard white dwarf (WD) explosion with a normal intrinsic luminosity, and this luminosity would be augmented by interaction of the ejecta with a H-/He-poor circumstellar medium. We find that none of the models tested easily explains SN~2009dc. With the 2\vgv\Msun\ WD model, our abundance analysis predicts small amounts of burning products in the intermediate-/high-velocity part of the ejecta ($v$\vgv{}$\gtrsim$\vgv{}9000\vgv{}\kms). However, in the original explosion simulations, where the nuclear energy release per unit mass is large, burned material is present at high velocities. This contradiction can only be resolved if asymmetries strongly affect the radiative transfer or if C-O WDs with masses significantly above 2\vgv\Msun\ exist. In a core-collapse scenario, low velocities of Fe-group elements are expected, but the abundance stratification in SN~2009dc seems ``SN-Ia-like''. The interaction-based model looks promising, and we have some speculations on possible progenitor configurations. However, radiation-hydrodynamics simulations will be needed to judge whether this scenario is realistic at all.
\end{abstract}


\section{Introduction}
\label{sec:introduction}

Type Ia supernovae (SNe Ia) are subject to vigorous research owing to their relevance to different fields of astrophysics. Their production via different paths of binary evolution is investigated in detail \citep[\eg][]{rui09,claeys10,podsiadlowski10a}, as is their impact on galaxies \citep[\eg][]{sca08,tang09}. Most prominently they are used today in cosmological studies \citep[\eg][]{per97,rie98,per99,ast06,woo07} to measure luminosity distances. SNe Ia show differences in their intrinsic luminosity. Observations have, however, shown that the luminosity of each object can be estimated from its light curve shape (\eg \citealt{phi93,phi99}), which is measurable independently of distance. Empirical calibration procedures exploit this, allowing for the construction of accurate SN Hubble diagrams.

The fact that SNe Ia constitute a relatively homogeneous SN subclass is usually explained by the homogeneity of their progenitor systems, which are supposedly carbon-oxygen white dwarfs (C-O WDs) in binary systems. In the ``classical'' single-degenerate Chandrasekhar-mass scenario, one WD explodes after having accreted matter from a companion star. A thermonuclear runaway occurs when the WD is close to the stability limit \citep{hil00}, \ie its Chandrasekhar mass ($M_\textrm{Ch,non-rot}$\vgv{}$\sim$\vgv{}1.4\vgv$\Msun$ without rotation). Numerical simulations of such explosions have been successful in explaining the ``leading-order'' homogeneity in the observations, and also part of the diversity among the objects \citep{maz07,kas09}. The most important variation in the properties of SNe Ia is caused by a different efficiency in producing \Nifs, which powers the optical light curve through the radioactive decays \Nifs\ $\rightarrow$ \Cofs\ $\rightarrow$ \Fefs.

Despite the success of standard explosion models, peculiar SNe Ia, discovered in increasing numbers (\eg \citealt{fil92bg,phi92,Li2003a,how06}), are difficult to explain thus far. Meanwhile, the precision of cosmological studies based on SNe Ia has significantly increased. In this situation, we have to clarify what mechanisms can produce SNe Ia in order to avoid errors in SN cosmology which could be caused by populations of peculiar SNe \citep[\cf \eg][]{fol10bt}. Within the ``classical'' single-degenerate Chandrasekhar-mass scenario, there is the possibility of having an ``supermassive'' white dwarf as progenitor, which is stabilised by rotation (\eg \citealt{steinmetz92,yoo05,how06,pfa10a,fink10diss,hachisu12a}, Fink \etal, in prep.). Also other explosion scenarios are re-considered to produce peculiar and even normal SNe Ia -- most prominently, the double-degenerate/WD-merger scenario \citep{pak10a} and the single-degenerate sub-Chandrasekhar scenario \citep{fink07a,fink10a,sim10}, where a He shell around the WD detonates and triggers the explosion of the star. Furthermore, there are relatively new, somewhat ``exotic'' SN~Ia models (\eg \citealt{bra09}).

On the observational side, one prominent SN Ia subclass waiting to be explained are the ``super-Chandrasekhar mass'' objects [SC SNe Ia -- \citet[][]{how06,hic07,sca10}, and references below; the designation refers to the Chandrasekhar mass for non-rotating WDs]. The early-time spectra of these SNe resemble those of other SNe~Ia while their luminosity is about twice the average \citep{how06}. According to Arnett's rule \citep{arn82}, \Nifs\ masses of more than a Chandrasekhar mass are required to explain the luminosity of some of these objects. A luminosity prediction from the light-curve shape, using the relation of \citet{phi99}, fails at least for some SC SNe \citep{tau11}. Possibly, part of the luminosity of SC SNe comes from energy sources other than \Nifs\ decay, such as interaction of the ejecta with circumstellar material. The more conventional paradigm, however, is that SC SNe Ia are more massive than normal objects (possibly the result of a rotating progenitor WD), and produce more \Nifs\ than a Chandrasekhar-mass explosion could do \citep{how06}. In any case, in-depth analyses of observations with radiative transfer models are needed to better understand SC objects.

In this work, we study the recent SC SN~2009dc, which reached a peak $M_\textrm{bol} \sim -\textrm{20}$ and exploded in a tidal bridge between two galaxies (UGC~10063/64, types S0/SBd). Observations with an extraordinarily good time coverage are available for this SN \citep{yam09,tan10,sil11,tau11}. At maximum light, SN~2009dc has spectral properties similar to SN~2003fg \citep{how06}, and therefore is representative at least for part of the SC objects. Other SC SNe are different to some degree. For example, SN~2007if \citep{sca10} shows higher-temperature spectra and somewhat higher line velocities (as also SN~2006gz, \citealt{hic07}). The reasons for this diversity are unclear, as is the theoretical interpretation of SC SNe. We use a radiative-transfer code to model the observed spectra of SN~2009dc, inferring ejecta densities and abundances. The method we employ (Abundance Tomography, \citealt{ste05}) has proved suitable for inferring the properties of quite diverse SNe (\eg \citealt{tan11}, \citealt{maz08eo}, \citealt{hachinger09}) from data taken in the photospheric phase, when the core of the ejecta is opaque to optical radiation. Assuming different density profiles and rise times, we infer a best-matching abundance structure, respectively, by fitting the observations with synthetic spectra. Comparing models based on different assumptions, we assess the applicability of different explosion scenarios by judging the fit quality and the physical consistency of the models.

Below (Sec.~\ref{sec:prerequisites}), we first give a relatively extended overview of our methodology as it is relevant to the present work. We then show a tomography model based on a density profile representing the explosion of a $\textrm{2.0}$\vgv$\Msun$ rotating white dwarf, and another model based on an empirically constructed $\textrm{3.0}$\vgv$\Msun$ density profile (Sections~\ref{sec:tomography-awd3det} and \ref{sec:tomography-exp}). Finally, we present a third tomography experiment with a more speculative set-up: we model the spectra assuming that the SN is a Chandrasekhar-mass explosion (of a non-rotating WD) where some light is ``externally'' generated through interaction of the ejecta with a surrounding medium (Sec.~\ref{sec:tomography-csm}). We discuss all our models in terms of consistency and theoretical interpretation (Sec.~\ref{sec:discussion}) before we conclude (Sec.~\ref{sec:conclusions}).

\section{Method and data}
\label{sec:prerequisites}

\subsection{Radiative-transfer code}
\label{sec:rt}

We use a 1D stationary Monte-Carlo radiative-transfer code (\citealt{abb85}, \citealt{maz93}, \citealt{luc99}, \citealt{maz00} and \citealt{ste05}) to calculate synthetic SN spectra from a given density and abundance profile.

The code computes the radiative transfer from an assumed photosphere through the SN ejecta at a given epoch. An initial density profile is used to calculate the densities within the ejecta, assuming homologous, force-free expansion. This expansion phase commences shortly after explosion in most SN models [\eg \citet{roepke05b}; \cf however Sec. \ref{sec:tomography-csm-density}], so that \vgv{} $r$\vgv{}$=$\vgv{}$v$\vgv{}$\times$\vgv{}$t$ \vgv{} holds for each particle at all times relevant to this work in good approximation ($r$: distance from the centre, $t$: time from explosion, $v$: velocity at the onset of homologous expansion). Radius and velocity can then be used interchangeably as spatial coordinates.

From the photosphere, which is located at an adjustable $v_{\textrm{ph}}$, continuous black body radiation [$I_{\nu}^{+}$\vgv{}$=$\vgv{}$B_{\nu}$$(T_{\textrm{ph}})$, where $I_{\nu}^{+}$ is the specific intensity in outward directions, and $B_{\nu}$$(T_{\textrm{ph}})$ the specific intensity according to the Planck law at temperature $T_\textrm{ph}$] is assumed to be emitted into the atmosphere. Notable deviations from this approximation to the flux in the inner layers appear in the red and infrared, where a discrepancy in the flux level of synthetic to observed spectra therefore is sometimes unavoidable.

The photon packets simulated undergo Thomson scattering and interactions with lines, which are treated in the Sobolev approximation. A downward branching scheme ensures a good approximation to the bound-bound emissivity. Radiative equilibrium is enforced by construction \citep{luc99} of the Monte-Carlo simulation. The excitation and ionisation state of the matter is calculated from the radiation field statistics using a modified nebular approximation \citep{maz93,maz00}. In this approximation, the gas state in each radial grid cell is mostly determined from a radiation temperature $T_\textrm{R}$ and a dilution factor $W$. $T_\textrm{R}$ corresponds to the mean frequency of the radiation field and $W$ parametrises its energy density (for given $T_\textrm{R}$). We iterate excitation, ionisation and the radiation field in turn until the $T_\textrm{R}$ values within the atmosphere are converged to the per cent level. Within these iterations, $T_\textrm{ph}$ is automatically adjusted in order to match a given bolometric output luminosity $L_\textrm{bol}$. Thus, the photospheric luminosity is adapted so as to compensate for reabsorption of radiation, which occurs when packets re-enter the photosphere. After the temperature iterations, the emergent spectrum is calculated by formal integration of the transfer equation \citep{luc99}, using a source function derived from the Monte Carlo statistics.

\subsection{Spectral models}

To summarise, a ``SN model'' from which our code calculates a synthetic spectrum is specified by the density profile, the abundance stratification, the luminosity $L_\textrm{bol}$, the photospheric velocity $v_\textrm{ph}$ and the time from explosion $t$. 

Spectral modelling means adjusting this model to obtain the best possible match of synthetic and observed spectrum; the  model parameters will then represent the SN properties. Usually, a specific density profile and rise time is assumed before starting to fit the spectra. Below, we briefly discuss how the various remaining parameters influence the radiative transfer and the spectra, and we describe our general strategy for fitting the observations, including the tomography technique.

\subsubsection{Abundances and their imprint on the spectrum}

The part of the abundance structure relevant for the SN spectrum varies with the epoch (see also Sec. \ref{sec:tomography-descr}). For each model, only the abundances above the photosphere (\ie at $v>v_\textrm{ph}$) are relevant. 

The abundances of intermediate-mass elements (IMEs, \eg Mg, Si, S, Ca) normally influence the strength of individual spectral features. In contrast, most Fe-group elements do not leave distinct features in the optical spectra. They rather influence the spectrum by modulating the line blocking in the UV and by a ``transfer'' of flux to the red [via branching, see \citet{maz00}]. The line blocking effect has the consequence of backwarming within the atmosphere (steepening of the temperature profile), exerting an indirect influence on line strengths of all elements via the ionisation and excitation state of the plasma. This state is not fixed by input parameters in our approach, but is determined such that it is consistent with the radiation field.

\subsubsection{Influence of $L_\textrm{bol}$ and $v_\textrm{ph}$}

$L_\textrm{bol}$ is the parameter controlling the flux level in the synthetic spectrum. For each epoch, it is adapted during the fitting process so as to match the overall flux in the optical part of the observed (and flux-calibrated) spectrum. Therefore, it is  easily constrained to a reasonable precision.

The photospheric velocity $v_\textrm{ph}$ has various, more complicated effects on the model. First, it sets a lower limit to the radii at which line absorption can occur in the simulation. Second, it determines the integrated optical depth of the atmosphere, which should be appropriate ($\tau \gtrsim \frac{2}{3}$ in the relevant wavelength range) even if the notion of a sharp photosphere emitting a black body is quite approximative in Type~I SNe \citep{sau06rf}. For our simulations, it is important that the photosphere is assumed deep enough such that the assumption of a blackbody-like emission at that depth is satisfied as well as possible. A very low $v_\textrm{ph}$, however, generally leads to a very blue radiation field at the lower boundary, as the emitting surface shrinks. When this is not fully compensated by attenuation above the photosphere, the model spectrum will be bluer. Within the freedom these constraints give us (roughly $\pm$\vgv{}1000\vgv{}\kms), we choose $v_\textrm{ph}$ such that we obtain an optimum fit.

During the fitting process, we assess and take into account additional effects of $L_\textrm{bol}$ and $v_\textrm{ph}$ on the strength of spectral lines via the plasma state. A lower $v_\textrm{ph}$ tends to make the radiation field bluer at least in parts of the atmosphere (\cf above). This results in higher radiation temperatures and an increased ionisation (and occupation of excited levels). A higher luminosity $L_\textrm{bol}$ makes the radiation field stronger and bluer, with the same result.

\subsubsection{Time $t$ -- sensitivity of the model on the rise time}
\label{sec:method-spectralmodels-timesensitivity}

The time from explosion $t$ must be determined from the respective observational epoch [usually given relative to $B$-band maximum ($B$ max.)] and the assumed\footnote{The rise time $t_\textrm{r}$ can only to some degree be constrained from the available early light curve data [\cf \citet{tau11} and Sec. \ref{sec:observations}].} rise time $t_\textrm{r}$, in which the SN rises to $B$ max. after explosion.

When the density distribution in velocity space is kept constant\footnote{Here, we mean that the time-invariant density distribution $\frac{\mathrm{d}m}{\mathrm{d}v}(v)$, \ie the explosion model, is kept constant.}, and $t_\textrm{r}$ is changed, the spectrum shows an overall shift of line velocities (mean blueshifts) for the following reasons:

\begin{list}{$\bullet \ \ \ \ $}{\setlength{\labelwidth}{3.0cm} \setlength{\leftmargin}{0.65cm}}
\item Models with a smaller $t$ (smaller $t_\textrm{r}$) usually have photospheres at higher $v_\textrm{ph}$. With the more compact ejecta, the higher $v_\textrm{ph}$ is necessary in order to avoid excessive backscattering and a too blue spectrum of the outgoing radiation at the lower boundary (when a high luminosity is emitted from a small photosphere). The higher $v_\textrm{ph}$ then leads to some increase of line velocities.\vspace{0.15cm}
\item The ejecta are more compact for a smaller $t_r$, since \linebreak $\,$$r$\vgv{}$=$\vgv{}$v$\vgv{}$\times$\vgv{}$t$$\,$, where $t$ is the sum of $t_r$ and the epoch (relative to $B$ maximum). Therefore, for a smaller $t_r$ the energy density of the radiation field inside the ejecta tends to be larger at any given velocity $v$ (with $L_\textrm{bol}$ kept constant). The zone where ionisation is low enough for line formation (\ie triply and especially doubly ionised species dominate) will be located at higher $v$. This then reflects in higher line velocities.
\end{list}

When reducing $t_\textrm{r}$ below realistic values, line formation would have to occur in the outermost zones of the ejecta. The low densities (and the strong ionisation due to the low densities) in these zones however disfavour the formation of prominent features such as \SiII\ $\lambda$6355. As a result, the spectra of an observed object can not be explained if too low a rise time is assumed.

\subsubsection{Fitting observed spectra}

The process of spectral fitting begins with choosing a density profile and a rise time, which subsequently remain unchanged. We then iteratively determine optimum values of those abundances which influence features on the one hand, and parameters that control the temperature structure and the line velocities on the other. We try to match the spectral features as well as the overall flux, but if this is not possible simultaneously, we prefer a better fit to the lines. In the present case, we sometimes allow the synthetic spectra to be too red if otherwise the model atmospheres would become too hot.

\subsubsection{Abundance Tomography}
\label{sec:tomography-descr}

The method of Abundance Tomography \citep{ste05} extends the approach of fitting one spectrum \citep{maz93}. It aims at inferring the abundance profile of a SN step by step from the outer to the inner layers (assuming a density structure and a rise time). To this end, one fits a time series of photospheric spectra, in which different layers of the ejecta leave their imprint. In the photospheric phase, some core of the ejecta is optically thick at UV and optical wavelengths. A spectrum is then only sensitive to the optically thin part of the ejecta. As time progresses, a larger and larger fraction of the ejecta becomes visible. Therefore, later spectra are sensitive to more central layers, while the influence of the outer layers on them is smaller.

In order to create a model for the outermost ejecta of an object, we fit the earliest spectrum available. The photospheric velocity, the luminosity and the abundances are optimised. In the case of SN 2009dc, we use three abundance zones above the earliest photosphere in order to have composition gradients in the outer ejecta, allowing for an optimum fit (\cf Sec. \ref{sec:tomography-awd3det-rt}).

The spectrum at the next epoch carries the imprint of the material in the outer envelope and additionally that of the layers which the photosphere has receded into in the meantime. As the outer abundance structure is already constrained, the new photospheric velocity and the abundances of the newly visible layers are now the primary fitting parameters. If the abundances of the outer layers have a significant impact on the quality of the fit, and no decent fit is possible with the values inferred earlier, those values are revised (normally by less than $\textrm{30}$\% of their initial value) to equally optimise the fit to all observations. The procedure is continued with later spectra until the complete time series is fitted.

Degeneracies, \ie equally good fits with different parameters, are the most important source of uncertainty in the fitting process. Abundances not well constrained by an early spectrum can sometimes be determined when fitting a later spectrum. The respective values may then undergo significant revision ($> \textrm{30}$\%). In a few cases, this a posteriori determination does not work. Then, we assume the unknown abundances to match typical nucleosynthesis patterns within SN Ia explosion models \citep{iwa99}. Most importantly, when Fe-group elements do not lead to the formation of individual features, the mix of Fe-group elements is set so as to resemble that in the outer, intermediate or inner zone of the \citet{iwa99} models as appropriate. Furthermore, in the intermediate zone of the ejecta, Si has been assumed to be the most abundant element, even when these zones contributed little to the formation of the spectral features.

\subsection{Study of SN~2009dc: concept}
\label{sec:modellingstrategy}

SN~2009dc is in many respects different to other SNe which Abundance Tomography has been applied to. In particular, as it is unknown which explosion model applies, the density profile and the rise time $t_\textrm{r}$ constitute major elements of uncertainty in our analysis. 

We reflect this in our modelling approach: we perform Abundance Tomography three times, using different density profiles which represent different explosion types. The aim is to compare the different models in the end and to assess their consistency. Before conducting our final tomography, respectively, we decide on an optimum rise time for each density profile.

\subsection{Observations of SN~2009dc}
\label{sec:observations}

We analyse the luminosity-calibrated spectra of SN~2009dc of \citet{tau11}. This data set samples SN~2009dc extremely well from 9.4\vgv{}d before $B$ maximum to late times. We do not model spectra later than 36.4\vgv{}d past $B$ maximum. At this epoch -- late for a SN Ia -- the photosphere recedes deep into the \Nifs-rich zone and our assumption of radiative equilibrium in the simulated atmosphere eventually breaks down.

As the spectra only show a slow evolution of line velocities, we expect a very moderate photospheric recession for all our models (which is confirmed by our actual calculations). This has an impact on the tomography method: if the abundance zone between two subsequent photospheres is too thin, the zone has little influence on the spectrum, which implies a large uncertainty in the abundance determination. In order to avoid this, we only model the spectra at $-\textrm{9.4}$\vgv{}d, $-\textrm{3.7}$\vgv{}d, $+\textrm{4.5}$\vgv{}d, $+\textrm{12.5}$\vgv{}d, $+\textrm{22.6}$\vgv{}d and $+\textrm{36.4}$\vgv{}d with respect to $B$ maximum, leaving out the other ones in between. 

Following \citet{tau11}, we generally assume an extinction towards SN~2009dc of
\begin{equation*}
E(B-V)_\textrm{tot} \sim E(B-V)_\textrm{host}+E(B-V)_\textrm{Gal} \sim \textrm{0.17\vgv{}mag}.
\end{equation*}
Because of the relatively short distance to the SN, we can add up the reddening values without taking redshift into account here. The Galactic extinction taken into account is \mbox{$E(B$\vgv{}$-$\vgv{}$V)_\textrm{Gal}$\vgv{}$=$\vgv{}0.07\vgv{}mag} \citep{sch98}, while \citet{bur82} give a lower value $E(B$\vgv{}$-$\vgv{}$V)_\textrm{Gal}$\vgv{}$=$\vgv{}0.04\vgv{}mag. The reddening due to the host galaxy carries an error of 0.08\vgv{}mag \citep{tau11}. Therefore, the lower-limit reddening within the error bars is 
\begin{equation*}
E(B-V)_\textrm{tot} \geq \textrm{0.06\vgv{}mag}.
\end{equation*}
While we generally use the standard value of 0.17\vgv{}mag, we repeat some analyses for a value of 0.06\vgv{}mag to test the influence on the results.

The rise time of SN~2009dc has been observationally estimated to be shorter than $\sim$\hspace{0.1em}$\textrm{30}$\vgv{}d and longer than $\sim$\hspace{0.1em}$\textrm{22}$\vgv{}d [we follow \citet{tau11} here who give a larger, more conservatively estimated range than \citet{sil11}].

\section{Tomography assuming an explosion of a supermassive WD}
\label{sec:tomography-awd3det}

Our first set of models is based on a density profile representative of an explosion of a rotating WD: we adopt the profile resulting from the ``AWD3det'' simulation of Fink \etal [in prep.; see also \citet{fink10diss}], who re-calculated the ``AWD3 detonation'' model of \citet{pfa10b} with improved detonation physics. The AWD3det model can be regarded prototypical for explosions of rotating, massive ($\textrm{2}$\vgv$\Msun$) white dwarfs. The thermonuclear flame in AWD3det propagates as a pure detonation (shock-driven, supersonic combustion wave) from the very beginning. A pure deflagration (subsonic combustion flame) in a massive, rotating WD would be less likely to explain SN~2009dc, as it does not produce much \Nifs\ (\mbox{$\sim$\hspace{0.1em}$\textrm{0.7}$\vgv$\Msun$}), and leads to a strongly mixed composition \citep{pfa10a}. Delayed-detonation models for these progenitors, in which the flame initially proceeds as a deflagration and later as a detonation (\cf \citealt{kho91}), produce roughly similar outcomes as AWD3det, as the initial deflagration in a rotating WD tends to be weak and burns little material \citep{fink10diss}. 

\begin{figure}
   \centering
   \vspace{0.15cm}
   \includegraphics[angle=270,width=7.7cm]{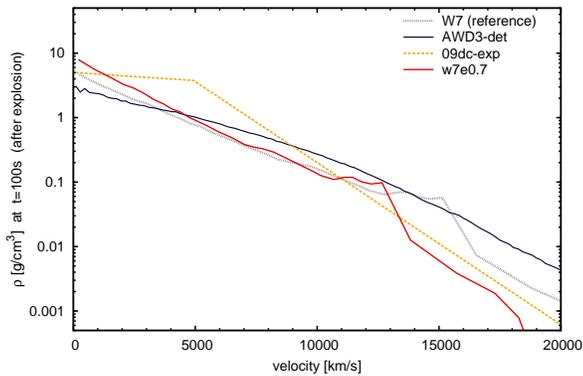}
    \caption{Density profiles used for Abundance Tomography (Sections \ref{sec:tomography-awd3det}, \ref{sec:tomography-exp} and \ref{sec:tomography-csm}), and W7 profile as a reference. The AWD3det model has been converted to 1D (\cf Sec.~\ref{sec:tomography-awd3det}); the 09dc-exp model is shown with its slope adapted to a rise time of 29\vgv{}d (\cf Sec.~\ref{sec:tomography-exp-rt}).}
   \label{fig:densityprofiles-tomography}
\end{figure}

In order to conduct a tomography analysis in spherical symmetry as usual, we had to reduce the AWD3det density field to an averaged 1D density profile by adding up the cell masses into radial velocity bins (Fig. \ref{fig:densityprofiles-tomography}). Asphericities are to some degree inherent in a rotating explosion model \citep{fink10diss}. We keep this in mind as a caveat for the discussion of our results.

\subsection{Rise time estimate}
\label{sec:tomography-awd3det-rt}

We determine an optimum rise time for our AWD3det-based models from the $-$9.4\vgv{}d spectrum of SN~2009dc. The earliest observation is most useful for this purpose, as it is most sensitive on the assumed rise time (the relative difference in $t$, the time from explosion onset, is largest at early epochs). 

The abundance structure of the early-time models comprises three abundance shells. The set up allows optimisation, but avoids an overparametrisation and remains sensible in terms of the chemical structure: the models have a zone of thickness 600\vgv{}\kms\ above the photosphere with some Fe-group elements (up to 15\% by mass; higher fractions are probably not realistic) and high IME mass fraction. Above this zone, there is a layer depleted in Fe-group elements but still rich in IMEs, with a thickness of $\sim$\hspace{0.1em}2000\vgv{}\kms. It causes relatively little backscattering, such that temperature drops quickly with radius and becomes favourable for the formation of prominent lines due to singly ionised IMEs. Still above, the atmosphere is assumed to consist of a C-O mix containing only traces of other elements (mass fractions generally $\lesssim$\hspace{0.1em}10\% of those in the zone below). The photospheric velocities are selected such that a reasonable temperature structure (\ie a large enough strength of \SiII\ lines) results and less than two thirds of the radiation packets are re-absorbed due to backscattering. Whenever in doubt, we choose the lowest possible photospheric velocity in order not to artificially suppress the formation of lines deep inside the ejecta.

As described in Sec \ref{sec:method-spectralmodels-timesensitivity}, the line velocities in the synthetic spectra change with $t_\textrm{r}$. We set up models for different $t_\textrm{r}$ and compared the positions of the synthetic and the observed \SiII\ $\lambda 6355$ feature\footnote{\SiII\ $\lambda \textrm{6355}$ velocities in normal SNe are known to be well fitted by our spectral models (\eg \citealt{ste05}).} (measured as in \citealt{hac06}). Precisely, the rise times we have tested are \mbox{$t_\textrm{r}=\textrm{21}$\vgv{}d}, \mbox{22.5\vgv{}d}, \mbox{25\vgv{}d}, \mbox{27.5\vgv{}d}, \mbox{30\vgv{}d}, \mbox{35\vgv{}d} and \mbox{40\vgv{}d} (only the most promising models have been finalised). For the four models which show the least mismatch $\Delta\lambda$ in the position of the observed and synthetic \SiII\ $\lambda 6355$ feature, $t_r$ and $\Delta\lambda$ values are given in Table \ref{tab:earlytimemodels-awd3}. These values sample the relation between $t_\textrm{r}$ and $\Delta\lambda$, to which we calculate a linear fit (least-squares method). We demand $\Delta\lambda = \textrm{0}$ and obtain from the linear-fit relation an optimum value $t_\textrm{r,opt}$ (see again Table \ref{tab:earlytimemodels-awd3}).

\begin{figure*}
   \centering
   \includegraphics[width=14.5cm]{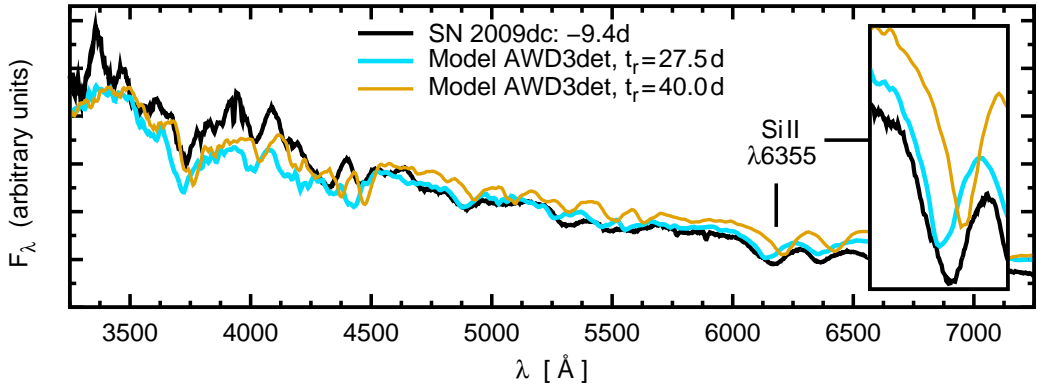}
   \caption{AWD3det-based models for the $-\textrm{9.4}$\vgv{}d spectrum of SN~2009dc for negligible progenitor metallicity and $E(B$\vgv{}$-$\vgv{}$V)$\vgv{}$=$\vgv{}0.17\vgv{}mag. We show the models for the most extreme $t_\textrm{r}$ values listed in Table~\ref{tab:earlytimemodels-awd3}. An optimum model would fit the velocity of the \SiII\ $\lambda \textrm{6355}$ line (mark; inset). With a very short rise time, the velocity (blueshift) of the synthetic feature is too high (blue graph); with a very long rise time, the velocity is too low (orange graph).}
   \label{fig:earlytimemodels-awd3}
\end{figure*}

Two example models, where one can clearly see the shift in the $\lambda 6355$ position with the assumed rise time, are shown in Fig. \ref{fig:earlytimemodels-awd3} together with the observed spectrum of SN~2009dc. A detailed discussion of the SN spectrum and our final optimum model based on AWD3det is given in the next Section (Sec. \ref{sec:tomography-awd3det-tomo}).

We first performed our rise time experiment assuming a total extinction of $E(B$\vgv{}$-$\vgv{}$V)$\vgv{}$=$\vgv{}0.17\vgv{}mag and negligible progenitor metallicity. The latter assumption, which proved optimum for the fit quality, has two main implications when modelling the intermediate and outer ejecta: First, there are no lower limits to the metal abundances in the progenitor WD (which we would have to respect when determining the optimum abundance structure). Second, no stable Fe is produced in the outer layers which undergo incomplete Si-burning in the explosion, as a pure C/O mix provides no excess neutrons \citep{iwa99}. Hence, there is no zone where stable Fe is the dominant iron-group element directly after the explosion. When the Fe-group mix could not be determined from the spectra, we therefore always assumed a mix of Fe-group elements which is dominated by \Nifs\ and its decay products and/or by $^{52}$Cr (\cf Sec. \ref{sec:tomography-descr}).

Afterwards, we repeated the analysis assuming solar metallicity and a lower-than-standard reddening \mbox{$E(B$\vgv{}$-$\vgv{}$V)_\textrm{tot}$\vgv{}$\sim$\vgv{}0.06\vgv{}mag} (see Sec.~\ref{sec:observations}), respectively. This enables us to estimate the influence of these assumptions on our study. The respective results are shown in \mbox{Table \ref{tab:earlytimemodels-awd3}} as well.

\begin{table}
\scriptsize
\caption{Rise time estimates from our spectral fits using the AWD3det density profile described in the text, for different values of the reddening $E(B-V)$ and metallicity. The second and third columns give, for each density profile, the four values of the rise time $t_{r}$ for which optimised models have been calculated, and the corresponding $v_\textrm{ph}$ in these models. The fourth column gives the wavelength deviation between the synthetic \SiII\ $\lambda \textrm{6355}$ feature and the observed one \mbox{$\Delta\lambda$\vgv{}$=$\vgv{}$\lambda_\textrm{syn}$\vgv{}$-$\vgv{}$\lambda_\textrm{obs}$} (with $\lambda_\textrm{obs}$\vgv{}$=$\vgv{}6165.6\vgv{}\AA). From a linear fit to the $\Delta\lambda$-$t_\textrm{r}$ relation, we calculate an optimum rise time $t_\textrm{r,opt}$ for each density profile.}
\label{tab:earlytimemodels-awd3}
\centering
\begin{tabular}{lcrrr}
$\rho$ profile $\!\!\!\!$ & $\!\!\!\!$ $t_{\textrm{r}}$ $\!\!\!\!$ & $\!\!\!\!\!\!$ $v_{\textrm{ph}}$ $\ \ $ & $\!\!\!\!$ $\Delta\lambda$ $\!$ & $\!\!\!\!$ $t_\textrm{r,opt}$ $\!\!$ \\	
$\!\!\!\!$ $\!\!\!\!$ & $\!\!\!\!$ [d] $\!\!\!\!$ & $\!\!\!\!$ [\kms] $\!\!\!\!$ & $\!\!\!\!\!\!$ [\AA] & $\!\!\!\!\!\!$ [d] $\!\,$  \\ 
\vspace{-0.50pt} & \vspace{-0.50pt} & \vspace{-0.50pt} & \vspace{-0.50pt} & \vspace{-0.50pt} \\ \hline
\multicolumn{5}{c}{$\!\!$standard: $E(B$\vgv{}$-$\vgv{}$V)=$\vgv0.17\vgv{}mag, prog. metallicity $Z$\vgv$\sim$\vgv0$\!\!$}									\\ \hline
AWD3det	&	27.5	&	9440	&	-24.9	&		\\
	&	30.0	&	8640	&	-3.7	&		\\
	&	35.0	&	7120	&	20.8	&		\\
	&	40.0	&	5950	&	42.1	&	31.5	\\ \hline
\multicolumn{5}{c}{$E(B$\vgv{}$-$\vgv{}$V)=$\vgv0.06\vgv{}mag, prog. metallicity $Z$\vgv$\sim$\vgv0}									\\ \hline
AWD3det	&	22.5	&	10610	&	-42.9	&		\\
	&	25.0	&	9100	&	-26.8	&		\\
	&	27.5	&	7860	&	1.9	&		\\
	&	30.0	&	6850	&	17.0	&	27.8	\\ \hline
\multicolumn{5}{c}{$E(B$\vgv{}$-$\vgv{}$V)=$\vgv0.17\vgv{}mag, solar prog. metallicity}									\\ \hline
AWD3det	&	27.5	&	10070	&	-28.5	&		\\
	&	30.0	&	9200	&	-8.0	&		\\
	&	35.0	&	7680	&	19.1	&		\\
	&	40.0	&	6560	&	38.7	&	32.1	\\ \hline
\end{tabular}
\end{table} 

For AWD3det, with standard assumptions, the optimum rise time turns out to be 31.5\vgv{}d (Table~\ref{tab:earlytimemodels-awd3}). Evidently, optimum rise times from measured $\Delta\lambda$ values will always carry some uncertainty, from the measurements as well as the modelling methods. With one step in our $t_\textrm{r}$ grid, the shift in line position (Table \ref{tab:earlytimemodels-awd3}) is, however, relatively large, so that we can crudely estimate the error in our $t_\textrm{r}$ determination to be equal or smaller than an average step in our $t_\textrm{r}$ grid (3\vgv{}d). We thus give the following limit on the rise time:
\begin{equation*}
  t_\textrm{r, 09dc/AWD3det} \gtrsim \textrm{28.5}\:\textrm{d.}
\end{equation*}

A very similar limit is found for solar progenitor metallicity. If one assumes a weaker reddening of \mbox{$E(B-V)\sim \textrm{0.06}$\vgv{}mag}, the optimum rise times for our models are lower. This is due to the lower luminosity, reducing the ionisation and allowing lines to form further inside in the ejecta (\ie line velocities as low as in the observations of SN~2009dc can be explained by models with relatively short rise times, \cf Sec. \ref{sec:method-spectralmodels-timesensitivity}). From Table~\ref{tab:earlytimemodels-awd3}, we conclude that for \mbox{$E(B-V)\sim \textrm{0.06}$\vgv{}mag} our rise time limit would be $t_\textrm{r,09dc/AWD3det} \gtrsim \textrm{24.8}\:\textrm{d}$. However, at present we see no compelling reason for preferring the low reddening over the one given by \citet{tau11}.

\subsection{Tomography}
\label{sec:tomography-awd3det-tomo}

\begin{figure*}
   \centering
   \includegraphics[width=14.5cm]{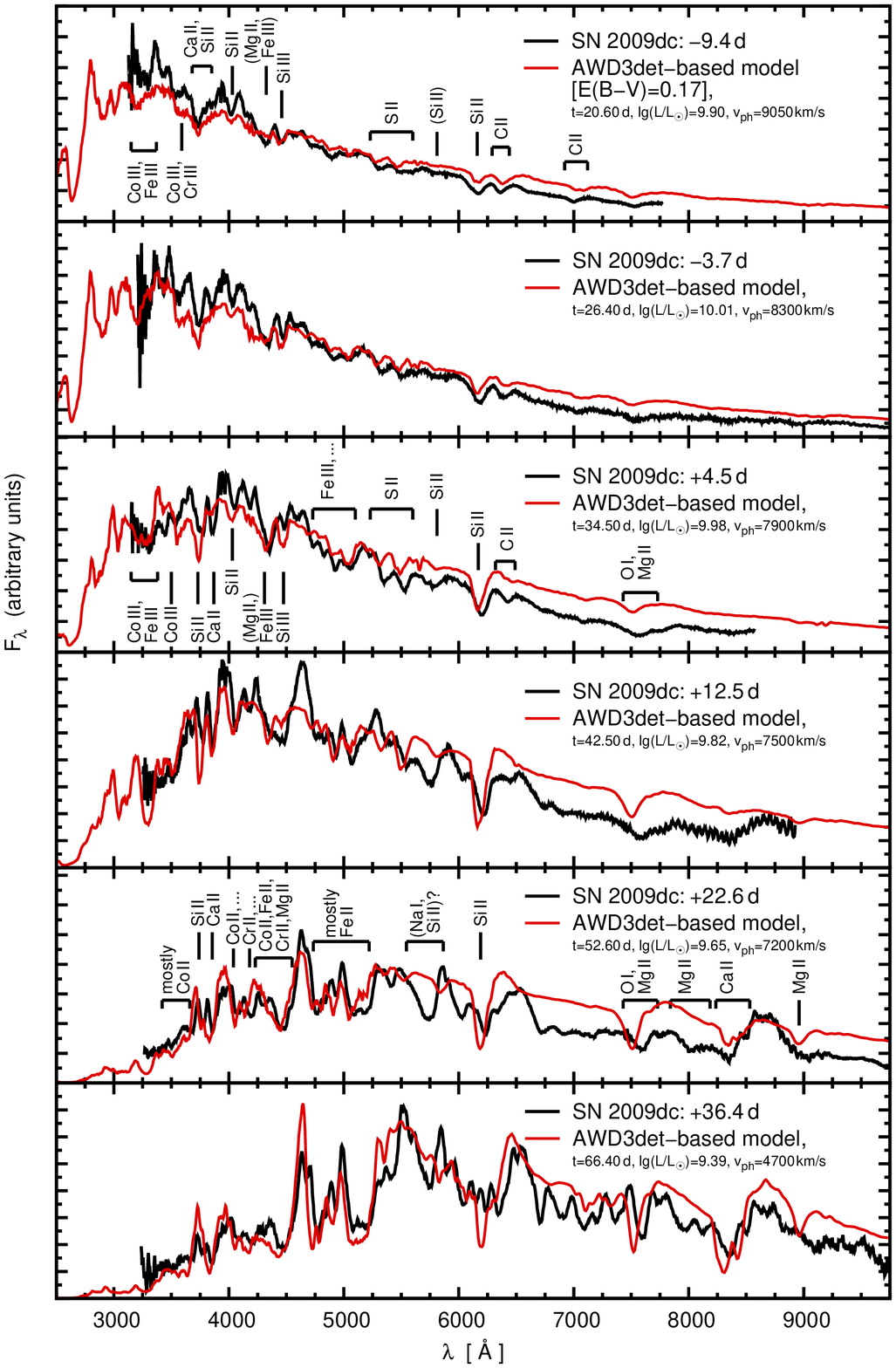}
   \caption{Sequence of spectral models (red graphs) for SN~2009dc, based on the AWD3det density profile, and observations (black graphs). Identifications are given for the most prominent features.}
   \label{fig:sequence-awd3-det}
\end{figure*}

The synthetic spectra based on AWD3det are shown in Fig.~\ref{fig:sequence-awd3-det}. The rise time we have used for the calculations is 30\vgv{}d. The optimum rise time inferred above is a bit longer, but 30\vgv{}d is the strict upper limit allowed by observations \citep{tau11}. The code-input parameters of our tomography models are compiled in Appendix~\ref{app:modelparameters-tomo}.

\subsubsection{$-\textrm{9.4}$\vgv{}d and $-\textrm{3.7}$\vgv{}d models}

The observed $-\textrm{9.4}$\vgv{}d spectrum of SN~2009dc extends down to 3000\vgv{}\AA, where the flux is unusually high for a SN Ia. Despite the high luminosity and blue colour, the optical spectrum shows strong signatures of singly ionised Fe-group and IME species. The similarly luminous SN~2007if does not show clear \SiII\ or \SII\ lines at this epoch \citep{sca10}. This points towards smaller IME abundances or perhaps towards a higher degree of ionisation in SN~2007if [\cf SN~1991T -- \citet{maz95} vs. SN~1999ee -- \citet{ham02}, where variations in ionisation are crucial].

In the case of SN~2009dc, an optimum fit to the spectra is achieved with three abundance zones in the outer ejecta (just as described in Section~\ref{sec:tomography-awd3det-rt}). From the outside inwards, these zones are unburned (\ie depleted in Fe-group elements and also IMEs to avoid Ca and Si high-velocity absorption), somewhat richer in IMEs (Si mass fraction in this model: 8\%) and finally in IMEs/Fe-group elements (Si mass fraction: 15\%, Fe-group mass fraction: 9\%), respectively. \CII\ $\lambda\textrm{6580}$ appears as a prominent line in the spectra. It is reproduced using surprisingly low mass fractions of C (3\% in the outermost zone, 2\% everywhere else), as the high luminosity of the SN leads to an ionisation state favouring highly excited levels of \CII.

Our synthetic spectra are generally too red. A bluer colour would have implied a high $T_\textrm{R}$ throughout the atmosphere. We cannot implement this, as a low enough $T_\textrm{R}$ is crucial for a reasonable ionisation balance, which is needed to match \eg the ratio between the \SiIII\ $\lambda$4563 and \SiII\ $\lambda$6355 features.

Within the first six days covered by the observations, the spectrum of SN~2009dc does not evolve very much. For both spectra, the photospheric velocities are low (9050\vgv{}\kms\ at $-\textrm{9.4}$\vgv{}d and 8300\vgv{}\kms\ at $-\textrm{3.7}$\vgv{}d) compared to normal SNe Ia before maximum. The most prominent lines due to \SiII\ and \SII\ are somewhat stronger at $-\textrm{3.7}$\vgv{}d, but are still formed in the same atmospheric layers as before. This means that a reduction of IMEs in the outer layers has not only an effect on the $-\textrm{9.4}$\vgv{}d spectrum, but also on later ones.

\subsubsection{$+\textrm{4.5}$\vgv{}d model}

The spectrum and the model have now changed considerably. Lines produced by IMEs still gain more strength, as temperatures drop in the outer layers of the atmosphere. The cores of the lines of singly ionised IMEs (such as \SiII) remain at relatively high velocities (as at the early epochs). Only the low-velocity wings of the absorptions are caused by the lower layers of the ejecta. In \SiII\ $\lambda 6355$ and in the \SII\ W-feature (observed around $\sim$\hspace{0.1em}$\textrm{5400}$\vgv{}\AA), these low-velocity wings can unfortunately not be reproduced, regardless of the abundances assumed. This may indicate a too strong ionisation in the lower layers of the model, or a too high photospheric velocity, which however is required in order not to obtain a hot lower boundary spectrum.

The \CII\ $\lambda$6580 feature in the observed spectrum has shifted to much lower velocities by this epoch. This clearly indicates the presence of C in the lower layers. The line's strength at high velocities, at the same time, is automatically reduced because of lower densities and temperatures (the line originates from a highly excited level). Another feature indicating a drop in temperature is \OI\ $\lambda$7773, which becomes deeper in the observations and in the model.

\subsubsection{$+\textrm{12.5}$\vgv{}d model}
\label{sec:tomo-awd3-det-p125}

The $+\textrm{12.5}$\vgv{}d spectrum shows a considerable drop in UV flux. Photospheric temperatures are lower, but the line blocking in the UV also increases. The Fe-group abundances at the photosphere are higher than for the previous spectra. Especially \CoII\ and \CrII\ are now effective in suppressing the flux around $\sim$\hspace{0.1em}$\textrm{3500}$\vgv{}\AA: the ionisation degree is lower than at the previous epoch, favouring recombination of \CoIII\ and \CrIII\ into the singly ionised state. Despite the decreasing ionisation, the synthetic \FeIII\ feature at $\sim$\hspace{0.1em}$\textrm{4300}$\vgv{}\AA\ is deep enough to reproduce the observed line, as significant amounts of \Nifs\ have decayed into \Fefs. No additional $^{54}$Fe is assumed to be present in the models.

The observed features are all in all reproduced, with some exceptions: at $\sim$\hspace{0.1em}$\textrm{4500}$\vgv{}\AA, there is a marked absorption and re-emission feature in the observations, which is reproduced only at later epochs in our models. The other exception is \SiII\ $\lambda \textrm{5972}$, which is too shallow. We suggest that \NaI\ D starts influencing the $\lambda \textrm{5972}$ feature around this epoch. Na is not included in our models, as past attempts to fit the feature have suffered from deviations in the ionisation of this element \citep{maz97bg}.

The \SiII\ $\lambda \textrm{6355}$ velocity shows a mismatch between models and data, which increases with epoch. The observed line moves towards the red (lower velocities) as time progresses, while the centre of the synthetic line does not change that much. We have already removed Si from the outermost layers as far as the earliest spectra permit. Part of the velocity drop in the observed spectra may actually be caused by a feature forming bluewards, which gets stronger with time and emits into the \SiII\ $\lambda \textrm{6355}$ region. Judging from our line list, this is probably a \FeII\ feature, but we do not reproduce it with our models. 

\subsubsection{$+\textrm{22.6}$\vgv{}d model}

At 22.6\vgv{}d past $B$ maximum, the photosphere recedes into a zone rich in Fe-group elements. The most prominent features in the spectrum are due to singly ionised Fe-group elements with very strong lines (marked in Fig.~\ref{fig:sequence-awd3-det}). The balance of these lines is mostly a matter of temperature and density structure.

\subsubsection{$+\textrm{36.4}$\vgv{}d model}

Remarkably, SN~2009dc still shows a photospheric spectrum long after maximum, with a $v_\textrm{ph}$ of 4700\vgv{}\kms\ at $+\textrm{36.4}$\vgv{}d. The quality of the spectral fits is relatively good given the late epoch. The slow evolution into the nebular phase indicates higher densities in the intermediate and inner part compared to normal SNe Ia.

The photospheric velocity at $+\textrm{36.4}$\vgv{}d is much lower than at $+\textrm{22.6}$\vgv{}d. This jump is, however, probably not physically significant, and rather an artefact of the photospheric approximation for such late epochs: an unusually low backscattering rate in the latest-epoch model shows that this approximation begins to break down here.

\subsubsection{Abundances}
\label{sec:tomo-awd3-det-abundances}

\begin{figure}
   \centering
   \vspace{-0.27cm}\includegraphics[angle=270,width=7.7cm]{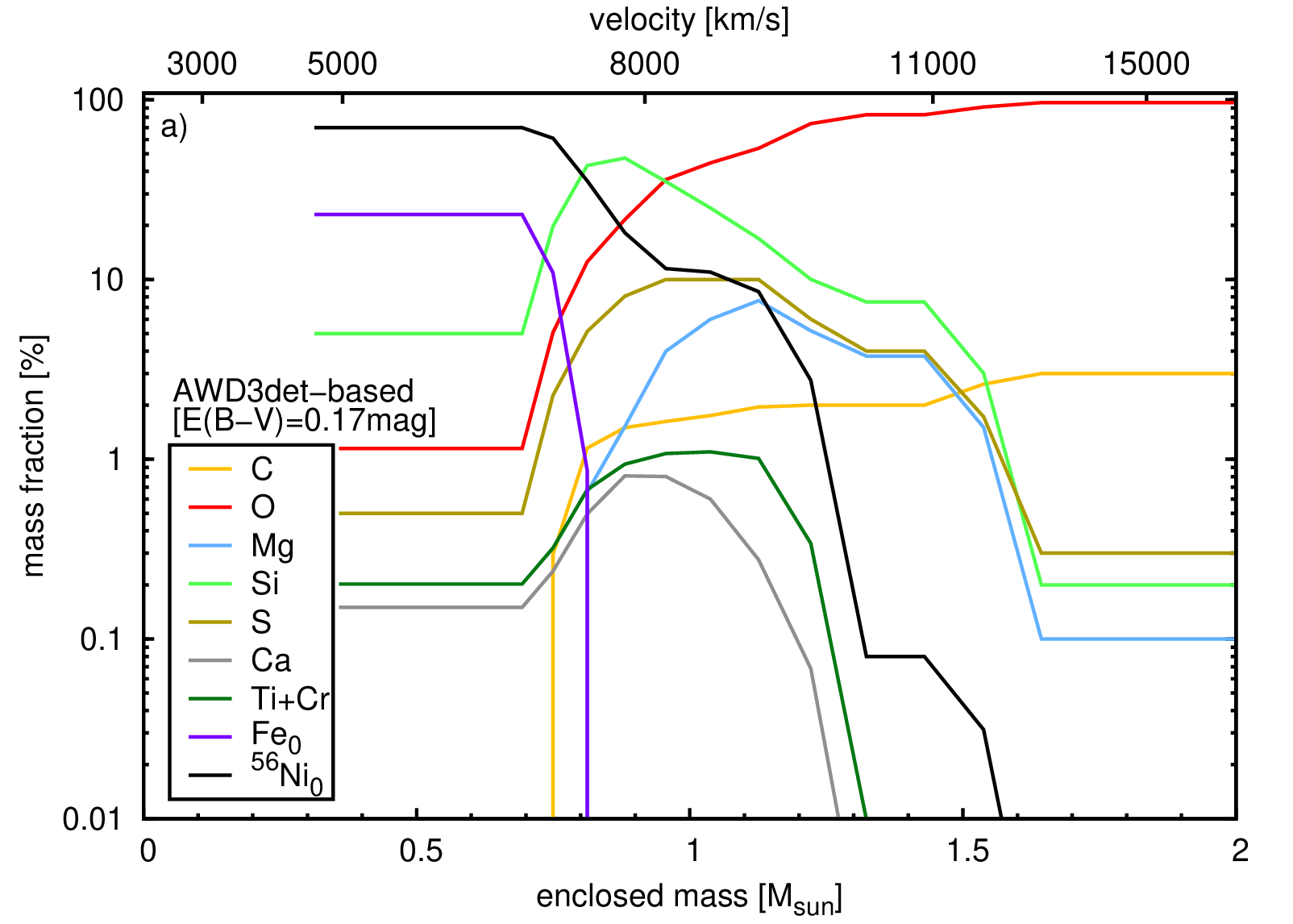}  \\[0.0cm]
   \includegraphics[angle=270,width=7.7cm]{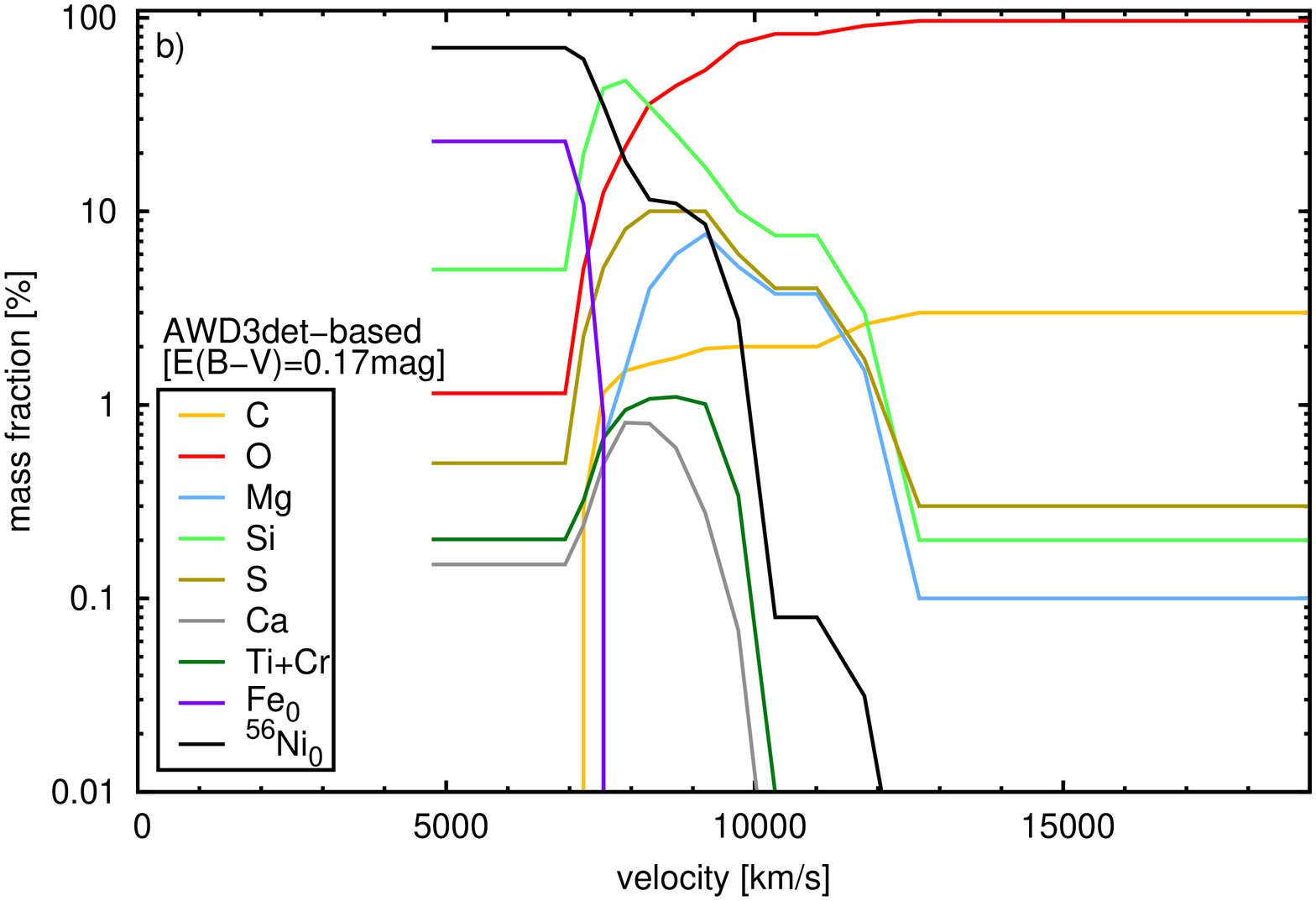}  \\[0.0cm]
   \hspace*{-0.4cm}\includegraphics[angle=270,width=8.05cm]{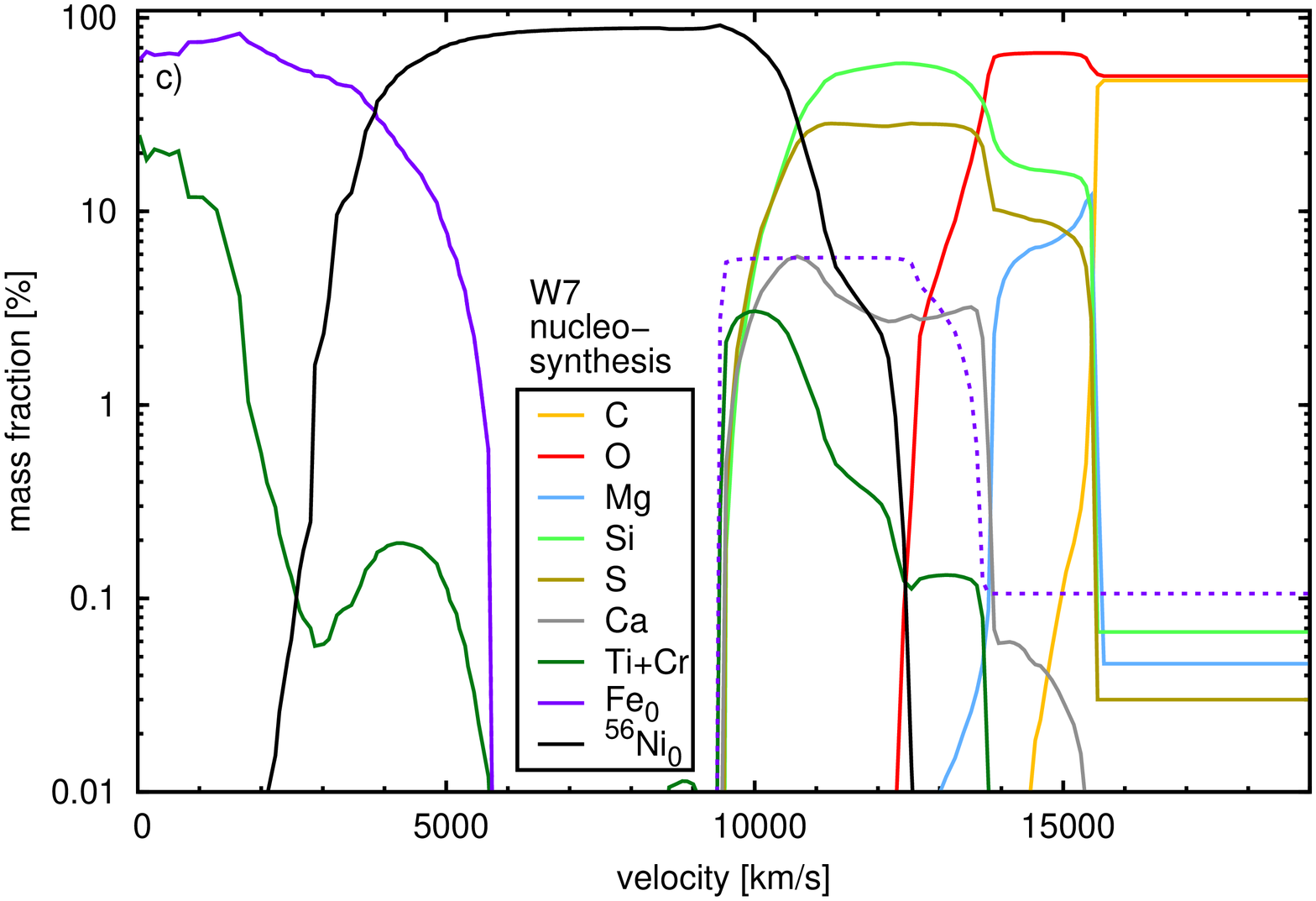} \\[0.0cm]
   \caption{Abundance Tomography of SN~2009dc based on the AWD3det density profile. The ``\Nifs$_0$'' and ``Fe$_0$'' lines represent the \Nifs\ and stable Fe abundances directly after explosion, respectively (\cf Appendix \ref{app:modelparameters-tomo}, Table \ref{tab:modelparameters}). \textit{Panel a)}: abundances in mass space. \textit{Panel b)}: same, in velocity space. In \textit{panel c)}, the abundances of the SN Ia explosion model W7 (\citealt{iwa99}; \cf \citealt{ste05}) for normal SNe Ia are plotted for comparison. When assuming a zero-metallicity progenitor [model W70, \citet{iwa99}], the most significant change is the absence of ``Fe$_0$'' in the outer layers, represented by the dashed part of the respective curve.}
   \label{fig:abundances-awd3-det}
\end{figure}

The abundance stratification of the AWD3det-based ejecta models is shown in Fig.~\ref{fig:abundances-awd3-det}.

The outer zone ($v$\vgv{}$\gtrsim$\vgv{}9650\vgv{}\kms) only contains small amounts of IMEs and Fe-group elements. This corresponds to the weak line blocking and the relatively shallow features in the early-time spectra. Due to the relatively high ionisation, carbon mass fractions of some per cent suffice to fit the observed \CII\ $\lambda \textrm{6580}$ line, although it is much stronger than in other SNe Ia. C must be present down to $\sim$\hspace{0.1em}$\textrm{7500}$\vgv{}\kms\ in order to explain the observed low velocity in the post-maximum $\lambda \textrm{6580}$ feature. The majority of the material in the weakly burned zones, however, consists of O.

Between $\sim$\hspace{0.1em}$\textrm{9000}$ and 7000\vgv{}\kms\ IMEs are highly abundant. The mass fraction of Fe-group elements in this layer, between 12\% and 58\%, is constrained by line blocking of the UV flux.

Although our model sequences extend to late epochs, we mostly explore regions of incomplete burning. At velocities \mbox{$<$\vgv{}7500\vgv{}\kms}, finally, Fe-group elements start to dominate. These zones are sampled by the $+\textrm{22.6}$\vgv{}d and $+\textrm{36.4}$\vgv{}d spectra. Due to degeneracy, the data can be fitted with different Fe-group abundances. We choose rather high mass fractions, consistent with the fact that we see some Fe-group elements already in the \mbox{$-\textrm{9.4}$\vgv{}d} spectrum. 
The mass fractions of ``\Nifs$_0$'' and ``Fe$_0$'' (abundances of \Nifs\ and stable Fe directly after explosion, respectively) together constitute 93\% by mass at the $+\textrm{36.4}$\vgv{}d photosphere. ``Fe$_0$'' is only present deep inside our model ejecta, as our mix of Fe-group elements has been set up (\cf Sec.~\ref{sec:tomography-descr}) to resemble the low-metallicity W7-based nucleosynthesis calculation W70 \citep{iwa99}. 

With respect to the W7/W70 nucleosynthesis (see Fig. \ref{fig:abundances-awd3-det}c), all burning products are found at much lower velocities in SN~2009dc. As a further difference, the ``abundance mixing'' is bit stronger than that of W7, as has also been found in tomography studies of other SNe Ia \citep{ste05}. In SN~2009dc, we see C in fairly deep layers, and need some Fe-group material in the outer zones to reproduce the spectra early on. IME material is present in the unburned and Fe-group layers. Except for the outer layers, the sensitivity of the spectra on the IME mass fraction is limited. Thus, the spectra can also be reproduced with some more down-mixing of O, at the expense of Si. Stricter limits on the O abundances in the lower layers may be deduced in further studies modelling the nebular spectra of SN~2009dc.

The cumulative chemical yields (total mass of C/O, IME and Fe-group elements) of all our models are given and discussed in Section \ref{sec:discussion-abundances}.

\subsection{Abundance profiles and assumptions on metallicity/reddening}
\label{sec:metallicityreddening}

\begin{figure}
   \centering
   \vspace{-0.27cm}
   \includegraphics[angle=270,width=7.7cm]{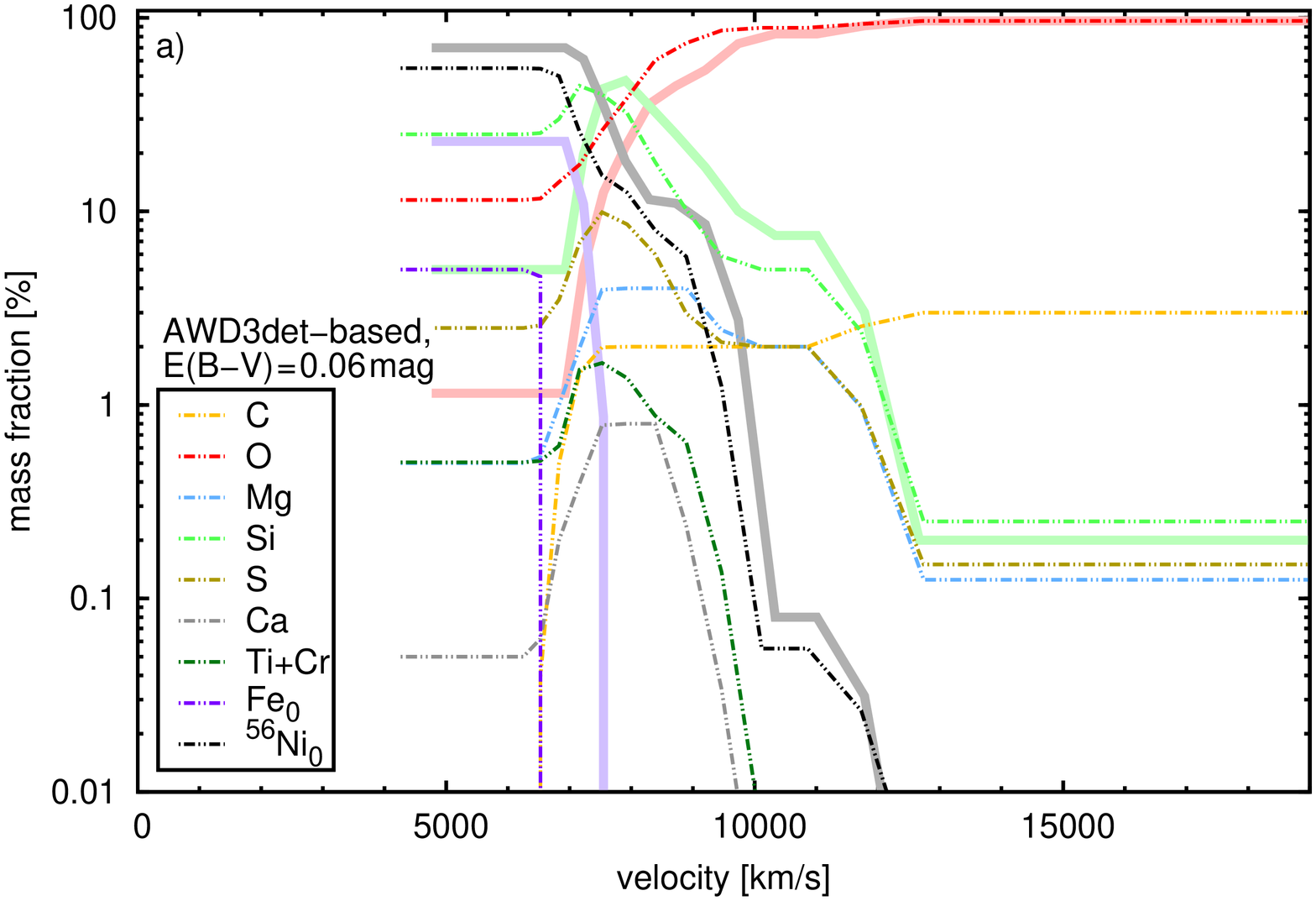} \\[0.0cm]
   \includegraphics[angle=270,width=7.7cm]{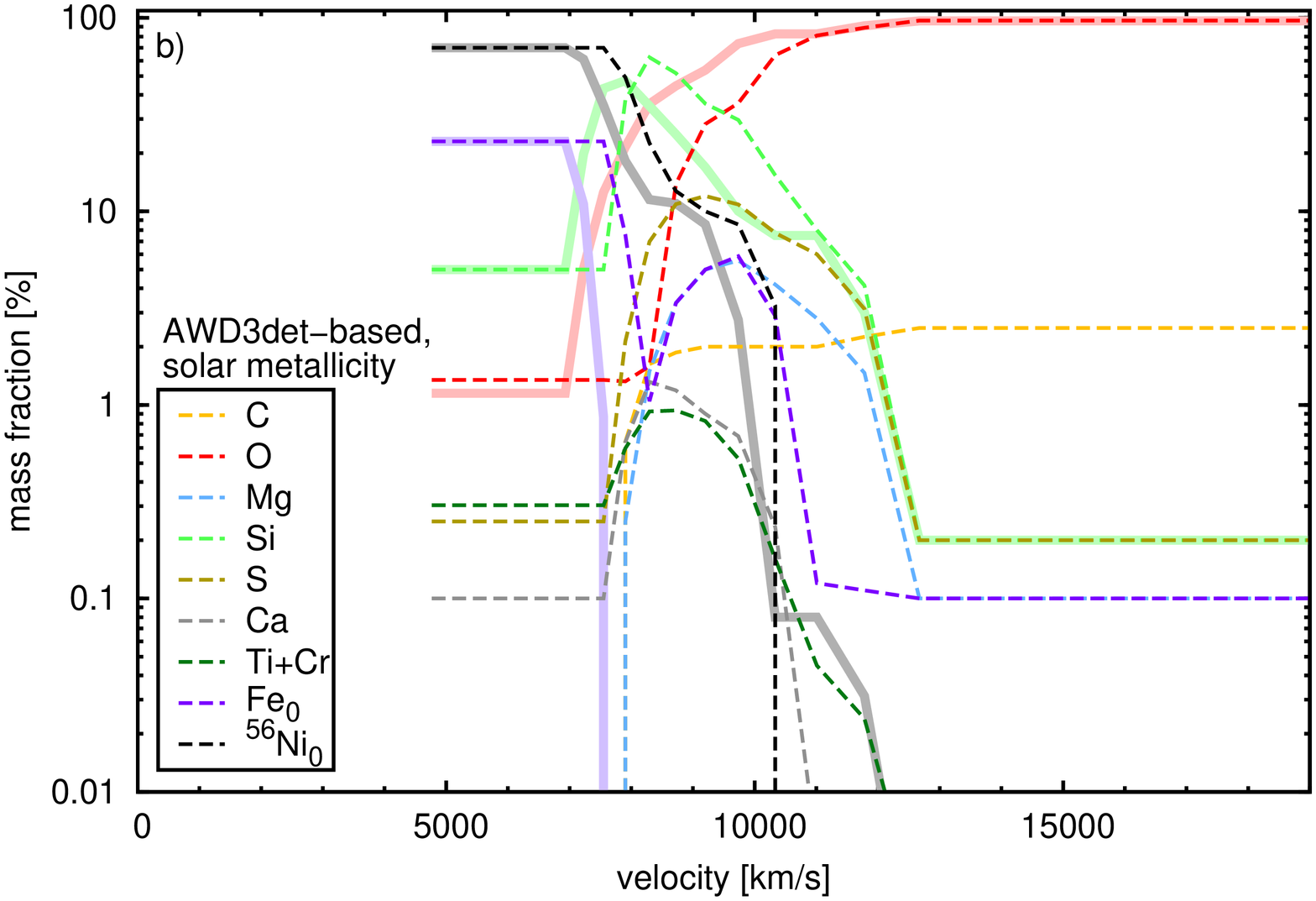} \\[0.0cm]
   \caption{Influence of reddening and metallicity on the AWD3det-based tomography (plots in velocity space). \textit{Panel a)}: abundances inferred for the lower limit reddening $E(B$\vgv{}$-$\vgv{}$V)$\vgv{}$\sim$\vgv{}0.06\vgv{}mag and low metallicity. \mbox{\textit{Panel b)}}: same for $E(B$\vgv{}$-$\vgv{}$V)$\vgv{}$\sim$\vgv{}0.17\vgv{}mag (normal), but solar metallicity (\ie higher metallicity than in our standard model). The abundances of O, Si, ``Fe$_0$'' and ``\Nifs$_0$'' in the standard  model [$E(B$\vgv{}$-$\vgv{}$V)$\vgv{}$\sim$\vgv{}0.17\vgv{}mag, low metallicity -- \cf Fig. \ref{fig:abundances-awd3-det}] are plotted as thick, lighter lines for comparison.}
   \label{fig:abundances-awd3-det-errors}
\end{figure}

We repeat the AWD3det-based tomography for a lower reddening $E(B-V)$\vgv{}$\sim$\vgv{}0.06\vgv{}mag and solar progenitor metallicity, respectively, in order to examine how our results change with these assumptions.

When using a lower reddening, the models have a lower luminosity and the radiation field in their interior is less intense, decreasing the ionisation. This holds even though we use a shorter optimum rise time (Sec. \ref{sec:tomography-awd3det-rt}) of only \mbox{28\vgv{}d} (which tends to increase radiation field densities, cf.\ Sec. \ref{sec:method-spectralmodels-timesensitivity}). The models contain more singly ionised material, which is effective in producing lines (IME) and blocking the UV flux (Fe-group). Thus, we reproduce the spectra with somewhat lower abundances of IME in the outer layers of the ejecta, and lower abundances of Fe-group elements. Qualitatively, however, the abundance stratification in velocity space remains the same (Fig.~\ref{fig:abundances-awd3-det-errors}a). The quality of the models is a bit better than with the high standard reddening, which tends to exacerbate the models' flux excess in the red.

A high (solar) metallicity, on the other hand, implies that there are lower limits to the abundances of Fe-group elements in the outer ejecta \citep{asp09}. Furthermore, there is a change in the typical nucleosynthesis patterns: significant amounts of $^{54}$Fe appear in layers with incomplete burning \citep{iwa99}. We reflect this, as far as possible, in the abundances of Fe-group elements in our models (Fig.~\ref{fig:abundances-awd3-det-errors}b). The resulting spectra have a slightly worse quality: the outer layers with solar Fe-group abundances block flux, heating up the atmosphere and increasing the flux excess in the red, where the radiation can finally escape. Photospheric velocities are quite a bit higher in order to keep the backscattering reasonable and the atmospheric temperatures low enough to form the spectral lines. This shifts the borders of some abundance zones. Still, the abundance stratification we infer is again qualitatively similar to the standard, low-metallicity case.

All in all, we see that our results are somewhat sensitive to reddening and metallicity, but the respective changes are not large enough to compromise our conclusions.

\section{Tomography with a core-collapse-like, empirically motivated density profile}
\label{sec:tomography-exp}

In this part, we conduct a tomography with an ``empirically motivated'' density profile (``09dc-exp''). This profile is constructed (Sections \ref{sec:tomography-exp-rt}, \ref{sec:tomography-exp-densprofile}) from two exponentials (\ie it is a piecewise exponential curve) with a flatter slope in the inner part. The outer part is inferred from the evolution of line velocities in the spectra, assuming that lines always form at a density roughly constant with time (see Appendix \ref{app:densityconstruction}).

Although we do not assume any explosion model in the beginning, the profile resembles density profiles found in simulations of core-collapse SNe with C-O star progenitors (\eg \citealt{iwamoto94,iwamoto00}). With an ejecta mass of $\sim$\hspace{0.1em}3\Msun, our models based on 09dc-exp can be seen as an approach to interpret SN~2009dc as a core-collapse explosion.

\subsection{Empirically inferred outer density profile and optimum rise time}
\label{sec:tomography-exp-rt}

\subsubsection{Grid of empirically inferred density profiles}

\begin{figure}
   \centering
   \includegraphics[angle=270,width=7.7cm]{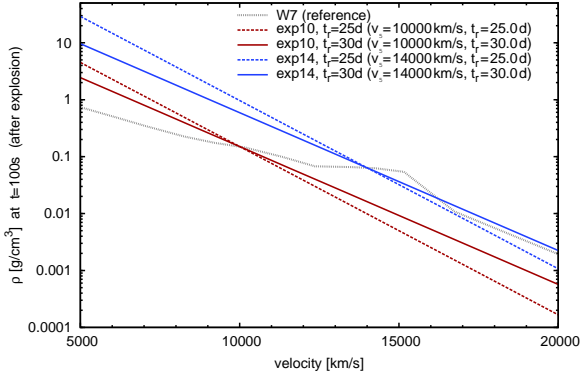}
   \caption{Examples of the density profiles for the outer ejecta constructed in Sec. \ref{sec:tomography-exp-rt}\vgv{}/\vgv{}Appendix \ref{app:densityconstruction}, and the W7 profile as reference. All profiles are plotted for a reference time \mbox{$t_0$\vgv{}=\vgv{}100\vgv{}s} after the explosion, although they have been created assuming different rise times $t_\textrm{r}$.}
   \label{fig:densityprofiles-exp-early}
\end{figure}

We have constructed a two-parameter grid of exponential density profiles for the outer layers of the ejecta (Appendix \ref{app:densityconstruction}; examples see Fig. \ref{fig:densityprofiles-exp-early}). One parameter is the absolute density scale, which we parametrise here in terms of the velocity $v_\times$ at which the profile first intersects the W7 model (\cf Fig. \ref{fig:densityprofiles-exp-early}). For this we allow values of 9000, 10000, 12000 or 14000\vgv{}\kms. The other parameter is the rise time (21\vgv{}d, 22.5\vgv{}d, 25\vgv{}d, 27.5\vgv{}d, 30\vgv{}d, 35\vgv{}d or 40\vgv{}d), with which the slope of the density profile inferred from the line velocity evolution changes somewhat, but not drastically (\cf Appendix \ref{app:densityconstruction}). The set of exponential profiles, which are given names reflecting $v_\times$ (\eg ``exp9''  matches W7 at $v_\times=\textrm{9000}$\hspace{0.25em}\kms), covers well the physical parameter space in which solutions for SN 2009dc will be found (see below).

For each $v_\times$, we infer an optimum rise time separately. In the end, we choose one combination of $(v_\times | t_r)$, which promises optimum results, in order to construct the full 09dc-exp density profile.

\subsubsection{Density/rise time test}

The optimum rise time for each $v_\times$ is inferred from the position mismatch of the \SiII\ $\lambda \textrm{6355}$ feature, comparing the observed spectra to the synthetic ones. This procedure has already been described in Section \ref{sec:tomography-awd3det-rt}. Again, the influence of solar metallicity, and of a lower-than-standard reddening is explored by repeating the modelling for these cases.

\begin{table}
\scriptsize
\caption{Rise time estimates from our spectral fits, using different exponential density profiles for the outer ejecta (see text), for different values of the reddening $E(B-V)$ and metallicity. The table is analogous to Table \ref{tab:earlytimemodels-awd3}, with the difference that several density profiles are tested, corresponding to different $v_\times$ (the slope of the density also changes with $t_\textrm{r}$, \cf Appendix \ref{app:densityconstruction}, however only slightly). The optimum rise time $t_\textrm{r,opt}$ is again calculated from a linearly fitted $\Delta\lambda$-$t_\textrm{r}$ relation.}
\label{tab:earlytimemodels-exp}
\centering
\begin{tabular}{lcrrr}
$\rho$ profile $\!\!\!\!$ & $\!\!\!\!$ $t_{\textrm{r}}$ $\!\!\!\!$ & $\!\!\!\!\!\!$ $v_{\textrm{ph}}$ $\ \ $ & $\!\!\!\!$ $\Delta\lambda$ $\!$ & $\!\!\!\!$ $t_\textrm{r,opt}$ $\!\!$ \\	
$\!\!\!\!$ $\!\!\!\!$ & $\!\!\!\!$ [d] $\!\!\!\!$ & $\!\!\!\!$ [\kms] $\!\!\!\!$ & $\!\!\!\!\!\!$ [\AA] & $\!\!\!\!\!\!$ [d] $\!\,$  \\ 
\vspace{-0.50pt} & \vspace{-0.50pt} & \vspace{-0.50pt} & \vspace{-0.50pt} & \vspace{-0.50pt} \\ \hline
\multicolumn{5}{c}{$\!\!$standard: $E(B$\vgv{}$-$\vgv{}$V)=$\vgv0.17\vgv{}mag, prog. metallicity $Z$\vgv$\sim$\vgv0$\!\!$}									\\ \hline
exp9	&	25.0	&	9600	&	$-\textrm{9.4}$	&		\\
	&	27.5	&	10060	&	$-\textrm{8.3}$	&		\\
	&	30.0	&	8580	&	12.4	&		\\
	&	35.0	&	7210	&	33.7	&	27.8	\\
exp10	&	25.0	&	10250	&	$-\textrm{10.3}$	&		\\
	&	27.5	&	9580	&	$-\textrm{4.5}$	&		\\
	&	30.0	&	8630	&	12.0	&		\\
	&	35.0	&	7250	&	30.4	&	27.7	\\
exp12	&	25.0	&	10610	&	$-\textrm{36.9}$	&		\\
	&	27.5	&	9520	&	$-\textrm{16.4}$	&		\\
	&	30.0	&	8600	&	5.5	&		\\
	&	35.0	&	7170	&	31.8	&	30.0	\\
exp14	&	27.5	&	10190	&	$-\textrm{28.5}$	&		\\
	&	30.0	&	9320	&	$-\textrm{5.2}$	&		\\
	&	35.0	&	7460	&	24.4	&		\\
	&	40.0	&	6090	&	42.1	&	31.6	\\ \hline
\multicolumn{5}{c}{$E(B$\vgv{}$-$\vgv{}$V)=$\vgv0.06\vgv{}mag, prog. metallicity  $Z$\vgv$\sim$\vgv0}	\\ \hline
exp9	&	22.5	&	11500	&	$-\textrm{10.1}$	&		\\
	&	25.0	&	8920	&	7.1	&		\\
	&	27.5	&	7900	&	20.3	&		\\
	&	30.0	&	7050	&	30.7	&	24.0	\\
exp10	&	21.0	&	10450	&	$-\textrm{19.5}$	&		\\
	&	22.5	&	10160	&	$-\textrm{18.3}$	&		\\
	&	25.0	&	8880	&	4.3	&		\\
	&	27.5	&	8000	&	17.0	&	24.7	\\
exp12	&	22.5	&	10310	&	$-\textrm{28.8}$	&		\\
	&	25.0	&	9200	&	$-\textrm{0.1}$	&		\\
	&	27.5	&	8310	&	12.9	&		\\
	&	30.0	&	7450	&	25.8	&	25.9	\\
exp14	&	27.5	&	9770	&	$-\textrm{23.1}$	&		\\
	&	30.0	&	8920	&	$-\textrm{6.1}$	&		\\
	&	35.0	&	7200	&	22.6	&		\\
	&	40.0	&	5350	&	49.4	&	31.3	\\ \hline
\multicolumn{5}{c}{$E(B$\vgv{}$-$\vgv{}$V)=$\vgv{}0.17\vgv{}mag, solar prog. metallicity}		\\ \hline
exp9	&	25.0	&	10200	&	$-\textrm{2.5}$	&		\\
	&	27.5	&	10920	&	$-\textrm{6.0}$	&		\\
	&	30.0	&	9190	&	6.5	&		\\
	&	35.0	&	7590	&	27.4	&	27.4	\\
exp10	&	25.0	&	10800	&	$-\textrm{29.5}$	&		\\
	&	27.5	&	10130	&	$-\textrm{8.9}$	&		\\
	&	30.0	&	9040	&	7.2	&		\\
	&	35.0	&	7650	&	29.5	&	29.5	\\
exp12	&	25.0	&	11040	&	$-\textrm{32.3}$	&		\\
	&	27.5	&	9980	&	$-\textrm{7.3}$	&		\\
	&	30.0	&	9030	&	6.2	&		\\
	&	35.0	&	7620	&	26.2	&	29.7	\\
exp14	&	27.5	&	10620	&	$-\textrm{29.8}$	&		\\
	&	30.0	&	9720	&	$-\textrm{13.9}$	&		\\
	&	35.0	&	8020	&	12.1	&		\\
	&	40.0	&	6540	&	41.8	&	32.7	\\ \hline
\end{tabular}
\begin{minipage}[c]{6.5cm}
\end{minipage}
\end{table} 

\begin{figure*}
   \centering
   \includegraphics[width=14.5cm]{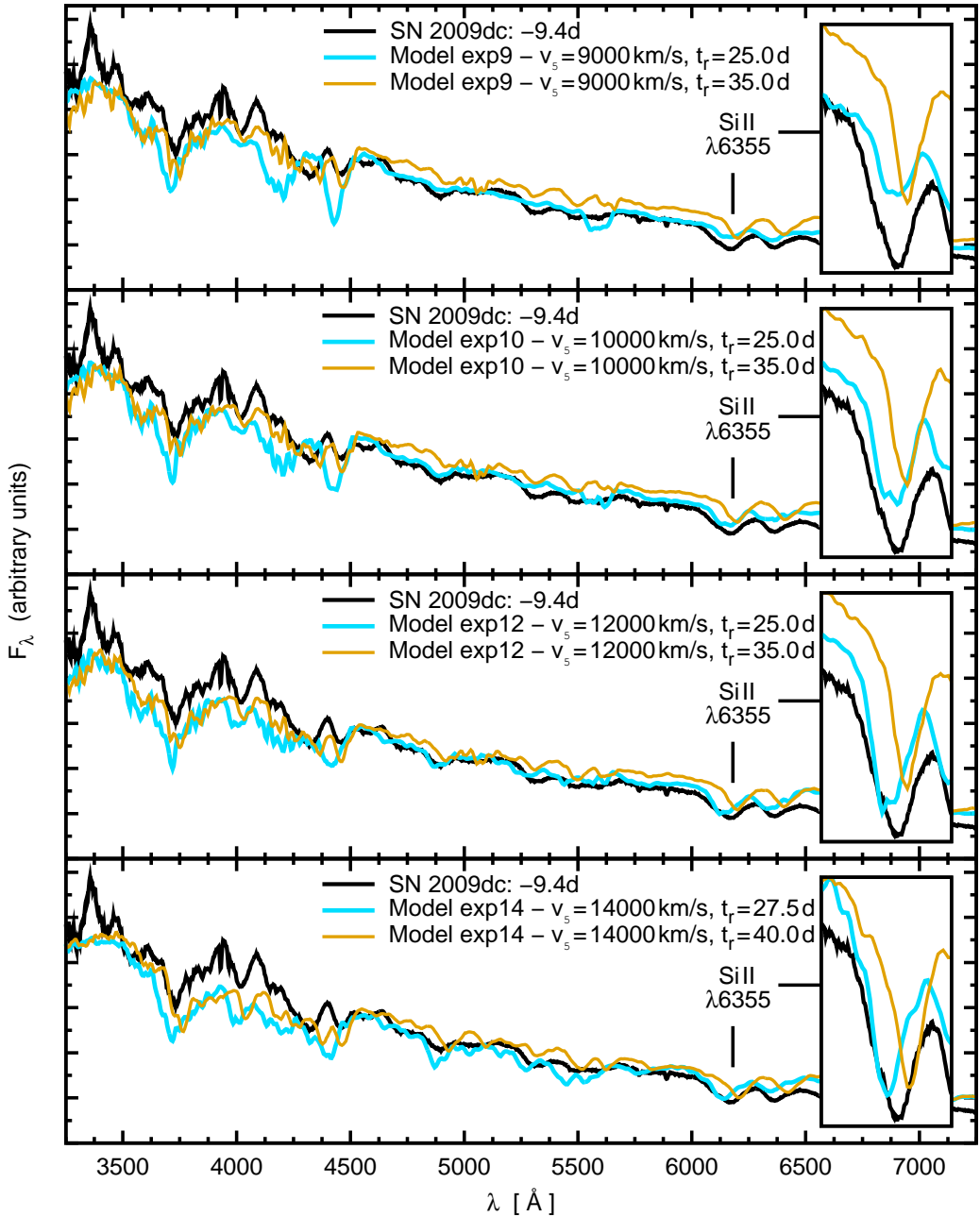}
   \caption{Models for the $-\textrm{9.4}$\vgv{}d spectrum of SN~2009dc based on the empirically inferred density profiles [analogous to Fig. \ref{fig:earlytimemodels-awd3}, for negligible progenitor metallicity and $E(B$\vgv{}$-$\vgv{}$V)$\vgv{}$=$\vgv{}0.17\vgv{}mag]. Each of the four panels contains two models for the same density parameter $v_\times$ and the most extreme $t_\textrm{r}$ values listed in Table~\ref{tab:earlytimemodels-exp}. An optimum model would fit the velocity of the \SiII\ $\lambda \textrm{6355}$ line (marks; insets). With a very short rise time, the velocity (blueshift) of the synthetic feature is too high (blue graphs); with a very long rise time, the velocity is too low (orange graphs). For the exp9-based model with $t_\textrm{r}$\vgv{}$=$\vgv{}25\vgv{}d, the synthetic \SiII\ $\lambda \textrm{6355}$ line lacks depth due to a low number density of \SiII\ ions (\cf text).}
   \label{fig:earlytimemodels-exp}
\end{figure*}

Table~\ref{tab:earlytimemodels-exp} gives -- for each assumed reddening, metallicity and $v_{\times}$ separately -- the measured line position mismatches between the observations and the four optimum models (for different $t_\textrm{r}$), as well as the resulting $t_\textrm{r,opt}$. In Figure~\ref{fig:earlytimemodels-exp}, we plot spectra for a selection of these models [for \mbox{$E(B-V)$\vgv{}$\sim$\vgv{}0.17\vgv{}mag} and negligible metallicity]. 

Using a smaller $v_\times$ for the empirical density leads to smaller densities throughout the envelope. Looking from the outer layers inwards, high densities are reached later, and the zone favoured for \SiII\ line formation tends to be deeper inside (which reduces line velocities). Thus, the $t_\textrm{r}$ values needed to reproduce the low line velocities of SN~2009dc are usually shorter for a low $v_\times$. If $v_\times$ is very low, the zone dense enough for line formation may be far inside where the radiation field is strong. This then favours \SiIII\, with the result that the \SiII\ $\lambda \textrm{6355}$ line cannot be well formed. For this reason, the exp9 models allow us to reproduce the observed \SiII\ line depths only with unrealistically high Si mass fractions $\gtrsim \textrm{70}\%$ (\cf \citealt{iwa99}), if at all (see $t_r = \textrm{25.0}$\vgv{}d model in Fig.~\ref{fig:earlytimemodels-exp}). We therefore expect the actual densities in the outer layers of SN~2009dc to be higher than those in the exp9 models. The exp14 models, on the other hand, need a long rise time to reproduce the \SiII\ $\lambda \textrm{6355}$ line (Table~\ref{tab:earlytimemodels-exp}) and have an excessively high opacity. Thus, they constitute an upper limit to the real densities.

With the intermediate $v_\times=\textrm{10000}$\vgv{}--\vgv{}12000\vgv{}\kms, we obtain the most reasonable models for the $-$9.4\vgv{}d spectrum.

\begin{figure*}
   \centering
   \includegraphics[width=14.5cm]{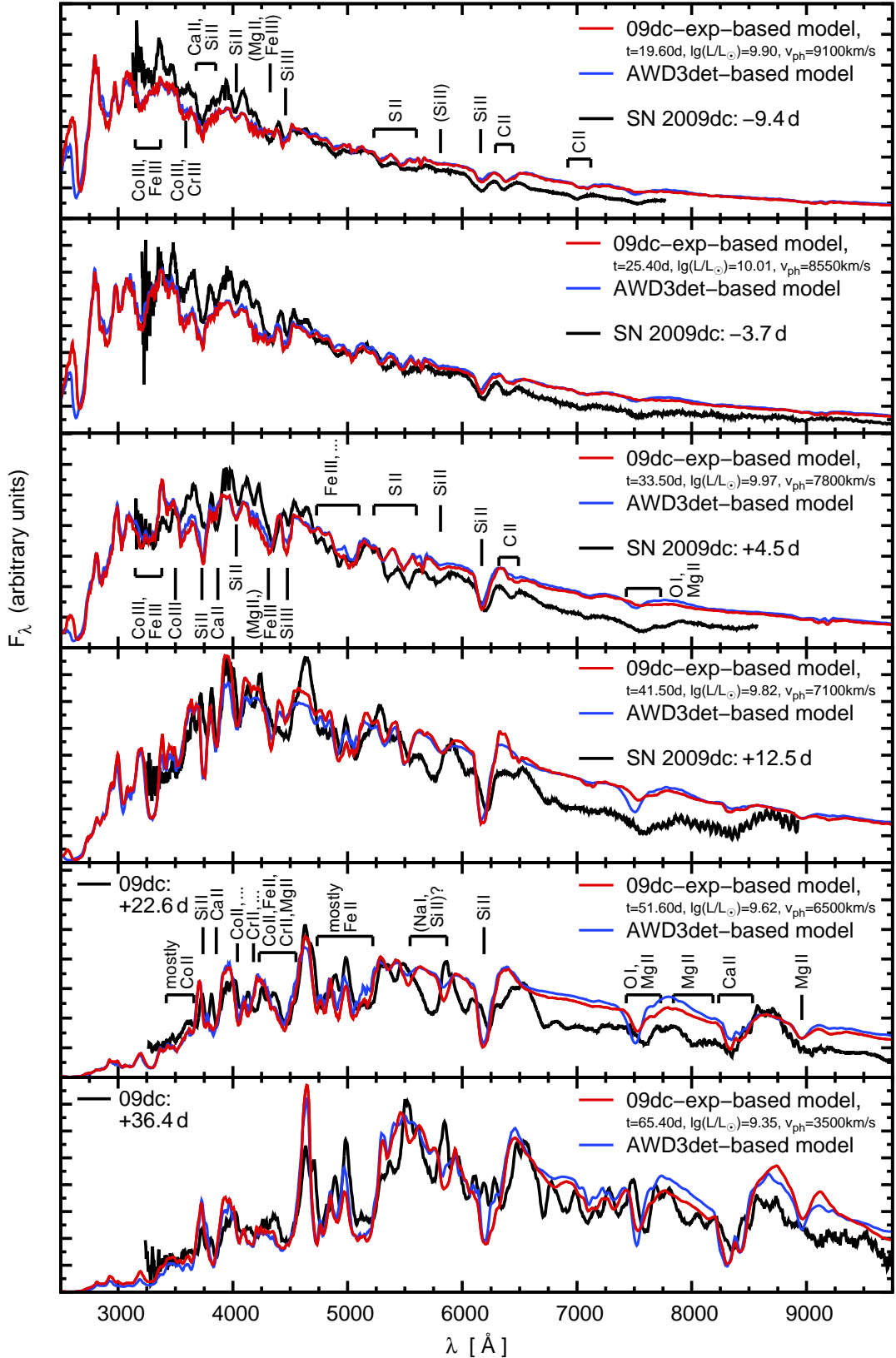}
   \caption{Sequence of spectral models for SN~2009dc, based on the 09dc-exp density profile (red graphs). The models based on the AWD3det density profile (\cf Fig. \ref{fig:sequence-awd3-det}) are plotted in blue for comparison, as well as the observations of SN~2009dc (black graphs). Identifications are given for prominent features.}
   \label{fig:sequence-exp11}
\end{figure*}

\subsection{Construction of the complete 09dc-exp profile}
\label{sec:tomography-exp-densprofile}

Therefore, we now conduct a tomography experiment with a ``09dc-exp'' density profile, whose outer part has a $v_\times$ of 11000\vgv\kms. As a rise time we use 29\vgv{}d, which is the mean of the optimum values for the exp10 and exp12 profiles (\mbox{28\vgv{}d} and \mbox{30\vgv{}d}, respectively, see above) under our standard assumptions [$E(B$\vgv{}$-$\vgv{}$V)=$\vgv{}0.17\vgv{}mag, low progenitor metallicity $Z$\vgv$\sim$\vgv0].

In Fig. \ref{fig:densityprofiles-tomography}, we have plotted the 09dc-exp density. To construct this profile, we have used the density profile with $v_\times =\textrm{11000}$\vgv\kms\ for the outer ejecta. In the inner ejecta, however, we assume the slope to be flatter, which is a common feature in SN explosion models (\eg \citealt{iwamoto94,iwa99,pak11}). This is done by attaching, below some transition velocity, an exponential inner part with a smaller slope parameter $b_\textrm{inner} = \textrm{0.1}\times b_\textrm{outer}$ (\cf parameter $b$ in Appendix \ref{app:densityconstruction}, Formula \ref{formula:rhoprofile}; here $b_\textrm{outer}=\textrm{5.79}\times\textrm{10}^{-9}$\vgv{}km$^{-1}$\quadbynine{}s). The transition velocity is chosen such that the total mass is $\textrm{3}$\vgv$\Msun$, minimally larger than the ejecta-mass estimate of \citet[][ their Sec. 5.1]{tau11}. The flatter slope in the inner zone leads to moderate core densities, comparable to those in core-collapse explosion models  [\eg \citet{woo95} -- from their Fig.~12~/~model~20 we obtain, for 100\hspace{0.25pt}s past explosion, densities of $\sim$\hspace{0.10em}2\hspace{0.25em}g$\,$cm$^{-3}$ at $v<$\hspace{0.25em}5000\hspace{0.25em}\kms].

\subsection{Tomography}
\label{sec:tomography-exp-tomo}

\subsubsection{Spectral models}

The optimum synthetic spectra calculated with the 09dc-exp profile closely resemble those obtained with the AWD3det model (both sets of spectra are shown in Fig. \ref{fig:sequence-exp11}). At early epochs (\mbox{$-\textrm{9.4}$\vgv{}d}, \mbox{$-\textrm{3.7}$\vgv{}d}, \mbox{$+\textrm{4.5}$\vgv{}d}), the low-velocity wing of \SiII\ $\lambda$6355 is reproduced better with 09dc-exp. This is due to the lower densities in the outer layers (Fig. \ref{fig:densityprofiles-tomography}), due to which backwarming is less efficient; temperatures are then lower throughout the atmosphere, so that some \SiII\ is present also in the inner parts (where temperatures are normally highest, so that only \SiIII\ exists).

Later on, the most significant difference between the models is the better fit to \OI\ $\lambda$7773 with 09dc-exp. The AWD3det model has too high density at high velocities, causing too much absorption in the blue wing of \OI\ $\lambda$7773.

In addition, the 09dc-exp model has smaller photospheric velocities at the latest epochs modelled ($+\textrm{22.6}$\vgv{}d and $+\textrm{36.4}$\vgv{}d). The backscattering rates for these models are therefore no longer inconsistent with the assumption of a photosphere, as they were for AWD3det.

\subsubsection{Abundances}

\begin{figure}   
   \centering
   \vspace{-0.27cm}
   \includegraphics[angle=270,width=7.7cm]{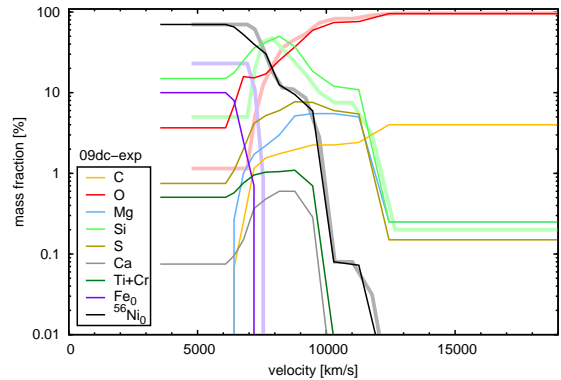}  \\[0.5cm]
   \caption{Abundance stratification of SN~2009dc from our analysis based on the 09dc-exp density profile. The abundances of O, Si, ``Fe$_0$'' and ``\Nifs$_0$'' in the AWD3det-based model [$E(B$\vgv{}$-$\vgv{}$V)$\vgv{}$\sim$\vgv{}0.17\vgv{}mag, low metallicity -- \cf Fig. \ref{fig:abundances-awd3-det}] are plotted as thick, lighter lines for comparison.}
   \label{fig:abundances-exp}
\end{figure}

The abundance structure inferred with the 09dc-exp profile (Fig.~\ref{fig:abundances-exp}) is similar to the AWD3det standard case, except for some shifts in the zone borders. Photospheric velocities are insignificantly higher in the beginning, which corresponds to the assumption of a smaller $t_\textrm{r}$. As mentioned above, the opposite holds after $B$ maximum.

Due to the different density profile, the integrated yields (total masses of unburned material, IME and Fe-group elements) change significantly with respect to the AWD3det-based models. This will be discussed further in Sec. \ref{sec:discussion-abundances}.

\section{Tomography assuming an external contribution to the luminosity}
\label{sec:tomography-csm}

\begin{figure*}
   \centering
   \includegraphics[width=14.5cm]{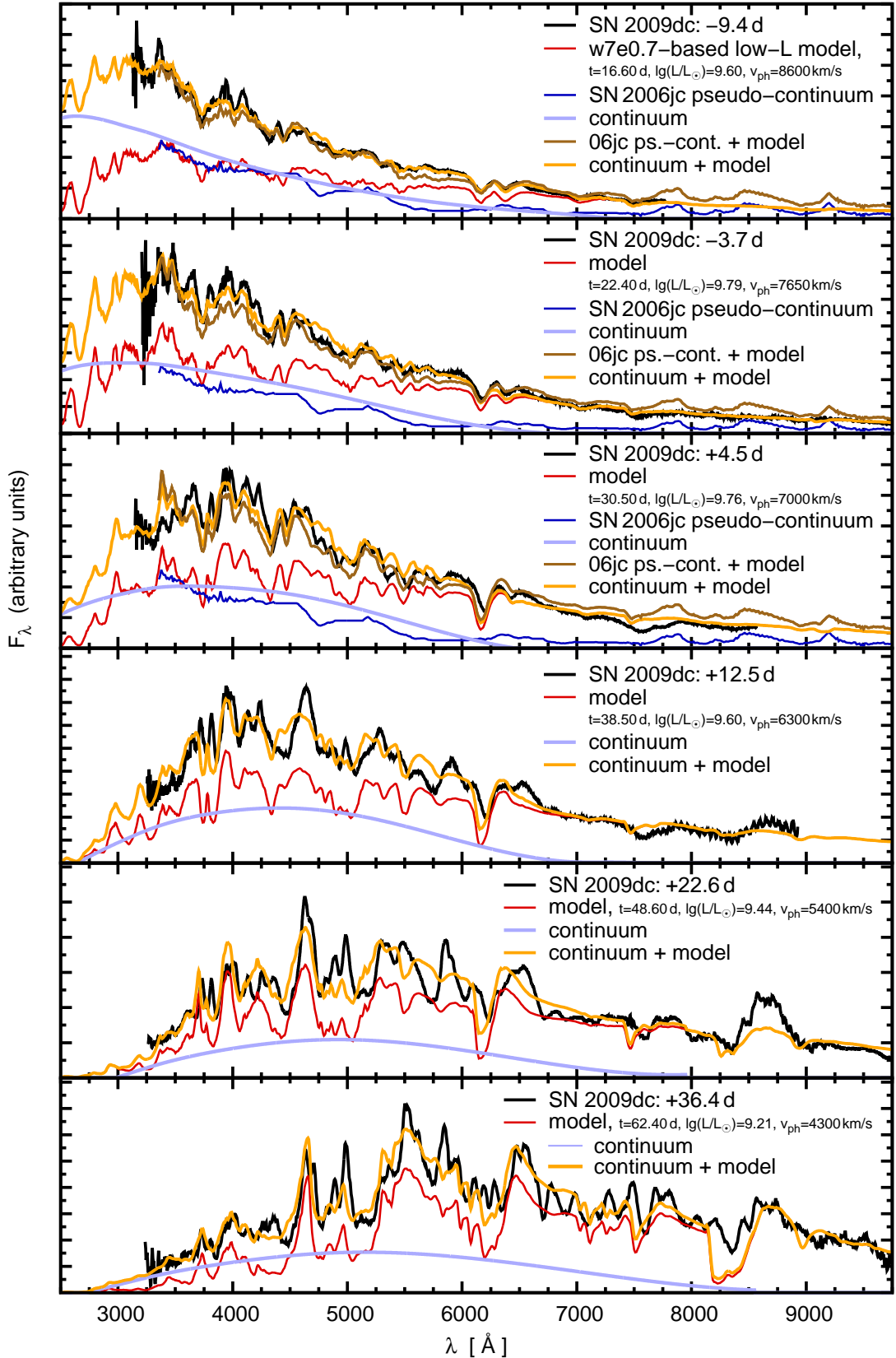}
\caption{09dc-int models for SN~2009dc. Each observed spectrum (black) is modelled as the sum (orange) of an intrinsic SN model spectrum (red) and a continuum (light blue), resulting from interaction with circumstellar material. The continua (see also text) are third order polynomials which are used only where their values are greater than zero and bluewards from their minimum (which occurs at $\gtrsim\textrm{7000}$\AA). For the earliest three epochs, the continua are very similar in shape to a post-maximum spectrum of the interaction-dominated SN~2006jc with the He emission lines removed (\citealt{pastorello07c}; see also text). Adding the SN~2006jc spectrum (scaled; dark blue lines) as a pseudo-continuum to our photospheric models therefore reproduces the observations very well (brown lines).}
   \label{fig:sequence-csm}
\end{figure*}

As already discussed by \citet{tau11}, an explosion of a single $M_\textrm{Ch,non-rot}$ WD is most probably inconsistent with SN~2009dc if all the observed light is generated internally by \Nifs\ decay. Such an explosion may, however, explain the SN if strong interaction of the ejecta with some circumstellar medium (CSM) generates some of the light received from the SN site. In this case, the SN needs to contain only a smaller amount of \Nifs, and the long rise time may not be the result of a large progenitor mass. Instead, it may be due to an increased optical depth caused by the swept up CSM, or due to a longer time-scale on which the interaction produces light.

In an approach inspired by \citet{ham03}, we test whether the spectra of SN~2009dc can be fitted with less luminous models, augmented by some continuous distribution of radiation [09dc-int (interaction) model]. For this experiment, we reduce the bolometric luminosity in the simulation by 50\% at $-\textrm{9.4}$\vgv{}d (with respect to the AWD3det-based model). At the subsequent epochs, we keep the absolute luminosity reduction the same ($\textrm{1.5}\times\textrm{10}^{43}$\vgv{}erg), assuming no significant increase of the continuum. From $+\textrm{12.5}$\vgv{}d on, we begin a slow transition towards normal luminosities, so that at $+\textrm{36.4}$\vgv{}d the simulation produces 67\% of the luminosity in the models presented before. The mismatch between the observations and the model spectra 
with reduced luminosity is automatically fitted with a third-order polynomial\footnote{The difference between the observed and the synthetic spectrum is fitted using a Levenberg-Marquardt method. The wavelength range for the fit is restricted to $<$\hspace{0.1em}7500\vgv{}\AA, so that the procedure is consistent among the spectra whose IR extent varies.} after each code run in the fitting process. The final model parameters lead to an optimum fit of the observed spectrum by the sum of the model and the automatically generated polynomial.

\subsection{Density profile and rise time}
\label{sec:tomography-csm-density}

The density profile used for these calculations is ``w7e0.7'' (Fig.~\ref{fig:densityprofiles-tomography}), a rescaled W7 model with Chandrasekhar mass, but a kinetic energy reduced to 70\% of the original value. The smaller kinetic energy favours line formation at low velocities and may be interpreted in terms of a deceleration of the ejecta by the CSM. It is technically implemented by rescaling each point in the W7 velocity-density grid according to:
\begin{eqnarray*}
v' & = & v_\textrm{W7}\cdot\left(\frac{E'_\textrm{k}}{E_{\textrm{k,W7}}}\right)^{1/2} \\
\rho'& = &\rho_\textrm{W7}\cdot\left(\frac{E'_\textrm{k}}{E_{\textrm{k,W7}}}\right)^{-3/2} 
\end{eqnarray*}
where ($v_\textrm{W7}$, $\rho_\textrm{W7}$) are the W7 values and $\frac{E'_\textrm{k}}{E_{\textrm{k,W7}}}=\textrm{0.7}$. 

The interaction-based model is more of an exploratory nature: as the scenario implies a deceleration of the ejecta when interacting with the CSM, the assumption of force-free expansion is not really accurate here; the force-free time evolution of the density assumed by the spectrum code will deviate to some degree from that of real interacting ejecta. We therefore want to draw only qualitative conclusions from this part of the study.

In line with this, we have refrained from inferring the rise time in such detail as in the other cases. Instead, we have taken advantage of the fact that the outer part of the w7e0.7 profile is quite similar to the 09dc-exp profile (\cf Fig.~\ref{fig:densityprofiles-tomography}), for which we had inferred a rise time of 29\vgv{}d (Sec. \ref{sec:tomography-exp-rt}). Starting with this value, we began modelling the earliest spectrum and obtained a somewhat too red \SiII\ $\lambda$6355 line. We then reduced the risetime of our model in crude steps of 1\vgv{}d. We finally obtained a good match of the observed and synthetic line for a rise time of  26\vgv{}d.

\subsection{Spectral models}

Our synthetic spectra for the 09dc-int model, with their respective intrinsic and continuum contributions, are shown in Fig.~\ref{fig:sequence-csm}. The fit to the observations is very good, especially at the early epochs where the continuum contribution now allows us to perfectly match the blue colour of the observed spectra. The polynomial continua become less luminous and redder with time. 

The most marked mismatch between synthetic and observed spectra remains in the \mbox{$\sim \textrm{5700}$\vgv{}\AA} feature at later epochs, where we have not attempted to fit the \NaI\ D line. Furthermore, the \SiII\ $\lambda \textrm{6355}$ line keeps too strong a blue wing at late epochs, even more than in the AWD3det and 09dc-exp models. This obviously depends on densities and ionisation in the models, and it is -- as other, minor mismatches -- not a major concern at the current stage.

We emphasise that our continua (as we have tested) can not well be fitted by blackbody curves. We speculate that the shape may be explained by emission of overlapping lines \eg of Fe-group elements. In order to test whether this is realistic, we tried to replace the polynomial continuum by a spectrum of the Type Ibn SN~2006jc [taken about three weeks past maximum light, \citet{pastorello07c}]. We cut out the He emission features from this spectrum\footnote{By this we mean inserting a constant flux at \mbox{4375\myto4550\vgv\AA}, \mbox{4900\myto5125\vgv\AA}, \mbox{5750\myto6050\vgv\AA},  \mbox{6650\myto6700\vgv\AA}, \mbox{7000\myto7150\vgv\AA} and \mbox{7200\myto7350\vgv\AA}.}. What remains is a pseudo-continuous flux distribution, which at the epoch chosen should be -- at least in part -- due to atomic lines excited by interaction of the SN ejecta with a CSM \citep[][]{foley07,chugai09,smith09c}. Indeed, the spectra of SN~2009dc at $-\textrm{9.4}$\vgv{}d, $-\textrm{3.7}$\vgv{}d and $+\textrm{4.5}$\vgv{}d are well reproduced by adding the ``modified'' SN~2006jc spectrum (multiplied by a scaling factor which is the same for all three epochs) to our photospheric model (Fig.~\ref{fig:sequence-csm}, upper three panels -- dark blue and brown graphs). At $+\textrm{22.6}$\vgv{}d and $+\textrm{36.4}$\vgv{}d, our continua are somewhat redder than any of the spectra of SN~2006jc presented by \citet{pastorello07c}. The pseudo-continuum in SN~2009dc obviously is generated by less and less energetic processes with time. To understand what information this gives on the ionisation and excitation state of the medium, and on the interaction process going on, more extensive radiation-hydrodynamics modelling will be needed.

\subsection{Abundances}

\begin{figure}   
   \centering
   \vspace{-0.27cm}
   \includegraphics[angle=270,width=7.7cm]{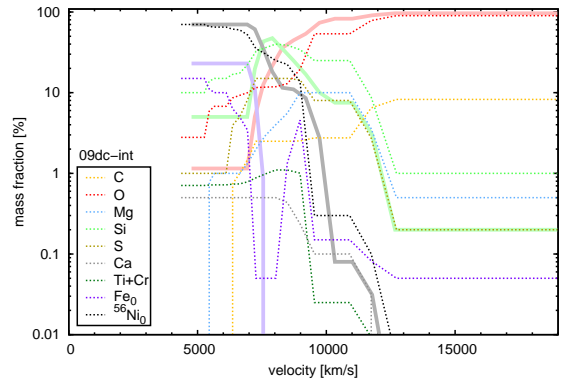}  \\[0.5cm]
   \caption{Abundance Tomography of SN~2009dc assuming Chandrasekhar-mass ejecta, with some of the light generated by interaction with a CSM: 09dc-int model (abundances plotted in velocity space). The abundances of O, Si, ``Fe$_0$'' and ``\Nifs$_0$'' in the AWD3det-based model [$E(B$\vgv{}$-$\vgv{}$V)$\vgv{}$\sim$\vgv{}0.17\vgv{}mag, low metallicity -- \cf Fig. \ref{fig:abundances-awd3-det}] are plotted as thick, lighter lines for comparison. A metallicity of $\textrm{0.5} \times Z_{\odot}$ was assumed when creating the model (\cf Sec. \ref{sec:metallicityreddening}).}
   \label{fig:abundances-csm}
\end{figure}

Despite the physical differences, the abundance structure (Fig.~\ref{fig:abundances-csm}) remains similar to the one obtained with the AWD3det model (again in velocity space -- not in terms of the integrated yields discussed later in Sec. \ref{sec:discussion-abundances}). The increase of burning products from the outside inwards is a bit steeper. Higher abundances are needed to fit the observed line depths as the continuum contribution makes the lines shallower in the final model (sum spectrum). We have assumed an intermediate metallicity here (half-solar) because in 09dc-int, according to test models we calculated, the reproduction of the spectra does not depend much on the metallicity assumed.

\section{Discussion}
\label{sec:discussion}

With our tomography models, we tested different possible explosion scenarios for SN~2009dc. Here, we comment on the viability of these, summarising the constraints under which they are consistent with our results. We separately discuss fully radioactive (Sec.~\ref{sec:discussion-explmodels-intrinsic}) and interaction-luminosity (Sec.~\ref{sec:discussion-explmodels-csm}) models. As a starting point, we discuss the limits on the rise time (Sec. \ref{sec:discussion-risetimelimit}) and on the chemical yields (Sec. \ref{sec:discussion-abundances}) of SN~2009dc which we obtain from our spectral models.

\subsection{Lower limit on the rise time of SN~2009dc}
\label{sec:discussion-risetimelimit}

In Sections \ref{sec:tomography-awd3det-rt}, \ref{sec:tomography-exp-rt}, and \ref{sec:tomography-csm-density}, we have inferred best-fit rise times for SN~2009dc assuming different density models, which well cover the physically sensible parameter space for the SN ejecta. Without assuming interaction, the optimum rise time inferred is lowest with the exp9 or exp10 profiles; the respective values constitute lower limits to the real rise time of the SN for this case. For a reddening of \mbox{$E(B$\vgv{}$-$\vgv{}$V)$\vgv{}$\sim$\vgv{}0.17\vgv{}mag} and negligible progenitor metallicity, we derive (from Table~\ref{tab:earlytimemodels-exp}):

\begin{equation*}
  t_\textrm{r, 09dc} \gtrsim \textrm{25}\:\textrm{d}
\end{equation*}

Here, a possible error of 3\vgv{}d (\cf Sec. \ref{sec:tomography-awd3det-rt}) has again been accounted for already. The lower limit also holds for solar progenitor metallicity, as the respective models need somewhat longer rise times. With a weaker extinction of \mbox{$E(B-V)\sim \textrm{0.06}$\vgv{}mag}, the rise time limit inferred from the models is $t_\textrm{r,09dc} \gtrsim \textrm{21}\:\textrm{d}$ (but, as already mentioned in Sec. \ref{sec:tomography-awd3det-rt}, there is no compelling reason for such an assumption). 

In the same manner, one can read an upper rise time limit of $\sim$\hspace{0.1em}36\vgv{}d from Table~\ref{tab:earlytimemodels-exp}); this limit is however less stringent than the respective observational limit [30\vgv{}d, \citet{tau11}].

In the interaction scenario, we found the optimum rise time to be 26\vgv{}d (Sec. \ref{sec:tomography-csm-density}), \ie within the given range.

\subsection{Limits on the integrated nucleosynthesis yields}
\label{sec:discussion-abundances}

The total masses of C/O, IME, Fe-group elements and \Nifs\ we find in our different tomography models are compiled in Table~\ref{tab:totalabundances}. The \Nifs\ masses shown do not include the \Nifs\ in the opaque core of the ejecta below the $+\textrm{36.4}$\vgv{}d photosphere. The mass of this core, which is expected to contain a significant additional amount of \Nifs, is also given in the Table.

\begin{table}
\scriptsize
\caption{Integrated yields of C/O, IME and Fe-group elements in the models. The Fe-group values include the respective amount of \Nifs. For the core below the $+\textrm{36.4}$\vgv{}d photosphere [mass $M(\textrm{core})$], we cannot constrain the abundances. We expect it to contain a significant amount of \Nifs.}
\label{tab:totalabundances}
\centering
\begin{tabular}{lccccc}
$\!\!\!\!\!$ model $\!\!\!\!$ & $\!\!\!\!\!$ $M(\textrm{C}/\textrm{O})$ $\!\!\!\!\!$ & $\!\!\!\!\!$ $M(\textrm{IME})$ $\!\!\!\!\!$ & $\!\!\!\!\!\!$ $M(\textrm{Fe-group})$ $\!\!\!\!\!\!$ & $\!\!\!\!\!$ $M(\textrm{\Nifs})$ $\!\!\!\!\!$ & $\!\!\!\!\!$ $M(\textrm{core})$ $\!\!\!\!\!$ \\
$\!\!\!\!\!\!\!\!$  $\!\!\!\!$ & $\!\!\!\!$ [$\Msun$] $\!\!\!\!$ & $\!\!\!\!$ [$\Msun$] $\!\!\!\!$ & $\!\!\!\!$ [$\Msun$] $\!\!\!\!$ & $\!\!\!\!$ [$\Msun$] $\!\!\!\!$ & $\!\!\!\!$ [$\Msun$] $\!\!\!\!$ \\ 
\vspace{-0.05cm} & \vspace{-0.05cm} & \vspace{-0.05cm} & \vspace{-0.05cm} & \vspace{-0.05cm} & \vspace{-0.05cm} \\ \hline
$\!\!\!\!$09dc-AWD3det & 0.92 & 0.28 & 0.50 & 0.39 & 0.30 \\
$\!\!\!\!$09dc-AWD3det, low $E(B-V)$  & 1.10 & 0.34 & 0.34 & 0.31  & 0.23 \\
$\!\!\!\!$09dc-AWD3det, solar metallicity    & 0.75 & 0.33 & 0.62 & 0.46 & 0.30 \\
$\!\!\!\!$09dc-exp  & 0.57 & 0.63 & 1.38 & 1.21 & 0.39 \\
$\!\!\!\!$09dc-int  & 0.38 & 0.31 & 0.36 & 0.31 & 0.35 \\
\hline 
\end{tabular}
\end{table}

\subsubsection{Mass of \Nifs: Arnett's rule vs. spectral analysis}
\label{sec:igemass}
\label{sec:w7decline}

The \Nifs\ mass allowed by each tomography model is an important indicator for the consistency with SN~2009dc, as we can compare it to what Arnett's rule \citep{arn82} requires to produce the light-curve maximum of SN~2009dc. This rule states that at maximum light the luminosity roughly equals the instantaneous energy input from radioactive decay:
\begin{equation*}
 L(t_\textrm{max})=M_{\textrm{Ni-56}}(t_\textrm{max})\times \zeta_{\textrm{Ni-56}} +M_{\textrm{Co-56}}(t_\textrm{max})\times \zeta_{\textrm{Co-56}}.
\end{equation*}
$\zeta_X$ is the energy gained from radioactive decay of an isotope $X$ per unit mass (of this isotope) and unit time. The time-dependent masses of \Nifs\ and \Cofs\ are calculated using the half-lives $t_{1/2,\textrm{Ni-56}}=\textrm{6.077}\:\textrm{d}$ and  $t_{1/2,\textrm{Co-56}}=\textrm{77.27}\:\textrm{d}$. Applied as in \citet{tau11}, the rule yields a \Nifs\ mass of $\textrm{2.0}\pm\textrm{0.7}$ or $\textrm{2.3}\pm\textrm{0.8}$\vgv$\Msun$ for an assumed rise time of 25\vgv{}d (our strict lower limit) or 30\vgv{}d (the observational upper limit), respectively. The error estimates include uncertainties in the distance, extinction and in applying Arnett's model (\cf \citealt{tau11}). These values for the \Nifs\ mass, which are significantly larger than the Chandrasekhar mass $M_\textrm{Ch,non-rot}$, may be inaccurate in case of large asymmetries in the supernova \citep{hillebrandt07a,sim07b}. As spectropolarimetry of SN~2009dc however does not indicate extreme global asymmetries \citep{tan10}, one can however say that Arnett's rule in combination with our rise time limit practically rules out a Chandrasekhar-mass progenitor if \Nifs\ is the only source of light in SN~2009dc.

The amount of \Nifs\ in our tomography models, which we can compare with the values given, is somewhat uncertain due to degeneracies especially at late epochs. We have chosen \Nifs\ abundances as high as possible. If we additionally assume the opaque core of the ejecta at $+\textrm{36.4}$\vgv{}d to be dominated by \Nifs, we can perform a consistency assessment which is conservative in the sense that we will not reject models when uncertain.

The smallest \Nifs\ mass ($\lesssim$\hspace{0.1em}$\textrm{0.7}$\vgv$\Msun$) is derived for the 09dc-int model. This model is a special case as it assumes part of the luminosity not to come from \Nifs. Comparing to the \Nifs\ mass required by Arnett's rule, one recognises that the interaction with the CSM would have to produce $\sim$\hspace{0.1em}$\textrm{2}/\textrm{3}$ of the maximum luminosity. This is somewhat more than we assumed in the models, but taking into account uncertainties the model will be more or less consistent.

For models assuming the luminosity to be intrinsic, we obtain direct constraints. The AWD3det based models only allow for a \Nifs\ mass $\lesssim$\hspace{0.1em}$\textrm{0.7}$\vgv$\Msun$, which is a severe problem if one wants to explain SN~2009dc. We further discuss this in Sec.~\ref{sec:awd3-det-consistency}. The 09dc-exp model has a larger mass concentrated in the central, \Nifs-rich layers. Therefore it contains up to $\textrm{1.6}$\vgv$\Msun$, which is consistent with Arnett's rule within the error bars.

\subsubsection{Minimum mass of directly synthesised Fe, IME and C/O in SN~2009dc}
\label{sec:coimemass}

Our models explore a range of explosion scenarios expected to explain SN~2009dc, which imply different density structures and different conditions for ionisation and excitation in the ejecta. We can therefore regard the minimum of the directly synthesised Fe (\Fefs\ from \Nifs\ decay excluded), IME and C/O masses found in our models as rough lower limits to the actual nucleosynthesis yields of the SN.

The amount of directly synthesised Fe (which is roughly the difference between the Fe-group and the \Nifs\ values in Table \ref{tab:totalabundances}) is practically insignificant (mostly about 0.1\Msun). For the intermediate-mass elements, we obtain
\begin{equation*}
  M_\textrm{IME,09dc} \gtrsim \textrm{0.3\vgv\Msun} ,
\end{equation*}
which is the IME mass in the 09dc-AWD3det model.

For the C/O mass, to which C contributes only little, we obtain from the 09dc-int model
\begin{equation*}
  M_{\textrm{C}+\textrm{O,09dc}} \gtrsim \textrm{0.4\vgv\Msun}.
\end{equation*}
The oxygen included here is needed for reproducing the observed \OI\ $\lambda\textrm{7773}$ feature.

\subsection{Candidate explosion models without CSM interaction}
\label{sec:discussion-explmodels-intrinsic}

\subsubsection{Rapidly rotating WD models}
\label{sec:awd3-det-consistency}

\begin{figure}
   \centering
   \includegraphics[angle=270,width=8.5cm]{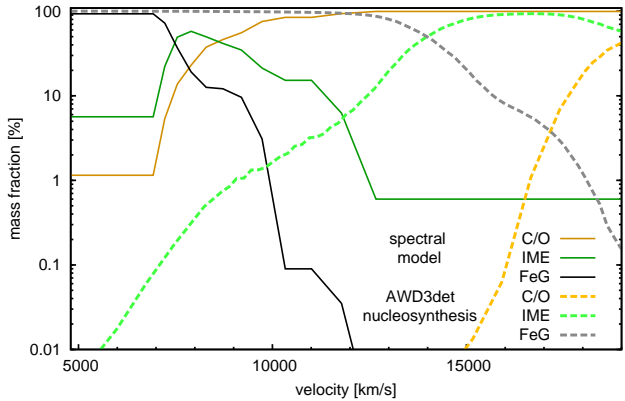}  \\[0.5cm]
   \caption{Comparison of the abundance stratification inferred from the spectra (using the AWD3det density) with the nucleosynthesis of the AWD3det model \citep{pfa10b,fink10diss}. We show the total abundances of three element groups (C/O, IME and FeG / Fe-group elements), as they are used in the AWD3det calculations. The AWD3det mass fractions have been obtained from the original 3D model by adding up the masses of C/O, IME and Fe-group elements, respectively, into radial velocity bins. The lower limit of the velocity axis corresponds to the $+\textrm{36.4}$\vgv{}d photosphere.}
   \label{fig:awd3-det-comparison}
\end{figure}

With our AWD3det-based models, we explored the possibility of explaining SN~2009dc as an explosion of a massive, rotating C-O white dwarf. Rotating white dwarfs with up to $\sim$\hspace{0.1em}$\textrm{2}$\vgv$\Msun$ may exist \citep{yoo05}, although there is no observational proof and the mass limit is a matter of debate [see the summarising discussion of \citet{sca10}]. \citet{hachisu12a} propose that even masses up to $\sim$\hspace{0.1em}2.5\vgv$\Msun$ may be realistic.

Assuming the AWD3det density, our spectral models do not allow enough \Nifs\ in the intermediate and outer (\ie high-velocity) zones of the ejecta in order to obtain the large \Nifs\ masses expected for SN~2009dc (cf. Sec.~\ref{sec:igemass}). Nucleosynthesis calculations for the hydrodynamical explosion model indicate a \Nifs\ mass of $\sim$\hspace{0.1em}$\textrm{1.5}$\vgv$\Msun$, which would be compatible with SN~2009dc. However, such efficient nucleosynthesis within the white dwarf implies (in terms of energetics) that burning products, in particular Fe-group elements, are ejected with high velocities, contradicting our models (Fig.~\ref{fig:awd3-det-comparison}). Also, the abundances of IME and C/O in the AWD3det simulation ($< \textrm{0.1}$\vgv\Msun, respectively) are lower than the limits given in Sec. \ref{sec:discussion-abundances}.

The major uncertainty in our analysis -- besides the maximum WD mass -- is that we assume spherical symmetry in our spectral analysis and in applying Arnett's rule. The AWD3det ejecta are significantly asymmetric (\citealt{fink10diss}; \cf \citealt{pfa10b}), possibly provoking viewing-angle effects in the spectrum. Another caveat is that explosions of extremely massive WDs \citep{hachisu12a} may produce the \Nifs\ mass for a bright light curve without having too-high-velocity Fe-group material. Therefore, we can not yet completely exclude SN~2009dc to be explained by a rotating WD model, although the factor $> \textrm{2}$ between the \Nifs\ mass in our AWD3det-based model and the \Nifs\ mass calculated with Arnett's rule (Sec.~\ref{sec:igemass}) seems difficult to overcome. In the end, 3D radiative transfer simulations of massive WD explosions may allow for a final judgement.

\subsubsection{Mergers of C-O white dwarfs}
\label{sec:discussion-nocsm-wdmergers}

Mergers of two C-O WDs are another channel for creating a ``super-Chandrasekhar-mass object'' which may undergo a thermonuclear explosion \citep[\eg][]{yoo07,pak10a,dan12a}. \citet{hic07} and \citet{sca10} have already considered merger scenarios to explain the SC SNe Ia 2006gz and 2007if.

For small mass ratios (\ie if one WD is much heavier than the other), accretion is a self-stabilising and relatively slow process \citep{dso06,mot07}: the separation of the white dwarfs gets larger when mass is accreted, stopping the accretion. Such systems may make ``normal'' Chandrasekhar-mass SNe~Ia if the primary (the heavier white dwarf) ever grows sufficiently.

For higher mass ratios, for which mass transfer is unstable, the most probable outcomes are a violent merger or an accretion-induced collapse. In a violent merger, the WDs are quickly disrupted or at least heavily deformed. During this process, the ignition of an off-center detonation is likely \citep{pak11}, and a SN~Ia explosion results. If conditions are not extreme for a violent merger (\ie no detonation is ignited), C will ignite off-centre after some time, and the primary will slowly be burned to an O-Ne-Mg WD. This WD will subsequently undergo accretion-induced collapse \citep{nom85aic,sai98}. Therefore, in order to explain a SN explosion with a non-violent merger, C ignition has to be avoided. This may be possible at least in a few configurations \citep{yoo07}. The primary may then grow towards its Chandrasekhar mass and explode.

In any case, if a ``merger SN~Ia'' is made by two WDs with significantly different mass, the \Nifs\ mass generated will not be larger than $M_\textrm{Ch,non-rot}$, as would be needed to explain the light curve of SN~2009dc: the material of the secondary will be too dilute to be converted into \Nifs\ when the explosion occurs (\cf \citealt{pakmor12a}). WD pairs of roughly equal mass and a total mass $\sim$\hspace{0.1em}2\Msun, in turn, are probably not too frequent (the exact occurrence rates are not exactly known). 

Therefore, we suggest that WD mergers will rather not produce an extreme \Nifs\ mass, as it is required for SN~2009dc without assuming CSM interaction. It seems more likely that a double-degenerate system could explain SN~2009dc if an extended, C/O-rich circumstellar medium was generated during the accretion/merging process, and some of the luminosity of SN~2009dc was contributed by an interaction of the ejecta with that medium (see Sec.~\ref{sec:discussion-explmodels-csm}).

\subsubsection{Core collapse and Type $\textrm{I}\verticalonehalf$ SNe}

The 09dc-exp model, which has an ejecta mass of $\textrm{3}$\vgv$\Msun$, corresponds to an explosion of an extended progenitor star instead of a WD. This would imply either core-collapse, or a thermonuclear explosion within a star \citep{erg74}. In any case, the explosion model has to produce an ejecta mass $> \textrm{2}$\vgv$\Msun$ and a \Nifs\ mass $\gtrsim \textrm{1.5}$\vgv$\Msun$ (\cf Sec. \ref{sec:discussion-abundances}).

\citet{ume08} discussed the possibility of producing \Nifs\ masses of some $\Msun$ in a core collapse. Their models, based on stars with main-sequence masses of \mbox{$M \geq \textrm{25}$\vgv$\Msun$}, yield \Nifs\ masses as we need them for SN~2009dc. However, they still contain H and He envelopes ($M_\textrm{H+He}\gtrsim \textrm{10}$\vgv$\Msun$), which would have to be removed by stripping or mass loss in order for the SN to potentially display a Type Ia spectrum. One characteristic feature of the (1D) models of \citet{ume08} is an abundance stratification (Fig. 5 of their paper) which shows similarities to our tomography results and generally to SN Ia models. O is not significantly mixed with \Nifs\ on macroscopic scales. This may disfavour the appearance of nebular O lines (which are normally present in core-collapse SNe), and therefore ensure compatibility with the non-detection of these lines in SC SNe Ia \citep{tau11}. It is necessary to explore whether envelope stripping would significantly alter the conclusions of \citet{ume08}, and whether the general structure of the model (abundances, densities) would remain the same in a multi-D core-collapse simulation, in which the explosion energy is deposited in the medium by neutrinos (\eg \citealt{mar09}) or acoustic waves \citep{bur07}.

Thermonuclear explosions within stars, on the other hand, have long been proposed but never been verified as a model for SNe \citep{erg74,ibe83}. Stars with a zero-age-main-sequence mass $\lesssim \textrm{8}$\vgv$\Msun$ form degenerate C-O cores after core He burning. After mass loss in the asymptotic-giant-branch (AGB) phase, the C-O core usually remains as a bare white dwarf. In the most massive AGB stars, however, the degenerate core may grow towards the Chandrasekhar limit, and an explosion may commence. The explosion would eject up to a Chandrasekhar mass of \Nifs, corresponding to the entire stellar core. The composition of the ejecta depends on the progenitor material and its nuclear reprocessing. In case only Fe-group material is produced in the core, a significant amount of C, O and IME would have to be produced in the envelope (\cf the limits in Sec. \ref{sec:coimemass}). \citet{ibe83} however suggested that H and He may be visible in the spectra of such SNe, and therefore dubbed them ``Type I\hspace{0.1em}\textonehalf'' explosions (\ie thermonuclear with a core-collapse-like display). This must be avoided, and it is not clear whether this is realistically possible \eg by nuclear reprocessing of the envelope. Small quantities especially of He may also be present without lines being detectable in the spectra \citep{hachinger12a}. In the end, the Type $\textrm{I}$\textonehalf\ scenario has to be simulated with recent hydrodynamics or radiation-hydro codes in order to make more definite predictions on its nucleosynthesis and observational outcome.

\subsection{Candidate explosion models with CSM interaction}
\label{sec:discussion-explmodels-csm}

Finally, we discuss a scenario which involves light generation by significant interaction of the ejecta with a CSM. The 09dc-int model (Sec.~\ref{sec:tomography-csm}) shows that the spectra can successfully be explained with light from \Nifs\ plus an ``external'' light contribution. 

The external contribution in our models is continuous, but does not have a blackbody shape. It may be explained by emitting lines which overlap: at the early epochs the pseudo-continuum is very similar to a spectrum of the interacting SN~Ibn 2006jc \citep{pastorello07c} some weeks past maximum light, with its He emission cut out. Such SN~Ibn spectra have been suggested to have a pseudo-continuum from overlapping emission lines of Fe-group elements (\citealt{foley07}, \citealt{chugai09}; \cf also \citealt{smith09c}).

Presumably, a significant optical depth of the emitting zones is needed to allow for ``partial thermalisation'' and generate a smooth flux distribution without major glitches. For the light of the SN to pass to the observer in such a scenario (as indicated by the relatively normal SN spectra), the emitting material has to be concentrated, \eg in clumps or in a torus, so that a major fraction of the lines of sight between SN and observer remain free.

With the additional light contribution, one may explain ``SC'' objects as an explosion of a WD with a mass of roughly $M_\textrm{Ch,non-rot}$. The CSM itself has probably to consist of C, O, or heavier elements, as no H and He lines are visible in the photospheric or nebular spectra of SC SNe Ia. Small enough quantities of He may again be present without making lines \citep{hachinger12a}. The CSM has to be sufficiently close to be reached by the ejecta within a timescale of 10\vgv{d} (\ie, for typical ejecta velocities of \mbox{$\sim$\hspace{0.1em}$\textrm{10000}$\vgv\kms}, at a distance of $\lesssim \textrm{10000}$\vgv\kms$ \times \textrm{10}$\vgv\textrm{d}\vgv$ \sim \textrm{10}^{15}$\vgv{cm}). The light production has to continue for several weeks; if this required sweeping up new material continuously, the extent of the CSM would have to be at least of the same order of magnitude.

It is worthwhile to think of progenitor systems possibly generating such a CSM. One model to consider are non-violent\footnote{There is only a very short period of mass transfer before a violent merger and there is no indication that it could generate an extended CSM \citep[\eg][]{pak10a,pak11,pakmor12a}.} WD mergers with unstable accretion. If the currently predicted scenario of burning to an O-Ne-Mg WD and accretion-induced collapse was avoided for such a system (\cf Sec. \ref{sec:discussion-nocsm-wdmergers}), the primary WD could explode when reaching $M_\textrm{Ch,non-rot}$ (or also a higher mass in case of fast rotation, for which however tidal locking would have to be inefficient). The extent of the CSM, which would have to be produced before the explosion, would probably have to be much larger than found in simulations of such systems by \citet{yoo07} and \citet{dan12a}. An alternative possibility would be accretion from a star which has lost most of its envelope, such that a H-deficient and maybe C/O-rich CSM could form. In any case, the explosion in an interaction-based scenario has to produce a significant, but not an extreme amount of \Nifs\ (just as a normal SN~Ia, \cf Sec. \ref{sec:tomography-csm}). 

The CSM interaction scenario has several appealing aspects. It may not only explain the luminosity of SC SNe Ia, but also a slow-down of the ejecta and therefore the low line velocities. The w7e0.7 density model allowed for good fits to the spectra (09dc-int models). In this model, 30\% of the kinetic energy of the explosion were assumed to be transferred to the circumstellar shells. Crudely taking the distribution of kinetic energy in the ejecta to be uniform, this would mean that the circumstellar material would constitute 30\% of the total mass (\ie\ $\textrm{0.6}$\vgv$\Msun$ when assuming a $M_\textrm{Ch,non-rot}$ primary).

Only a small part of the energy lost to the CSM (which in total is roughly $\textrm{0.4}\times\textrm{10}^{51}$\vgv{}erg, \ie  $\textrm{30\%}$ of $E_{\textrm{k},W7}$), would have to be converted into luminosity with time: the radiated energy differs only by some $\textrm{10}^{49}$\vgv{}erg between normal SNe Ia and SC objects. Eventually the complete CSM would be dragged with the expanding ejecta, and the interaction would become less violent. The contribution to the luminosity would then be expected to gradually vanish, which is consistent with the continuum getting less intense in our models.

A relatively strong argument against an interaction model is the lack of direct evidence in SN~2009dc; in Type IIn and Ibn SNe, narrow H/He line emission is observed as a sign of interaction. Moreover, the light-curve of SN~2009dc after the photospheric phase shows a slope compatible with \Cofs\ decay for a relatively long time \citep{sil11}, suggesting that all light may be generated by radioactivity. Whether all this really constitutes a problem for an interaction scenario must finally be judged on the basis of detailed radiation-hydrodynamics calculations. Also, population synthesis studies and explosion simulations should clarify which progenitor systems for this scenario may exist in significant numbers.

\section{Summary and conclusions}
\label{sec:conclusions}

We have conducted an analysis of SN~2009dc, a representative example of the poorly understood SC SN Ia subclass. These SNe are characterised by an extremely high luminosity, and by spectral lines typical of SNe Ia, with normal or even low velocities. The photospheric phase lasts unusually long, indicating dense ejecta. Using a radiative-transfer code to model spectra observed up to one month after $B$ maximum, we have inferred the abundance structure of the SN [``Abundance Tomography'', \citet{ste05}], assuming different density profiles. Thus we have tested different explosion scenarios for consistency.

With a density profile representing an explosion of a massive, rotating WD [\mbox{``AWD3det''} model, (\citealt{fink10diss} and Fink \etal, in prep.) -- an improved version of the ``AWD3 detonation'' model of \citet{pfa10b}], the abundance profile we infer allows Fe-group elements only at velocities $\lesssim$\hspace{0.1em}9000\vgv{}\kms. As only \mbox{$\textrm{1.1}$\vgv$\Msun$} of material are ejected at such low velocities, the \Nifs\ mass allowed with this model is low. In spherical symmetry, as assumed for the spectral models, Arnett's rule for the light curve \citep{arn82} demands the \Nifs\ mass of the SN to be $\sim$\hspace{0.1em}$\textrm{2.0}$\vgv\Msun\ (for \mbox{$t_r=\textrm{25}$\vgv{}d}). This discrepancy leads us to disfavour a rotating single WD model for SN~2009dc. However, the model may be rescued if either the explosion is strongly asymmetric [rendering our initial assumptions wrong -- \citet{hillebrandt07a,sim07b}, but see \citet{tan10}] or if explosions of C-O WDs more massive than $\textrm{2.0}$\vgv$\Msun$ existed \citep{hachisu12a}, for which the ejecta will be slower if the amount of burning products is not increased proportionally to the mass.

An alternative scenario may be core-collapse or a thermonuclear explosion within a star (Type \mbox{I\hspace{0.1em}\textonehalf} SN). This allows for a larger ejecta mass, leading to lower ejecta velocities. Thus, a larger \Nifs\ mass becomes compatible with the spectra. We showed this by conducting a tomography with an empirically constructed $\textrm{3}$\vgv$\Msun$ density profile. Although the integrated masses of \Nifs, IME and C/O are different from the AWD3det case, the abundance structure in velocity space remains similar. O dominates at \mbox{$v \gtrsim$\vgv{}9000\vgv{}\kms}, intermediate-mass elements below, and Fe-group elements including \Nifs, in the core. This stratification would have to be reproduced by a valid explosion model.

Finally, we explored whether a tomography of SN~2009dc can be performed assuming a normal SN Ia explosion enshrouded by a CSM (\cf \citealt{sca10}). The interaction of the ejecta with the CSM would in this scenario produce additional luminosity, distinguishing ``SC'' SNe Ia from normal ones. Thus, the exploding WD would not have to produce excessive amounts of \Nifs, and would not need to be supermassive. In order to test this model, we calculated synthetic spectra with an intrinsic SN luminosity reduced by \mbox{33\vgv{}$-$\vgv{}50\%} (with respect to the observed luminosity), using a W7 ($M_\textrm{Ch,non-rot}$) SN Ia density profile scaled to 70\% of its original kinetic energy. At each epoch, the difference between observed spectrum and model was fitted with a third-order polynomial, representing the luminosity contribution by interaction. This way, we obtained an outstanding reproduction of the observed spectra. The polynomial-fit ``continuum'' we used turned out to be very similar to an interaction-dominated spectrum of the Type Ibn SN~2006jc, apart from the He emission in that SN. In the context of an interaction scenario, the absence of H and He lines in SN~2009dc can only be explained if the companion is H-deficient and He-deficient (or not showing He lines because of the ionisation/excitation conditions). The scenario may thus correspond to a merger of two C-O WDs, if both a violent merger and an accretion-induced collapse, which currently are considered the most likely outcomes, can be avoided. Another possible progenitor system may be a C-O WD with a donor star which has lost parts of its envelope.

We conclude that the popular rotating / supermassive WD scenario is only one of several candidate models for explaining ``SC'' SNe Ia. While there are problems with this model, it might work with extremely massive WDs or when the explosions are strongly asymmetric. The alternatives have been largely unexplored in terms of hydrodynamical modelling. Therefore we hope that the present work encourages modellers to conduct ab-initio hydrodynamics and radiative-transfer simulations of all the scenarios we have discussed. The abundance profiles we have inferred may serve as a guideline for these simulations.

\section*{ACKNOWLEDGEMENTS}
This work was supported in part by the DFG-TCRC 33 ``The Dark Universe'', the DFG Emmy Noether programme (RO 3676/1-1) and the  programme ASI-INAF I/009/10/0. We thank J.~M.~M. Pfannes, J.~C. Niemeyer and W. Schmidt for providing their model data. Furthermore, we thank F.~K. R\"opke for helpful comments on the manuscript, K. Nomoto for discussions, and A. Pastorello for providing us the data of  SN~2006jc. We thank all others who contributed making good suggestions, in addition to everybody already mentioned.

\bibliographystyle{mn2e}
\bibliography{09dc.bib}

\appendix

\section{Empirically motivated density profiles}
\label{app:densityconstruction}
\label{app:linemeasurements}

We construct part of the density profiles we use in our study independent of explosion models (\cf Sec. \ref{sec:tomography-exp-densprofile}). The idea is to infer the outer density profile from the observed time evolution of the velocities at which prominent spectral lines form. Such velocities are measurable as absorption blueshifts\footnote{For the actual measurements, we use the procedure in \citet{hac06}, \ie\ we determine a pseudo-continuum and take the wavelength of maximum fractional depth as the blueshifted position.} in the spectra. In this appendix, we describe in some detail the method to infer the outer density, and present the results.

As an initial step, we make the (very approximate) assumption for spectral lines to \textit{form at a density constant over time} and exploit the fact that the ejecta are in force-free, homologous expansion (\cf Sec. \ref{sec:prerequisites}, $r$\vgv{}$=$\vgv{}$v$\vgv{}$\times$\vgv{}$t\,$). With these assumptions, the evolution of line velocities is directly related to the radial density profile. The density at a space-time point $(v_2,t_2)$ with respect to the density at another point $(v_1,t_1)$ can be calculated from the radial density profile. In turn, knowledge on the ratio of the density at two points $(v_1,t_1)$ and $(v_2,t_2)$ allows us to constrain the shape of the density profile [\ie $\rho(v_2,t_0)$ can be calculated from $\rho(v_1,t_0)$ for arbitrary $t_0$]. If we have velocities of line formation $v_{1\textrm{...}n}$ at times $t_{1\textrm{...}n}$, and assume a density ratio of 1.0 \textit{(constant density)} at these points, we obtain
\begin{eqnarray}
\label{formula:densitycalc}
\rho(v_i,t_i) \!\! & \!\!=\!\! & \!\! \rho(v_1,t_1) \ \textrm{ (constant density) } \ \textrm{ and } \nonumber \\
\rho(v_i,t_1) \!\! & \!\!=\!\! & \!\! \rho(v_i,t_i) \times (t_i / t_1)^3 \ \textrm{ (force-free, hom. expansion)} \nonumber \\
\Rightarrow \rho(v_i,t_1) \!\! & \!\!=\!\! & \!\! \rho(v_1,t_1) \times (t_i / t_1)^3
\end{eqnarray}
Here, the time offsets $t_i$ from explosion onset depend on the assumed rise time $t_\textrm{r}$. For each spectrum, $t_i$ is calculated by adding the assumed rise time $t_\textrm{r}$ to the observed epoch relative to $B$ maximum. We thus obtain from our measurements a two-parameter family of models with $t_\textrm{r}$ and the absolute density scale [here appearing in the form of $\rho(v_1,t_1)$] as free parameters.

We perform line velocity measurements in all spectra observed up to five days past $B$ max. With this choice, we avoid the epochs at which the velocities of some lines become constant (cf. \citealt{tau11}) as the photosphere moves below the zone where the respective elements are abundant. We use several prominent absorption features due to different species (\SiII\ $\lambda\lambda \textrm{4130,6355}$, \SiIII\ $\lambda \textrm{4563}$, \CII\ $\lambda \textrm{6580}$). Taking into account various ions with different ionisation and excitation energies and evaluating the average velocity, we make sure that the zone of line formation we infer does not track the excitation/ionisation conditions, but the density (as far as possible).

The measured velocities are given in Table~\ref{tab:rhofromvel}. Exactly speaking, blueshifts measured in spectra give us velocity components along the line of sight. The corresponding radial velocity of line formation is somewhat higher. Thus, we calculate  \textit{radial velocities of line formation} by multiplying the average line velocities with an estimated factor\footnote{For a hypothetical feature forming in a thin zone directly above the photosphere, the curvature effect which makes the difference from line-of-sight velocities to radial velocity is most extreme: absorption is possible at line-of-sight velocities from 0\vgv\kms up to the radial velocity. The radial velocity then differs by a factor of 1.33 from the average line-of-sight velocity (calculated integrating over the lines of sight, assuming no limb darkening).} of 1.25 (see Table~\ref{tab:rhofromvel}). These radial velocities are used as $v_i$ in Equation~(\ref{formula:densitycalc}).

\begin{table}
\scriptsize
\caption{Velocities measured in the \SiII\ $\lambda\lambda \textrm{4130,6355}$, \SiIII\ $\lambda \textrm{4563}$ and \CII\ $\lambda \textrm{6580}$ lines in the observed spectra of SN~2009dc up to 5\vgv{}d past $B$ max. We give the individual velocities, their average and a radial velocity of line formation $v_\textrm{LF}$. The values $v_\textrm{LF}$ are the $v_i$ used in equation~(\ref{formula:densitycalc}).}
\label{tab:rhofromvel}
\centering
\begin{tabular}{ccccccc}
$\!\!$ \makebox[1.0cm][c]{Epoch (rel.} $\!\!\!\!$ & $\!\!\!\!$ $v$ $\!\!\!\!$ & $\!\!\!\!$ $v$ $\!\!\!\!$ & $\!\!\!\!$ $v$ $\!\!\!\!$ & $\!\!\!\!$ $v$ $\!\!\!\!$ & $\!\!\!\!$ $v$ $\!\!\!\!$ & $\!\!\!\!$ $v_\textrm{LF}$ $\!\!\!\!$  \\	
\vspace{-0.80pt} & \vspace{-0.80pt} & \vspace{-0.80pt} & \vspace{-0.80pt} & \vspace{-0.80pt} & \vspace{-0.80pt} & \vspace{-0.80pt} \\
$\!\!$ \makebox[1.0cm][c]{to $B$ max.)} $\!\!\!\!$ & $\!\!\!\!\!$ $(\textrm{\SiII\ }\lambda\textrm{4130})$	$\!\!\!\!\!$ & $\!\!\!\!\!$ $(\textrm{\SiII\ }\lambda\textrm{6355})$ $\!\!\!\!\!$ & $\!\!\!\!\!$ $(\textrm{\SiIII\ }\lambda\textrm{4563})$	$\!\!\!\!\!$ & $\!\!\!\!\!$ $(\textrm{\CII\ }\lambda\textrm{6580})$ $\!\!\!\!\!$ & 	$\!\!\!\!$ (line avg.) $\!\!\!\!$ & 	$\!\!\!\!$ (radial) $\!\!\!\!$ \\	
\vspace{-0.60pt} & \vspace{-0.60pt} & \vspace{-0.60pt} & \vspace{-0.60pt} & \vspace{-0.60pt} & \vspace{-0.60pt} & \vspace{-0.60pt} \\
$\!\!$ \makebox[1.0cm][c]{\protect[d\protect]} $\!\!\!\!$ & $\!\!\!\!$ [\kms] $\!\!\!\!$ & $\!\!\!\!$ [\kms] $\!\!\!\!$ & $\!\!\!\!$ [\kms] $\!\!\!\!$ & $\!\!\!\!$ [\kms] $\!\!\!\!$ & $\!\!\!\!$ [\kms] $\!\!\!\!$ & $\!\!\!\!\!\!$ [\kms] $\!\!\!\!\!\!$ \\ 
\vspace{-0.80pt} & \vspace{-0.80pt} & \vspace{-0.80pt} & \vspace{-0.80pt} & \vspace{-0.80pt} & \vspace{-0.80pt} & \vspace{-0.80pt} \\ \hline
$-\textrm{9.4}$	$\!\!\!\!$ & $\!\!\!\!$	8298.9	$\!\!\!\!$ & $\!\!\!\!$		9071.6	$\!\!\!\!$ & $\!\!\!\!$	7061.8	$\!\!\!\!$ & $\!\!\!\!$		10142.0	$\!\!\!\!$ & $\!\!\!\!$	8643.6	$\!\!\!\!$ & $\!\!\!\!$		10804.5	\\
$-\textrm{8.5}$	$\!\!\!\!$ & $\!\!\!\!$	7906.2	$\!\!\!\!$ & $\!\!\!\!$		9241.8	$\!\!\!\!$ & $\!\!\!\!$	6665.3	$\!\!\!\!$ & $\!\!\!\!$		10168.7	$\!\!\!\!$ & $\!\!\!\!$	8495.5	$\!\!\!\!$ & $\!\!\!\!$		10619.4	\\
$-\textrm{7.5}$	$\!\!\!\!$ & $\!\!\!\!$	7628.1	$\!\!\!\!$ & $\!\!\!\!$		8754.1	$\!\!\!\!$ & $\!\!\!\!$	6457.1	$\!\!\!\!$ & $\!\!\!\!$	\phao	9963.1	$\!\!\!\!$ & $\!\!\!\!$	8200.6	$\!\!\!\!$ & $\!\!\!\!$		10250.8	\\
$-\textrm{3.7}$	$\!\!\!\!$ & $\!\!\!\!$	7219.0	$\!\!\!\!$ & $\!\!\!\!$		8260.8	$\!\!\!\!$ & $\!\!\!\!$	5853.6	$\!\!\!\!$ & $\!\!\!\!$	\phao	9517.6	$\!\!\!\!$ & $\!\!\!\!$	7712.7	$\!\!\!\!$ & $\!\!\!\!$	\phao	9640.9	\\
2.4/2.5	$\!\!\!\!$ & $\!\!\!\!$	6224.3	$\!\!\!\!$ & $\!\!\!\!$		7757.0	$\!\!\!\!$ & $\!\!\!\!$	5333.7	$\!\!\!\!$ & $\!\!\!\!$	\phao	7374.8	$\!\!\!\!$ & $\!\!\!\!$	6672.5	$\!\!\!\!$ & $\!\!\!\!$	\phao	8340.6	\\
4.5	$\!\!\!\!$ & $\!\!\!\!$	6021.9	$\!\!\!\!$ & $\!\!\!\!$		7482.9	$\!\!\!\!$ & $\!\!\!\!$	5197.8	$\!\!\!\!$ & $\!\!\!\!$	\phao	7191.3	$\!\!\!\!$ & $\!\!\!\!$	6473.4	$\!\!\!\!$ & $\!\!\!\!$	\phao	8091.8	\\
\hline
\end{tabular}
\end{table} 

In order to write the inferred density profile (for the outer ejecta) in a compact form and to avoid kinks in the profiles, which may result from measurement uncertainties, we fit exponentials 
\begin{equation}
\label{formula:rhoprofile}
\rho(v,t) = a \times  \exp(-b v ) \times (t_0 / t)^3
\end{equation}
to the measured values $\rho(v_i,t_i)$. Here, $t_0$ is again an arbitrarily chosen reference time. As the observations do not constrain the absolute density scale (\ie $a$ in this formula), the fit only gives a meaningful $b$ which depends on the assumed rise time $t_\textrm{r}$ ($t$\vgv{}$=$\vgv{}$t_\textrm{r}$\vgv{}$+$\vgv{}epoch rel. to $B$ max.). For obtaining an actual density profile, $t_\textrm{r}$ and finally $a$ have to be fixed. We therefore construct a grid of density profiles with different $t_\textrm{r}$ and $a$ (absolute density scales). We assume rise times $t_\textrm{r}$ of \mbox{21\vgv{}d}, \mbox{22.5\vgv{}d}, \mbox{25\vgv{}d}, \mbox{27.5\vgv{}d}, \mbox{30\vgv{}d}, \mbox{35\vgv{}d} and \mbox{40\vgv{}d} and choose $a$ such that our models intersect the W7 ``standard'' SN Ia model\footnote{W7 is evaluated for the reference time $t_0$ we use in Equation (\ref{formula:rhoprofile}).} \citep{nom84w7} at 9000, 10000, 12000 or 14000\vgv{}\kms. As the W7 density is relatively flat in this velocity range, a smaller intersection velocity $v_\times$ results in a lower $a$ and lower overall densities. This is illustrated in Fig.~\ref{fig:densityprofiles-exp-early}, where examples of our density profiles are plotted. Our grid of profiles, which are given names reflecting $v_\times$ (\eg ``exp9''  matches W7 at $v_\times=\textrm{9000}$\vgv{}\kms), is chosen so as to cover the physical parameter space in which solutions for SN 2009dc will be found (see main text).

Table \ref{tab:empiricaldensityparameters} gives the factors $a$ (for a given $t_0$) and slopes $b$ of our empirical density profiles (\cf Equation \ref{formula:rhoprofile}). The value of $b$ only depends on the rise time (assumed to be 21\vgv{}d, 22.5\vgv{}d, 25\vgv{}d, 27.5\vgv{}d, 30\vgv{}d, 35\vgv{}d or 40\vgv{}d). The parameter $a$ depends on the assumed rise time as well, but furthermore on the choice of $v_\times$ (9000, 10000, 12000 or 14000\vgv{}\kms), at which the profile intersects W7 \citep{nom84w7}. We have chosen the reference time $t_0$ to be 31.5\vgv{}d.

\begin{table}
\scriptsize
\caption{Parameters $a$ and $b$ of our empirical, exponential density models ``exp...'' for different choices of $v_\times$ (exp9: $v_\times=\textrm{9000}$\vgv\kms{}, exp10: $v_\times=\textrm{10000}$\vgv\kms{}, ...) and assumed rise times $t_r$. The $a$ values are given for a reference time of $t_0=\textrm{31.5\vgv{}d}$.}
\label{tab:empiricaldensityparameters}
\centering
\begin{tabular}{cccccc}
$\!\!\!\!\!\!$ Assumed $t_r$ $\!\!\!\!$ & $\!\!\!\!$  $a$ (exp9 model) $\!\!\!\!\!\!$ & $a$ (exp10) & $a$ (exp12) & $a$ (exp14) & $b$ (all models) \\ 
\vspace{-0.60pt} & \vspace{-0.60pt} & \vspace{-0.60pt} & \vspace{-0.60pt} & \vspace{-0.60pt} & \vspace{-0.60pt} \\
 $[$d$]$ & \multicolumn{2}{c}{$\!\!\!\![$$\textrm{10}^{-12}$g\vgv{}cm$^{-3}$$]\!\!\!\!$} & \multicolumn{2}{c}{$\!\!\!\![$$\textrm{10}^{-12}$g\vgv{}cm$^{-3}$$]\!\!\!\!$} & $\!\!\!\![$$\textrm{10}^{-9}$s\vgv{}km$^{-1}$$]\!\!\!\!$ \\ 
\vspace{-0.80pt} & \vspace{-0.80pt} & \vspace{-0.80pt} & \vspace{-0.80pt} & \vspace{-0.80pt} & \vspace{-0.80pt} \\ \hline
    21.0   $\!\!\!\!$ & $\!\!\!\!$      15.39   $\!\!\!\!$ & $\!\!\!\!$      28.33   $\!\!\!\!$ & $\!\!\!\!$      77.21   $\!\!\!\!$ & $\!\!\!\!$     324.8   $\!\!\!\!$ & $\!\!\!\!$      8.25   \\ 
    22.5   $\!\!\!\!$ & $\!\!\!\!$      $\phantom{\textrm{1}}$8.93   $\!\!\!\!$ & $\!\!\!\!$      15.48   $\!\!\!\!$ & $\!\!\!\!$      37.38   $\!\!\!\!$ & $\!\!\!\!$     139.3   $\!\!\!\!$ & $\!\!\!\!$      7.64   \\ 
    25.0   $\!\!\!\!$ & $\!\!\!\!$       $\phantom{\textrm{1}}$4.21   $\!\!\!\!$ & $\!\!\!\!$       $\phantom{\textrm{1}}$6.71   $\!\!\!\!$ & $\!\!\!\!$      13.70   $\!\!\!\!$ & $\!\!\!\!$      $\phantom{\textrm{1}}$43.2   $\!\!\!\!$ & $\!\!\!\!$      6.81   \\ 
    27.5   $\!\!\!\!$ & $\!\!\!\!$       $\phantom{\textrm{1}}$2.29   $\!\!\!\!$ & $\!\!\!\!$       $\phantom{\textrm{1}}$3.42   $\!\!\!\!$ & $\!\!\!\!$       $\phantom{\textrm{1}}$6.10   $\!\!\!\!$ & $\!\!\!\!$      $\phantom{\textrm{1}}$16.8   $\!\!\!\!$ & $\!\!\!\!$      6.13   \\ 
    30.0   $\!\!\!\!$ & $\!\!\!\!$       $\phantom{\textrm{1}}$1.39   $\!\!\!\!$ & $\!\!\!\!$       $\phantom{\textrm{1}}$1.96   $\!\!\!\!$ & $\!\!\!\!$       $\phantom{\textrm{1}}$3.14   $\!\!\!\!$ & $\!\!\!\!$       $\phantom{\textrm{11}}$7.7   $\!\!\!\!$ & $\!\!\!\!$      5.58   \\ 
    35.0   $\!\!\!\!$ & $\!\!\!\!$       $\phantom{\textrm{1}}$0.65   $\!\!\!\!$ & $\!\!\!\!$       $\phantom{\textrm{1}}$0.83   $\!\!\!\!$ & $\!\!\!\!$      $\phantom{\textrm{1}}$1.12   $\!\!\!\!$ & $\!\!\!\!$       $\phantom{\textrm{11}}$2.3   $\!\!\!\!$ & $\!\!\!\!$      4.72   \\ 
    40.0   $\!\!\!\!$ & $\!\!\!\!$       $\phantom{\textrm{1}}$0.37   $\!\!\!\!$ & $\!\!\!\!$       $\phantom{\textrm{1}}$0.44   $\!\!\!\!$ & $\!\!\!\!$      $\phantom{\textrm{1}}$0.53   $\!\!\!\!$ & $\!\!\!\!$       $\phantom{\textrm{11}}$1.0   $\!\!\!\!$ & $\!\!\!\!$      4.09   \\ 
\hline 
\end{tabular}
\end{table}

\section{Parameters of the Abundance Tomography models}
\label{app:modelparameters-tomo}

The code input parameters for our Abundance Tomography models (Sections \ref{sec:tomography-awd3det}, \ref{sec:tomography-exp} and \ref{sec:tomography-csm}) are listed in Table~\ref{tab:modelparameters}. Apart from the abundances, the code takes as input the photospheric velocity $v_\textrm{ph}$, the time from explosion $t$ and the bolometric luminosity $L_\textrm{bol}$. In addition to the input values, Table~\ref{tab:modelparameters} also gives the calculated temperature of the photospheric black body, $T_\textrm{ph}$, for each model.

\begin{table*}
\scriptsize
\caption{Parameters of the models. The abundance values, which are only given for significant elements, apply between the photospheric velocity $v_\textrm{ph}$ of the respective epoch and $v_\textrm{ph}$ of the previous epoch. The outer shells (``epoch'' designations ``oo'', ``o'') improve the fit to the $-\textrm{9.4}$\vgv{}d spectrum (\cf main text). For these shells,  $v_\textrm{ph}$ is not a photospheric velocity, but the velocity of the lower boundary of the respective zone.}
\label{tab:modelparameters}
\centering
\begin{tabular}{lrrrrcccccccccccc}
Model	$\!\!\!\!$ & $\!\!$	Epochs\vspace{0.2cm}	$\!\!\!\!\!\!$ & 	$\mathrm{lg}\!\left(\frac{L_\textrm{bol}}{\Lsun}\right)$	$\!\!\!\!\!\!\!$ & $\!\!$	$v_{\textrm{lowbnd}}$ $\!\!\!$ & $\!\!\!\!$	$T_\textrm{ph}$ &	\multicolumn{10}{c}{Element abundances (mass fractions)}  \\
\vspace{-0.80pt} & \vspace{-0.80pt} & \vspace{-0.80pt} & \vspace{-0.80pt} & \vspace{-0.80pt} & \vspace{-0.80pt} & \vspace{-0.80pt} & \vspace{-0.80pt} & \vspace{-0.80pt} & \vspace{-0.80pt} & \vspace{-0.80pt} & \vspace{-0.80pt} & \vspace{-0.80pt} & \vspace{-0.80pt} & \vspace{-0.80pt}  \\
$\!\!\!\!$ &  $\!\!\!\!$ [d] $\!\!$ & $\!\!\!\!$ $\!\!\!\!$ & $\!\!$	[\kms] $\!\!\!\!\!$  & $\!\!\!\!$ [K] $\!$ & $\!\!\!\!$	$X$(C)	$\!\!\!\!$ & $\!\!\!\!$	$X$(O)	$\!\!\!\!$ & $\!\!\!\!$	$X$(Mg)	$\!\!\!\!$ & $\!\!\!\!$	$X$(Si)	$\!\!\!\!$ & $\!\!\!\!$	$X$(S)	$\!\!\!\!$ & $\!\!\!\!$	$X$(Ca)	$\!\!\!\!$ & $\!\!\!\!$	$X$(Ti)	$\!\!\!\!$ & $\!\!\!\!$	$X$(Cr)	$\!\!\!\!$ & $\!\!\!\!$	$X$(Fe)$_{0}{}^{\textrm{a)}}$	$\!\!\!\!$ & $\!\!\!\!$	$X$(\Nifs)$_{0}{}^{\textrm{a)}}$ \\ 
\vspace{-0.60pt} & \vspace{-0.60pt} & \vspace{-0.60pt} & \vspace{-0.60pt} & \vspace{-0.60pt} & \vspace{-0.60pt} & \vspace{-0.60pt} & \vspace{-0.60pt} & \vspace{-0.60pt} & \vspace{-0.60pt} & \vspace{-0.60pt} & \vspace{-0.60pt} & \vspace{-0.60pt} & \vspace{-0.60pt} & \vspace{-0.60pt}  \\ \hline
09dc-AWD3det $\!\!\!\!$ & $\!\!\!\!$ oo $\!\!\!\!$ & $\!\!\!\!$ {-} $\!\!\!\!$ & $\!\!\!\!$ 11700 $\!\!\!\!$ & $\!\!\!\!$ {-} $\!\!\!\!$ & $\!\!\!\!$ 0.03 $\!\!\!\!$ & $\!\!\!\!$ 0.96 $\!\!\!\!$ & $\!\!\!\!$ 0.00 $\!\!\!\!$ & $\!\!\!\!$ 0.00 $\!\!\!\!$ & $\!\!\!\!$ 0.00 $\!\!\!\!$ & $\!\!\!\!$ 0.0000 $\!\!\!\!$ & $\!\!\!\!$ 0.0000 $\!\!\!\!$ & $\!\!\!\!$ 0.0000 $\!\!\!\!$ & $\!\!\!\!$ 0.0000 $\!\!\!\!$ & $\!\!\!\!$ 0.0000 $\!\!\!\!$ & $\!\!\!\!$  $\!\!\!\!$ & $\!\!\!\!$  \\ 
$t_{r}=\textrm{30.0}$\hspace{0.25em}d $\!\!\!\!$ & $\!\!\!\!$ o $\!\!\!\!$ & $\!\!\!\!$ {-} $\!\!\!\!$ & $\!\!\!\!$ 9650 $\!\!\!\!$ & $\!\!\!\!$ {-} $\!\!\!\!$ & $\!\!\!\!$ 0.02 $\!\!\!\!$ & $\!\!\!\!$ 0.83 $\!\!\!\!$ & $\!\!\!\!$ 0.04 $\!\!\!\!$ & $\!\!\!\!$ 0.07 $\!\!\!\!$ & $\!\!\!\!$ 0.04 $\!\!\!\!$ & $\!\!\!\!$ 0.0000 $\!\!\!\!$ & $\!\!\!\!$ 0.0000 $\!\!\!\!$ & $\!\!\!\!$ 0.0001 $\!\!\!\!$ & $\!\!\!\!$ 0.0000 $\!\!\!\!$ & $\!\!\!\!$ 0.0008 $\!\!\!\!$ & $\!\!\!\!$  $\!\!\!\!$ & $\!\!\!\!$  \\ 
{} $\!\!\!\!$ & $\!\!\!\!$ -9 $\!\!\!\!$ & $\!\!\!\!$ 9.90 $\!\!\!\!$ & $\!\!\!\!$ 9050 $\!\!\!\!$ & $\!\!\!\!$ 14264 $\!\!\!\!$ & $\!\!\!\!$ 0.02 $\!\!\!\!$ & $\!\!\!\!$ 0.56 $\!\!\!\!$ & $\!\!\!\!$ 0.08 $\!\!\!\!$ & $\!\!\!\!$ 0.15 $\!\!\!\!$ & $\!\!\!\!$ 0.10 $\!\!\!\!$ & $\!\!\!\!$ 0.0020 $\!\!\!\!$ & $\!\!\!\!$ 0.0009 $\!\!\!\!$ & $\!\!\!\!$ 0.0090 $\!\!\!\!$ & $\!\!\!\!$ 0.0000 $\!\!\!\!$ & $\!\!\!\!$ 0.0800 $\!\!\!\!$ & $\!\!\!\!$  $\!\!\!\!$ & $\!\!\!\!$  \\ 
{} $\!\!\!\!$ & $\!\!\!\!$ -4 $\!\!\!\!$ & $\!\!\!\!$ 10.01 $\!\!\!\!$ & $\!\!\!\!$ 8300 $\!\!\!\!$ & $\!\!\!\!$ 13931 $\!\!\!\!$ & $\!\!\!\!$ 0.02 $\!\!\!\!$ & $\!\!\!\!$ 0.45 $\!\!\!\!$ & $\!\!\!\!$ 0.06 $\!\!\!\!$ & $\!\!\!\!$ 0.25 $\!\!\!\!$ & $\!\!\!\!$ 0.10 $\!\!\!\!$ & $\!\!\!\!$ 0.0060 $\!\!\!\!$ & $\!\!\!\!$ 0.0010 $\!\!\!\!$ & $\!\!\!\!$ 0.0100 $\!\!\!\!$ & $\!\!\!\!$ 0.0000 $\!\!\!\!$ & $\!\!\!\!$ 0.1100 $\!\!\!\!$ & $\!\!\!\!$  $\!\!\!\!$ & $\!\!\!\!$  \\ 
{} $\!\!\!\!$ & $\!\!\!\!$ +5 $\!\!\!\!$ & $\!\!\!\!$ 9.98 $\!\!\!\!$ & $\!\!\!\!$ 7900 $\!\!\!\!$ & $\!\!\!\!$ 11519 $\!\!\!\!$ & $\!\!\!\!$ 0.01 $\!\!\!\!$ & $\!\!\!\!$ 0.27 $\!\!\!\!$ & $\!\!\!\!$ 0.02 $\!\!\!\!$ & $\!\!\!\!$ 0.45 $\!\!\!\!$ & $\!\!\!\!$ 0.10 $\!\!\!\!$ & $\!\!\!\!$ 0.0100 $\!\!\!\!$ & $\!\!\!\!$ 0.0005 $\!\!\!\!$ & $\!\!\!\!$ 0.0100 $\!\!\!\!$ & $\!\!\!\!$ 0.0000 $\!\!\!\!$ & $\!\!\!\!$ 0.1200 $\!\!\!\!$ & $\!\!\!\!$  $\!\!\!\!$ & $\!\!\!\!$  \\ 
{} $\!\!\!\!$ & $\!\!\!\!$ +13 $\!\!\!\!$ & $\!\!\!\!$ 9.82 $\!\!\!\!$ & $\!\!\!\!$ 7500 $\!\!\!\!$ & $\!\!\!\!$ 9224 $\!\!\!\!$ & $\!\!\!\!$ 0.01 $\!\!\!\!$ & $\!\!\!\!$ 0.15 $\!\!\!\!$ & $\!\!\!\!$ 0.01 $\!\!\!\!$ & $\!\!\!\!$ 0.50 $\!\!\!\!$ & $\!\!\!\!$ 0.06 $\!\!\!\!$ & $\!\!\!\!$ 0.0060 $\!\!\!\!$ & $\!\!\!\!$ 0.0002 $\!\!\!\!$ & $\!\!\!\!$ 0.0080 $\!\!\!\!$ & $\!\!\!\!$ 0.0000 $\!\!\!\!$ & $\!\!\!\!$ 0.2500 $\!\!\!\!$ & $\!\!\!\!$  $\!\!\!\!$ & $\!\!\!\!$  \\ 
{} $\!\!\!\!$ & $\!\!\!\!$ +23 $\!\!\!\!$ & $\!\!\!\!$ 9.65 $\!\!\!\!$ & $\!\!\!\!$ 7200 $\!\!\!\!$ & $\!\!\!\!$ 7591 $\!\!\!\!$ & $\!\!\!\!$ 0.01 $\!\!\!\!$ & $\!\!\!\!$ 0.08 $\!\!\!\!$ & $\!\!\!\!$ 0.00 $\!\!\!\!$ & $\!\!\!\!$ 0.30 $\!\!\!\!$ & $\!\!\!\!$ 0.04 $\!\!\!\!$ & $\!\!\!\!$ 0.0030 $\!\!\!\!$ & $\!\!\!\!$ 0.0000 $\!\!\!\!$ & $\!\!\!\!$ 0.0040 $\!\!\!\!$ & $\!\!\!\!$ 0.0250 $\!\!\!\!$ & $\!\!\!\!$ 0.5500 $\!\!\!\!$ & $\!\!\!\!$  $\!\!\!\!$ & $\!\!\!\!$  \\ 
{} $\!\!\!\!$ & $\!\!\!\!$ +36 $\!\!\!\!$ & $\!\!\!\!$ 9.39 $\!\!\!\!$ & $\!\!\!\!$ 4700 $\!\!\!\!$ & $\!\!\!\!$ 7291 $\!\!\!\!$ & $\!\!\!\!$ 0.00 $\!\!\!\!$ & $\!\!\!\!$ 0.01 $\!\!\!\!$ & $\!\!\!\!$ 0.00 $\!\!\!\!$ & $\!\!\!\!$ 0.05 $\!\!\!\!$ & $\!\!\!\!$ 0.01 $\!\!\!\!$ & $\!\!\!\!$ 0.0015 $\!\!\!\!$ & $\!\!\!\!$ 0.0000 $\!\!\!\!$ & $\!\!\!\!$ 0.0020 $\!\!\!\!$ & $\!\!\!\!$ 0.2300 $\!\!\!\!$ & $\!\!\!\!$ 0.7000 $\!\!\!\!$ & $\!\!\!\!$  $\!\!\!\!$ & $\!\!\!\!$  \\ 
09dc-AWD3det $\!\!\!\!$ & $\!\!\!\!$ oo $\!\!\!\!$ & $\!\!\!\!$ {-} $\!\!\!\!$ & $\!\!\!\!$ 11700 $\!\!\!\!$ & $\!\!\!\!$ {-} $\!\!\!\!$ & $\!\!\!\!$ 0.03 $\!\!\!\!$ & $\!\!\!\!$ 0.96 $\!\!\!\!$ & $\!\!\!\!$ 0.00 $\!\!\!\!$ & $\!\!\!\!$ 0.00 $\!\!\!\!$ & $\!\!\!\!$ 0.00 $\!\!\!\!$ & $\!\!\!\!$ 0.0000 $\!\!\!\!$ & $\!\!\!\!$ 0.0000 $\!\!\!\!$ & $\!\!\!\!$ 0.0000 $\!\!\!\!$ & $\!\!\!\!$ 0.0000 $\!\!\!\!$ & $\!\!\!\!$ 0.0000 $\!\!\!\!$ & $\!\!\!\!$  $\!\!\!\!$ & $\!\!\!\!$  \\ 
$E(B-V)=\textrm{0.06}$\vgv{}mag $\!\!\!\!$ & $\!\!\!\!$ o $\!\!\!\!$ & $\!\!\!\!$ {-} $\!\!\!\!$ & $\!\!\!\!$ 9300 $\!\!\!\!$ & $\!\!\!\!$ {-} $\!\!\!\!$ & $\!\!\!\!$ 0.02 $\!\!\!\!$ & $\!\!\!\!$ 0.89 $\!\!\!\!$ & $\!\!\!\!$ 0.02 $\!\!\!\!$ & $\!\!\!\!$ 0.05 $\!\!\!\!$ & $\!\!\!\!$ 0.02 $\!\!\!\!$ & $\!\!\!\!$ 0.0000 $\!\!\!\!$ & $\!\!\!\!$ 0.0000 $\!\!\!\!$ & $\!\!\!\!$ 0.0001 $\!\!\!\!$ & $\!\!\!\!$ 0.0000 $\!\!\!\!$ & $\!\!\!\!$ 0.0006 $\!\!\!\!$ & $\!\!\!\!$  $\!\!\!\!$ & $\!\!\!\!$  \\ 
$t_{r}=\textrm{27.8}$\vgv{}d $\!\!\!\!$ & $\!\!\!\!$ $-\textrm{9}$ $\!\!\!\!$ & $\!\!\!\!$ 9.72 $\!\!\!\!$ & $\!\!\!\!$ 8700 $\!\!\!\!$ & $\!\!\!\!$ 14178 $\!\!\!\!$ & $\!\!\!\!$ 0.02 $\!\!\!\!$ & $\!\!\!\!$ 0.76 $\!\!\!\!$ & $\!\!\!\!$ 0.04 $\!\!\!\!$ & $\!\!\!\!$ 0.09 $\!\!\!\!$ & $\!\!\!\!$ 0.03 $\!\!\!\!$ & $\!\!\!\!$ 0.0015 $\!\!\!\!$ & $\!\!\!\!$ 0.0006 $\!\!\!\!$ & $\!\!\!\!$ 0.0055 $\!\!\!\!$ & $\!\!\!\!$ 0.0000 $\!\!\!\!$ & $\!\!\!\!$ 0.0550 $\!\!\!\!$ & $\!\!\!\!$  $\!\!\!\!$ & $\!\!\!\!$  \\ 
{} $\!\!\!\!$ & $\!\!\!\!$ -4 $\!\!\!\!$ & $\!\!\!\!$ 9.81 $\!\!\!\!$ & $\!\!\!\!$ 8000 $\!\!\!\!$ & $\!\!\!\!$ 13308 $\!\!\!\!$ & $\!\!\!\!$ 0.02 $\!\!\!\!$ & $\!\!\!\!$ 0.60 $\!\!\!\!$ & $\!\!\!\!$ 0.04 $\!\!\!\!$ & $\!\!\!\!$ 0.18 $\!\!\!\!$ & $\!\!\!\!$ 0.06 $\!\!\!\!$ & $\!\!\!\!$ 0.0080 $\!\!\!\!$ & $\!\!\!\!$ 0.0008 $\!\!\!\!$ & $\!\!\!\!$ 0.0080 $\!\!\!\!$ & $\!\!\!\!$ 0.0000 $\!\!\!\!$ & $\!\!\!\!$ 0.0800 $\!\!\!\!$ & $\!\!\!\!$  $\!\!\!\!$ & $\!\!\!\!$  \\ 
{} $\!\!\!\!$ & $\!\!\!\!$ +5 $\!\!\!\!$ & $\!\!\!\!$ 9.79 $\!\!\!\!$ & $\!\!\!\!$ 7350 $\!\!\!\!$ & $\!\!\!\!$ 11227 $\!\!\!\!$ & $\!\!\!\!$ 0.02 $\!\!\!\!$ & $\!\!\!\!$ 0.27 $\!\!\!\!$ & $\!\!\!\!$ 0.04 $\!\!\!\!$ & $\!\!\!\!$ 0.40 $\!\!\!\!$ & $\!\!\!\!$ 0.10 $\!\!\!\!$ & $\!\!\!\!$ 0.0080 $\!\!\!\!$ & $\!\!\!\!$ 0.0015 $\!\!\!\!$ & $\!\!\!\!$ 0.0150 $\!\!\!\!$ & $\!\!\!\!$ 0.0000 $\!\!\!\!$ & $\!\!\!\!$ 0.1500 $\!\!\!\!$ & $\!\!\!\!$  $\!\!\!\!$ & $\!\!\!\!$  \\ 
{} $\!\!\!\!$ & $\!\!\!\!$ +13 $\!\!\!\!$ & $\!\!\!\!$ 9.66 $\!\!\!\!$ & $\!\!\!\!$ 7000 $\!\!\!\!$ & $\!\!\!\!$ 9104 $\!\!\!\!$ & $\!\!\!\!$ 0.01 $\!\!\!\!$ & $\!\!\!\!$ 0.18 $\!\!\!\!$ & $\!\!\!\!$ 0.02 $\!\!\!\!$ & $\!\!\!\!$ 0.45 $\!\!\!\!$ & $\!\!\!\!$ 0.07 $\!\!\!\!$ & $\!\!\!\!$ 0.0040 $\!\!\!\!$ & $\!\!\!\!$ 0.0005 $\!\!\!\!$ & $\!\!\!\!$ 0.0150 $\!\!\!\!$ & $\!\!\!\!$ 0.0000 $\!\!\!\!$ & $\!\!\!\!$ 0.2500 $\!\!\!\!$ & $\!\!\!\!$  $\!\!\!\!$ & $\!\!\!\!$  \\ 
{} $\!\!\!\!$ & $\!\!\!\!$ +23 $\!\!\!\!$ & $\!\!\!\!$ 9.51 $\!\!\!\!$ & $\!\!\!\!$ 6650 $\!\!\!\!$ & $\!\!\!\!$ 7518 $\!\!\!\!$ & $\!\!\!\!$ 0.01 $\!\!\!\!$ & $\!\!\!\!$ 0.14 $\!\!\!\!$ & $\!\!\!\!$ 0.01 $\!\!\!\!$ & $\!\!\!\!$ 0.30 $\!\!\!\!$ & $\!\!\!\!$ 0.04 $\!\!\!\!$ & $\!\!\!\!$ 0.0020 $\!\!\!\!$ & $\!\!\!\!$ 0.0001 $\!\!\!\!$ & $\!\!\!\!$ 0.0060 $\!\!\!\!$ & $\!\!\!\!$ 0.0000 $\!\!\!\!$ & $\!\!\!\!$ 0.5000 $\!\!\!\!$ & $\!\!\!\!$  $\!\!\!\!$ & $\!\!\!\!$  \\ 
{} $\!\!\!\!$ & $\!\!\!\!$ +36 $\!\!\!\!$ & $\!\!\!\!$ 9.26 $\!\!\!\!$ & $\!\!\!\!$ 4200 $\!\!\!\!$ & $\!\!\!\!$ 7346 $\!\!\!\!$ & $\!\!\!\!$ 0.00 $\!\!\!\!$ & $\!\!\!\!$ 0.11 $\!\!\!\!$ & $\!\!\!\!$ 0.01 $\!\!\!\!$ & $\!\!\!\!$ 0.25 $\!\!\!\!$ & $\!\!\!\!$ 0.03 $\!\!\!\!$ & $\!\!\!\!$ 0.0005 $\!\!\!\!$ & $\!\!\!\!$ 0.0001 $\!\!\!\!$ & $\!\!\!\!$ 0.0050 $\!\!\!\!$ & $\!\!\!\!$ 0.0500 $\!\!\!\!$ & $\!\!\!\!$ 0.5500 $\!\!\!\!$ & $\!\!\!\!$  $\!\!\!\!$ & $\!\!\!\!$  \\ 
09dc-AWD3det $\!\!\!\!$ & $\!\!\!\!$ oo $\!\!\!\!$ & $\!\!\!\!$ {-} $\!\!\!\!$ & $\!\!\!\!$ 11800 $\!\!\!\!$ & $\!\!\!\!$ {-} $\!\!\!\!$ & $\!\!\!\!$ 0.03 $\!\!\!\!$ & $\!\!\!\!$ 0.97 $\!\!\!\!$ & $\!\!\!\!$ 0.00 $\!\!\!\!$ & $\!\!\!\!$ 0.00 $\!\!\!\!$ & $\!\!\!\!$ 0.00 $\!\!\!\!$ & $\!\!\!\!$ 0.0000 $\!\!\!\!$ & $\!\!\!\!$ 0.0000 $\!\!\!\!$ & $\!\!\!\!$ 0.0000 $\!\!\!\!$ & $\!\!\!\!$ 0.0010 $\!\!\!\!$ & $\!\!\!\!$ 0.0000 $\!\!\!\!$ & $\!\!\!\!$  $\!\!\!\!$ & $\!\!\!\!$  \\ 
solar progen. metallicity $\!\!\!\!\!\!\!$ & $\!\!\!\!$ o $\!\!\!\!$ & $\!\!\!\!$ {-} $\!\!\!\!$ & $\!\!\!\!$ 10300 $\!\!\!\!$ & $\!\!\!\!$ {-} $\!\!\!\!$ & $\!\!\!\!$ 0.02 $\!\!\!\!$ & $\!\!\!\!$ 0.81 $\!\!\!\!$ & $\!\!\!\!$ 0.03 $\!\!\!\!$ & $\!\!\!\!$ 0.08 $\!\!\!\!$ & $\!\!\!\!$ 0.06 $\!\!\!\!$ & $\!\!\!\!$ 0.0001 $\!\!\!\!$ & $\!\!\!\!$ 0.0001 $\!\!\!\!$ & $\!\!\!\!$ 0.0003 $\!\!\!\!$ & $\!\!\!\!$ 0.0012 $\!\!\!\!$ & $\!\!\!\!$ 0.0000 $\!\!\!\!$ & $\!\!\!\!$  $\!\!\!\!$ & $\!\!\!\!$  \\ 
$t_{r}=\textrm{30}$\vgv{}d $\!\!\!\!$ & $\!\!\!\!$ $-\textrm{9}$ $\!\!\!\!$ & $\!\!\!\!$ 9.88 $\!\!\!\!$ & $\!\!\!\!$ 9700 $\!\!\!\!$ & $\!\!\!\!$ 13511 $\!\!\!\!$ & $\!\!\!\!$ 0.02 $\!\!\!\!$ & $\!\!\!\!$ 0.42 $\!\!\!\!$ & $\!\!\!\!$ 0.06 $\!\!\!\!$ & $\!\!\!\!$ 0.25 $\!\!\!\!$ & $\!\!\!\!$ 0.10 $\!\!\!\!$ & $\!\!\!\!$ 0.0053 $\!\!\!\!$ & $\!\!\!\!$ 0.0001 $\!\!\!\!$ & $\!\!\!\!$ 0.0030 $\!\!\!\!$ & $\!\!\!\!$ 0.0650 $\!\!\!\!$ & $\!\!\!\!$ 0.0750 $\!\!\!\!$ & $\!\!\!\!$  $\!\!\!\!$ & $\!\!\!\!$  \\ 
{} $\!\!\!\!$ & $\!\!\!\!$ -4 $\!\!\!\!$ & $\!\!\!\!$ 9.99 $\!\!\!\!$ & $\!\!\!\!$ 8750 $\!\!\!\!$ & $\!\!\!\!$ 13544 $\!\!\!\!$ & $\!\!\!\!$ 0.02 $\!\!\!\!$ & $\!\!\!\!$ 0.28 $\!\!\!\!$ & $\!\!\!\!$ 0.05 $\!\!\!\!$ & $\!\!\!\!$ 0.36 $\!\!\!\!$ & $\!\!\!\!$ 0.12 $\!\!\!\!$ & $\!\!\!\!$ 0.0090 $\!\!\!\!$ & $\!\!\!\!$ 0.0003 $\!\!\!\!$ & $\!\!\!\!$ 0.0080 $\!\!\!\!$ & $\!\!\!\!$ 0.0500 $\!\!\!\!$ & $\!\!\!\!$ 0.1000 $\!\!\!\!$ & $\!\!\!\!$  $\!\!\!\!$ & $\!\!\!\!$  \\ 
{} $\!\!\!\!$ & $\!\!\!\!$ +5 $\!\!\!\!$ & $\!\!\!\!$ 9.97 $\!\!\!\!$ & $\!\!\!\!$ 8300 $\!\!\!\!$ & $\!\!\!\!$ 11098 $\!\!\!\!$ & $\!\!\!\!$ 0.02 $\!\!\!\!$ & $\!\!\!\!$ 0.02 $\!\!\!\!$ & $\!\!\!\!$ 0.02 $\!\!\!\!$ & $\!\!\!\!$ 0.65 $\!\!\!\!$ & $\!\!\!\!$ 0.10 $\!\!\!\!$ & $\!\!\!\!$ 0.0143 $\!\!\!\!$ & $\!\!\!\!$ 0.0003 $\!\!\!\!$ & $\!\!\!\!$ 0.0100 $\!\!\!\!$ & $\!\!\!\!$ 0.0200 $\!\!\!\!$ & $\!\!\!\!$ 0.1500 $\!\!\!\!$ & $\!\!\!\!$  $\!\!\!\!$ & $\!\!\!\!$  \\ 
{} $\!\!\!\!$ & $\!\!\!\!$ +13 $\!\!\!\!$ & $\!\!\!\!$ 9.82 $\!\!\!\!$ & $\!\!\!\!$ 8000 $\!\!\!\!$ & $\!\!\!\!$ 8921 $\!\!\!\!$ & $\!\!\!\!$ 0.01 $\!\!\!\!$ & $\!\!\!\!$ 0.01 $\!\!\!\!$ & $\!\!\!\!$ 0.01 $\!\!\!\!$ & $\!\!\!\!$ 0.60 $\!\!\!\!$ & $\!\!\!\!$ 0.04 $\!\!\!\!$ & $\!\!\!\!$ 0.0120 $\!\!\!\!$ & $\!\!\!\!$ 0.0002 $\!\!\!\!$ & $\!\!\!\!$ 0.0080 $\!\!\!\!$ & $\!\!\!\!$ 0.0010 $\!\!\!\!$ & $\!\!\!\!$ 0.3000 $\!\!\!\!$ & $\!\!\!\!$  $\!\!\!\!$ & $\!\!\!\!$  \\ 
{} $\!\!\!\!$ & $\!\!\!\!$ +23 $\!\!\!\!$ & $\!\!\!\!$ 9.66 $\!\!\!\!$ & $\!\!\!\!$ 7800 $\!\!\!\!$ & $\!\!\!\!$ 7306 $\!\!\!\!$ & $\!\!\!\!$ 0.01 $\!\!\!\!$ & $\!\!\!\!$ 0.01 $\!\!\!\!$ & $\!\!\!\!$ 0.00 $\!\!\!\!$ & $\!\!\!\!$ 0.40 $\!\!\!\!$ & $\!\!\!\!$ 0.02 $\!\!\!\!$ & $\!\!\!\!$ 0.0060 $\!\!\!\!$ & $\!\!\!\!$ 0.0001 $\!\!\!\!$ & $\!\!\!\!$ 0.0060 $\!\!\!\!$ & $\!\!\!\!$ 0.0500 $\!\!\!\!$ & $\!\!\!\!$ 0.5000 $\!\!\!\!$ & $\!\!\!\!$  $\!\!\!\!$ & $\!\!\!\!$  \\ 
{} $\!\!\!\!$ & $\!\!\!\!$ +36 $\!\!\!\!$ & $\!\!\!\!$ 9.39 $\!\!\!\!$ & $\!\!\!\!$ 4700 $\!\!\!\!$ & $\!\!\!\!$ 7331 $\!\!\!\!$ & $\!\!\!\!$ 0.00 $\!\!\!\!$ & $\!\!\!\!$ 0.01 $\!\!\!\!$ & $\!\!\!\!$ 0.00 $\!\!\!\!$ & $\!\!\!\!$ 0.05 $\!\!\!\!$ & $\!\!\!\!$ 0.00 $\!\!\!\!$ & $\!\!\!\!$ 0.0010 $\!\!\!\!$ & $\!\!\!\!$ 0.0000 $\!\!\!\!$ & $\!\!\!\!$ 0.0030 $\!\!\!\!$ & $\!\!\!\!$ 0.2300 $\!\!\!\!$ & $\!\!\!\!$ 0.7000 $\!\!\!\!$ & $\!\!\!\!$  $\!\!\!\!$ & $\!\!\!\!$  \\ 
\hline
09dc-exp $\!\!\!\!$ & $\!\!\!\!$ oo $\!\!\!\!$ & $\!\!\!\!$ {-} $\!\!\!\!$ & $\!\!\!\!$ 11700 $\!\!\!\!$ & $\!\!\!\!$ {-} $\!\!\!\!$ & $\!\!\!\!$ 0.04 $\!\!\!\!$ & $\!\!\!\!$ 0.95 $\!\!\!\!$ & $\!\!\!\!$ 0.00 $\!\!\!\!$ & $\!\!\!\!$ 0.00 $\!\!\!\!$ & $\!\!\!\!$ 0.00 $\!\!\!\!$ & $\!\!\!\!$ 0.0000 $\!\!\!\!$ & $\!\!\!\!$ 0.0000 $\!\!\!\!$ & $\!\!\!\!$ 0.0000 $\!\!\!\!$ & $\!\!\!\!$ 0.0000 $\!\!\!\!$ & $\!\!\!\!$ 0.0000 $\!\!\!\!$ & $\!\!\!\!$  $\!\!\!\!$ & $\!\!\!\!$  \\ 
$t_{r}=\textrm{29.0}$\hspace{0.25em}d $\!\!\!\!$ & $\!\!\!\!$ o $\!\!\!\!$ & $\!\!\!\!$ {-} $\!\!\!\!$ & $\!\!\!\!$ 9700 $\!\!\!\!$ & $\!\!\!\!$ {-} $\!\!\!\!$ & $\!\!\!\!$ 0.02 $\!\!\!\!$ & $\!\!\!\!$ 0.74 $\!\!\!\!$ & $\!\!\!\!$ 0.06 $\!\!\!\!$ & $\!\!\!\!$ 0.12 $\!\!\!\!$ & $\!\!\!\!$ 0.06 $\!\!\!\!$ & $\!\!\!\!$ 0.0000 $\!\!\!\!$ & $\!\!\!\!$ 0.0000 $\!\!\!\!$ & $\!\!\!\!$ 0.0001 $\!\!\!\!$ & $\!\!\!\!$ 0.0000 $\!\!\!\!$ & $\!\!\!\!$ 0.0008 $\!\!\!\!$ & $\!\!\!\!$  $\!\!\!\!$ & $\!\!\!\!$  \\ 
{} $\!\!\!\!$ & $\!\!\!\!$ -9 $\!\!\!\!$ & $\!\!\!\!$ 9.90 $\!\!\!\!$ & $\!\!\!\!$ 9100 $\!\!\!\!$ & $\!\!\!\!$ 14319 $\!\!\!\!$ & $\!\!\!\!$ 0.02 $\!\!\!\!$ & $\!\!\!\!$ 0.56 $\!\!\!\!$ & $\!\!\!\!$ 0.06 $\!\!\!\!$ & $\!\!\!\!$ 0.20 $\!\!\!\!$ & $\!\!\!\!$ 0.08 $\!\!\!\!$ & $\!\!\!\!$ 0.0036 $\!\!\!\!$ & $\!\!\!\!$ 0.0008 $\!\!\!\!$ & $\!\!\!\!$ 0.0080 $\!\!\!\!$ & $\!\!\!\!$ 0.0000 $\!\!\!\!$ & $\!\!\!\!$ 0.0750 $\!\!\!\!$ & $\!\!\!\!$  $\!\!\!\!$ & $\!\!\!\!$  \\ 
{} $\!\!\!\!$ & $\!\!\!\!$ -4 $\!\!\!\!$ & $\!\!\!\!$ 10.01 $\!\!\!\!$ & $\!\!\!\!$ 8550 $\!\!\!\!$ & $\!\!\!\!$ 13670 $\!\!\!\!$ & $\!\!\!\!$ 0.02 $\!\!\!\!$ & $\!\!\!\!$ 0.39 $\!\!\!\!$ & $\!\!\!\!$ 0.06 $\!\!\!\!$ & $\!\!\!\!$ 0.35 $\!\!\!\!$ & $\!\!\!\!$ 0.08 $\!\!\!\!$ & $\!\!\!\!$ 0.0060 $\!\!\!\!$ & $\!\!\!\!$ 0.0010 $\!\!\!\!$ & $\!\!\!\!$ 0.0100 $\!\!\!\!$ & $\!\!\!\!$ 0.0000 $\!\!\!\!$ & $\!\!\!\!$ 0.0900 $\!\!\!\!$ & $\!\!\!\!$  $\!\!\!\!$ & $\!\!\!\!$  \\ 
{} $\!\!\!\!$ & $\!\!\!\!$ +5 $\!\!\!\!$ & $\!\!\!\!$ 9.97 $\!\!\!\!$ & $\!\!\!\!$ 7800 $\!\!\!\!$ & $\!\!\!\!$ 11910 $\!\!\!\!$ & $\!\!\!\!$ 0.02 $\!\!\!\!$ & $\!\!\!\!$ 0.25 $\!\!\!\!$ & $\!\!\!\!$ 0.03 $\!\!\!\!$ & $\!\!\!\!$ 0.50 $\!\!\!\!$ & $\!\!\!\!$ 0.06 $\!\!\!\!$ & $\!\!\!\!$ 0.0060 $\!\!\!\!$ & $\!\!\!\!$ 0.0005 $\!\!\!\!$ & $\!\!\!\!$ 0.0100 $\!\!\!\!$ & $\!\!\!\!$ 0.0000 $\!\!\!\!$ & $\!\!\!\!$ 0.1250 $\!\!\!\!$ & $\!\!\!\!$  $\!\!\!\!$ & $\!\!\!\!$  \\ 
{} $\!\!\!\!$ & $\!\!\!\!$ +13 $\!\!\!\!$ & $\!\!\!\!$ 9.82 $\!\!\!\!$ & $\!\!\!\!$ 7100 $\!\!\!\!$ & $\!\!\!\!$ 9969 $\!\!\!\!$ & $\!\!\!\!$ 0.01 $\!\!\!\!$ & $\!\!\!\!$ 0.15 $\!\!\!\!$ & $\!\!\!\!$ 0.02 $\!\!\!\!$ & $\!\!\!\!$ 0.40 $\!\!\!\!$ & $\!\!\!\!$ 0.05 $\!\!\!\!$ & $\!\!\!\!$ 0.0045 $\!\!\!\!$ & $\!\!\!\!$ 0.0003 $\!\!\!\!$ & $\!\!\!\!$ 0.0100 $\!\!\!\!$ & $\!\!\!\!$ 0.0000 $\!\!\!\!$ & $\!\!\!\!$ 0.3500 $\!\!\!\!$ & $\!\!\!\!$  $\!\!\!\!$ & $\!\!\!\!$  \\ 
{} $\!\!\!\!$ & $\!\!\!\!$ +23 $\!\!\!\!$ & $\!\!\!\!$ 9.62 $\!\!\!\!$ & $\!\!\!\!$ 6500 $\!\!\!\!$ & $\!\!\!\!$ 8206 $\!\!\!\!$ & $\!\!\!\!$ 0.00 $\!\!\!\!$ & $\!\!\!\!$ 0.16 $\!\!\!\!$ & $\!\!\!\!$ 0.01 $\!\!\!\!$ & $\!\!\!\!$ 0.25 $\!\!\!\!$ & $\!\!\!\!$ 0.02 $\!\!\!\!$ & $\!\!\!\!$ 0.0015 $\!\!\!\!$ & $\!\!\!\!$ 0.0001 $\!\!\!\!$ & $\!\!\!\!$ 0.0075 $\!\!\!\!$ & $\!\!\!\!$ 0.0250 $\!\!\!\!$ & $\!\!\!\!$ 0.5250 $\!\!\!\!$ & $\!\!\!\!$  $\!\!\!\!$ & $\!\!\!\!$  \\ 
{} $\!\!\!\!$ & $\!\!\!\!$ +36 $\!\!\!\!$ & $\!\!\!\!$ 9.35 $\!\!\!\!$ & $\!\!\!\!$ 3500 $\!\!\!\!$ & $\!\!\!\!$ 9205 $\!\!\!\!$ & $\!\!\!\!$ 0.00 $\!\!\!\!$ & $\!\!\!\!$ 0.04 $\!\!\!\!$ & $\!\!\!\!$ 0.00 $\!\!\!\!$ & $\!\!\!\!$ 0.15 $\!\!\!\!$ & $\!\!\!\!$ 0.01 $\!\!\!\!$ & $\!\!\!\!$ 0.0008 $\!\!\!\!$ & $\!\!\!\!$ 0.0001 $\!\!\!\!$ & $\!\!\!\!$ 0.0050 $\!\!\!\!$ & $\!\!\!\!$ 0.1000 $\!\!\!\!$ & $\!\!\!\!$ 0.7000 $\!\!\!\!$ & $\!\!\!\!$  $\!\!\!\!$ & $\!\!\!\!$  \\ 
\hline
09dc-int $\!\!\!\!$ & $\!\!\!\!$ oo $\!\!\!\!$ & $\!\!\!\!$ {-} $\!\!\!\!$ & $\!\!\!\!$ 11600 $\!\!\!\!$ & $\!\!\!\!$ {-} $\!\!\!\!$ & $\!\!\!\!$ 0.08 $\!\!\!\!$ & $\!\!\!\!$ 0.90 $\!\!\!\!$ & $\!\!\!\!$ 0.01 $\!\!\!\!$ & $\!\!\!\!$ 0.01 $\!\!\!\!$ & $\!\!\!\!$ 0.00 $\!\!\!\!$ & $\!\!\!\!$ 0.0000 $\!\!\!\!$ & $\!\!\!\!$ 0.0000 $\!\!\!\!$ & $\!\!\!\!$ 0.0000 $\!\!\!\!$ & $\!\!\!\!$ 0.0005 $\!\!\!\!$ & $\!\!\!\!$ 0.0001 $\!\!\!\!$ & $\!\!\!\!$  $\!\!\!\!$ & $\!\!\!\!$  \\ 
$t_{r}=\textrm{26.0}$\hspace{0.25em}d $\!\!\!\!$ & $\!\!\!\!$ o $\!\!\!\!$ & $\!\!\!\!$ {-} $\!\!\!\!$ & $\!\!\!\!$ 9200 $\!\!\!\!$ & $\!\!\!\!$ {-} $\!\!\!\!$ & $\!\!\!\!$ 0.03 $\!\!\!\!$ & $\!\!\!\!$ 0.54 $\!\!\!\!$ & $\!\!\!\!$ 0.10 $\!\!\!\!$ & $\!\!\!\!$ 0.25 $\!\!\!\!$ & $\!\!\!\!$ 0.08 $\!\!\!\!$ & $\!\!\!\!$ 0.0010 $\!\!\!\!$ & $\!\!\!\!$ 0.0001 $\!\!\!\!$ & $\!\!\!\!$ 0.0001 $\!\!\!\!$ & $\!\!\!\!$ 0.0015 $\!\!\!\!$ & $\!\!\!\!$ 0.0030 $\!\!\!\!$ & $\!\!\!\!$  $\!\!\!\!$ & $\!\!\!\!$  \\ 
{} $\!\!\!\!$ & $\!\!\!\!$ -9 $\!\!\!\!$ & $\!\!\!\!$ 9.60 $\!\!\!\!$ & $\!\!\!\!$ 8600 $\!\!\!\!$ & $\!\!\!\!$ 14284 $\!\!\!\!$ & $\!\!\!\!$ 0.03 $\!\!\!\!$ & $\!\!\!\!$ 0.16 $\!\!\!\!$ & $\!\!\!\!$ 0.10 $\!\!\!\!$ & $\!\!\!\!$ 0.35 $\!\!\!\!$ & $\!\!\!\!$ 0.15 $\!\!\!\!$ & $\!\!\!\!$ 0.0025 $\!\!\!\!$ & $\!\!\!\!$ 0.0010 $\!\!\!\!$ & $\!\!\!\!$ 0.0100 $\!\!\!\!$ & $\!\!\!\!$ 0.0500 $\!\!\!\!$ & $\!\!\!\!$ 0.1500 $\!\!\!\!$ & $\!\!\!\!$  $\!\!\!\!$ & $\!\!\!\!$  \\ 
{} $\!\!\!\!$ & $\!\!\!\!$ -4 $\!\!\!\!$ & $\!\!\!\!$ 9.79 $\!\!\!\!$ & $\!\!\!\!$ 7650 $\!\!\!\!$ & $\!\!\!\!$ 14475 $\!\!\!\!$ & $\!\!\!\!$ 0.03 $\!\!\!\!$ & $\!\!\!\!$ 0.12 $\!\!\!\!$ & $\!\!\!\!$ 0.04 $\!\!\!\!$ & $\!\!\!\!$ 0.40 $\!\!\!\!$ & $\!\!\!\!$ 0.15 $\!\!\!\!$ & $\!\!\!\!$ 0.0050 $\!\!\!\!$ & $\!\!\!\!$ 0.0010 $\!\!\!\!$ & $\!\!\!\!$ 0.0100 $\!\!\!\!$ & $\!\!\!\!$ 0.0005 $\!\!\!\!$ & $\!\!\!\!$ 0.2500 $\!\!\!\!$ & $\!\!\!\!$  $\!\!\!\!$ & $\!\!\!\!$  \\ 
{} $\!\!\!\!$ & $\!\!\!\!$ +5 $\!\!\!\!$ & $\!\!\!\!$ 9.76 $\!\!\!\!$ & $\!\!\!\!$ 7000 $\!\!\!\!$ & $\!\!\!\!$ 11736 $\!\!\!\!$ & $\!\!\!\!$ 0.03 $\!\!\!\!$ & $\!\!\!\!$ 0.12 $\!\!\!\!$ & $\!\!\!\!$ 0.02 $\!\!\!\!$ & $\!\!\!\!$ 0.33 $\!\!\!\!$ & $\!\!\!\!$ 0.15 $\!\!\!\!$ & $\!\!\!\!$ 0.0050 $\!\!\!\!$ & $\!\!\!\!$ 0.0008 $\!\!\!\!$ & $\!\!\!\!$ 0.0080 $\!\!\!\!$ & $\!\!\!\!$ 0.0005 $\!\!\!\!$ & $\!\!\!\!$ 0.3500 $\!\!\!\!$ & $\!\!\!\!$  $\!\!\!\!$ & $\!\!\!\!$  \\ 
{} $\!\!\!\!$ & $\!\!\!\!$ +13 $\!\!\!\!$ & $\!\!\!\!$ 9.60 $\!\!\!\!$ & $\!\!\!\!$ 6300 $\!\!\!\!$ & $\!\!\!\!$ 9502 $\!\!\!\!$ & $\!\!\!\!$ 0.01 $\!\!\!\!$ & $\!\!\!\!$ 0.09 $\!\!\!\!$ & $\!\!\!\!$ 0.01 $\!\!\!\!$ & $\!\!\!\!$ 0.17 $\!\!\!\!$ & $\!\!\!\!$ 0.05 $\!\!\!\!$ & $\!\!\!\!$ 0.0050 $\!\!\!\!$ & $\!\!\!\!$ 0.0004 $\!\!\!\!$ & $\!\!\!\!$ 0.0070 $\!\!\!\!$ & $\!\!\!\!$ 0.0500 $\!\!\!\!$ & $\!\!\!\!$ 0.6000 $\!\!\!\!$ & $\!\!\!\!$  $\!\!\!\!$ & $\!\!\!\!$  \\ 
{} $\!\!\!\!$ & $\!\!\!\!$ +23 $\!\!\!\!$ & $\!\!\!\!$ 9.44 $\!\!\!\!$ & $\!\!\!\!$ 5400 $\!\!\!\!$ & $\!\!\!\!$ 8164 $\!\!\!\!$ & $\!\!\!\!$ 0.00 $\!\!\!\!$ & $\!\!\!\!$ 0.07 $\!\!\!\!$ & $\!\!\!\!$ 0.01 $\!\!\!\!$ & $\!\!\!\!$ 0.15 $\!\!\!\!$ & $\!\!\!\!$ 0.01 $\!\!\!\!$ & $\!\!\!\!$ 0.0050 $\!\!\!\!$ & $\!\!\!\!$ 0.0002 $\!\!\!\!$ & $\!\!\!\!$ 0.0070 $\!\!\!\!$ & $\!\!\!\!$ 0.1000 $\!\!\!\!$ & $\!\!\!\!$ 0.6500 $\!\!\!\!$ & $\!\!\!\!$  $\!\!\!\!$ & $\!\!\!\!$  \\ 
{} $\!\!\!\!$ & $\!\!\!\!$ +36 $\!\!\!\!$ & $\!\!\!\!$ 9.21 $\!\!\!\!$ & $\!\!\!\!$ 4300 $\!\!\!\!$ & $\!\!\!\!$ 7164 $\!\!\!\!$ & $\!\!\!\!$ 0.00 $\!\!\!\!$ & $\!\!\!\!$ 0.03 $\!\!\!\!$ & $\!\!\!\!$ 0.00 $\!\!\!\!$ & $\!\!\!\!$ 0.10 $\!\!\!\!$ & $\!\!\!\!$ 0.01 $\!\!\!\!$ & $\!\!\!\!$ 0.0050 $\!\!\!\!$ & $\!\!\!\!$ 0.0001 $\!\!\!\!$ & $\!\!\!\!$ 0.0070 $\!\!\!\!$ & $\!\!\!\!$ 0.1500 $\!\!\!\!$ & $\!\!\!\!$ 0.7000 $\!\!\!\!$ & $\!\!\!\!$  $\!\!\!\!$ & $\!\!\!\!$  \\ 
\hline
\multicolumn{17}{l}{\parbox[t]{0.81\textwidth}{${}^{\textrm{a)}}$ The abundances of Fe, Co and Ni in our models are assumed to be the sum of \Nifs\ and the elements produced in the decay chain (\Cofs\ and \Fefs) on the one hand, and non-radioactive, directly synthesised Fe (including also progenitor Fe) on the other hand. Thus, they are conveniently given in terms of the \Nifs\ mass fraction at $t=\textrm{0}$ [$X($\Nifs$)_0$], the abundance of stable Fe at $t=\textrm{0}$ [$X(\textrm{Fe})_0$], and the time from explosion onset $t$.}}\\
\end{tabular} 
\end{table*}

\end{document}